**High-rank connectivity scaffolds support precision and generalisation in recurrent neural networks**

**Authors:** Ian Hawes[1,4], Matthew F. Nolan[1,2,3*]

[1]Institute for Neuroscience and Cardiovascular Research, [2]Simons Initiative for the Developing Brain, [3]Centre for Statistics, [4]Institute for Machine Learning, University of Edinburgh, Edinburgh, United Kingdom. * Corresponding author

## Abstract

The functions of higher-rank connectivity in recurrent neural networks (RNNs) remain poorly understood. Previous work has shown that low-rank connectivity can generate low-dimensional dynamics in trained RNNs, but whether higher-rank structure makes distinct computational contributions is unknown. Here we analyse RNNs trained to locate rewards by integrating continuously varying speed inputs in one- and two-dimensional spatial tasks. We find that dominant low-dimensional dynamics encode task locations, and can be causally manipulated to instruct behavioural outcomes. However, after decomposing the underlying circuitry we found that while low-rank connectivity accounts for the low-dimensional dynamics, accurate performance and generalisation to novel speed distributions requires high-rank connectivity. We demonstrate that this is achieved through distributed signalling that corrects errors in low-dimensional location representations. Thus, combined perturbation-, representation-, and circuit-level analyses demonstrate a novel mechanism for robust spatial computation and show how high-rank connectivity in neural circuits can provide a scaffold that supports precision and generalisation.

## Introduction

Low-dimensional structure provides a tractable route to mechanistic explanations for neural computations. In trained artificial and biological recurrent neural networks (RNNs), population dynamics often evolve along low-dimensional trajectories aligned with task-relevant variables[1–8]. This observation has motivated a mechanistic framework in which structured low-rank connectivity generates low-dimensional dynamics that implement computation[9–15]. This body of work has shown how mixed selectivity at the level of individual units can coexist with interpretable population-level dynamics[1,16–19], and has provided tractable accounts of memory, decision-making and flexible sensorimotor behaviour[9–15]. Low-rank recurrent structure has therefore become a central framework for connecting connectivity, dynamics and computation (Fig. 1a).

What remains less clear is the role of higher-rank connectivity, and whether any such role generalises beyond specific task settings. Biological and artificial task-optimised recurrent networks are not purely low-rank, yet the remaining higher-rank structure is often treated as residual complexity on which the dominant low-dimensional dynamics unfold[9–15]. This is a reasonable approximation when the goal is to explain the main task manifold, and in many relatively simple supervised tasks involving categorisation, discrete action selection, and timing, low-rank descriptions appear sufficient to capture the dominant computation[9–15]. However, this leaves unresolved whether higher-rank connectivity also makes distinct functional contributions in more complex tasks, especially when performance depends not only on generating the correct latent dynamics, but also on maintaining accuracy under variable inputs and generalising beyond the training distribution.

We address this question in spatial integration tasks that require path integration[20,21], where a network estimates location without external cues by integrating velocity-related signals over time (Fig. 1b)[22–28]. Path integration is central to spatial navigation, but it is also representative of a broader class of computations in which small integration errors can accumulate and compromise behaviour, including perceptual estimation[29–31], evidence accumulation[32,33], oculomotor control[34–37], and motor timing[38]. In mammals, successful path integration depends on the medial entorhinal cortex (MEC)[20,22,23,39,40], where neural activity includes grid-cell and ramping representations of position[20,41–47]. Related normative and task-optimised network models have shown that training on navigation and integration problems can give rise to grid-like, ramp-like and other MEC-relevant activity profiles without such tuning being imposed by hand[46,48–53]. These studies have explained how low-dimensional latent structure can support navigation-related computation, but mechanisms that support accuracy and generalisation under changing input statistics remain unclear.

Here we ask whether the recurrent structure that explains the dominant low-dimensional dynamics of a continuous integration task is also sufficient to explain accuracy in the face of noisy inputs and generalisation in response to novel input distributions. We show that trained RNNs solving one- and two-dimensional path integration tasks are organised by a core/scaffold decomposition of recurrent connectivity. In both settings, the dominant activity dynamics are low-dimensional, but accurate localisation and generalisation to speed distributions not encountered during training depend on additional higher-rank connectivity modes that provide distributed error correction signals. Our results assign a specific computational role to higher-rank recurrent connectivity, arguing against the view that it is simply residual complexity around an interpretable low-dimensional mechanism. More broadly, by combining causal perturbations with analyses of representations and connectivity, our study reconciles population-level and circuit-level views of computation and identifies a recurrent circuit architecture that supports interpretable low-dimensional dynamics and accurate, generalisable behaviour.

## Results

### RNNs learn to solve location memory tasks using path integration and cue-based strategies

We evaluated agents in one- and two dimensional tasks. In the 1D task, the most efficient strategy to maximise reward is to move along a linear track to a reward zone, stop to trigger a reward, and then move to the end of the track to initiate a new trial[22] (Fig. 1c). If cues marking the rewarded area are available, then stop decisions can be made by learning the association between the cue and receipt of a reward. In the absence of cues, the best strategy is to track location by path integration, and then compare the current location estimate with a learned estimate of the reward location. Mice solve this task using both strategies and learning the task requires neurons in the MEC[22]. The 2D task requires navigating around more complex geometry, but the most efficient strategy is again to move directly to the reward location and then navigate to the end point (Fig. 1d).

For conceptual clarity we focus initially on the 1D task, which we pose as a reinforcement learning problem, with an RNN-based agent's inputs and actions mapping onto those available to mice performing the experimental task (Supplementary Fig. 1). Thus, the agent receives scalar observations about movement speed, along with binary signals for whether the current location is on the track or in the black box, whether the current location is in the reward zone and whether a reward has been delivered. The reward zone signal was unavailable on 50% of the trials to simulate experimental 'uncued' trials. On each time step the agent chooses between stop and go actions. Go actions cause the agent to advance, with movement speed drawn from a rectified Gaussian distribution.

When trained using the proximal policy optimisation algorithm[54] agents learned to locate the reward zone in the presence or absence of cues (Supplementary Fig. 1). Accurate reward localisation was maintained on trials with speed distributions not experienced during training, demonstrating an ability to generalise and confirming that agents had learned a path integration strategy rather than a timing strategy (Extended Data Fig. 1). We focus here on a representative agent but evaluate results on 162 models that vary in their architecture (Supplementary Table 1; Supplementary Fig. 2).

**Distributed low-dimensional task representations emerge with learning**

We asked if dynamics relevant to solving the location memory task are captured by the principal components (PCs) of the RNN's activity[2,4,6,55]. During learning, the manifold defined by the first 2 PCs developed a looping structure that mapped onto track locations (Fig. 2a; Extended Data Fig. 2; Extended Data Fig. 3a). Information about movement speed, which is provided as an input to the network, was represented throughout training, but the ordering of the speed-related PC changed with learning (Fig. 2b). The learned spatial manifold defined by the first two PCs was similar on cued and uncued trials (Extended Data Fig. 3b), and on trials with different speed distributions (Supplementary Fig. 3), and was sufficient to accurately decode the agent's location (Extended Data Fig. 3c), indicating that a common location-tracking mechanism is engaged across trial types and movement speeds. Similar low-dimensional representations were observed across model seeds and hyperparameters (Supplementary Fig. 4; Supplementary Fig. 5) and using non-linear dimensionality reduction (Supplementary Fig. 6). Consistent with the network generating distributed representations, most PCs were influenced by many neurons and most neurons contributed to multiple PCs (Fig. 2c), with individual neurons typically exhibiting mixed-selective tuning to task variables such as location and speed. Neurons contributing to the top two PCs typically had ramp-like firing profiles that were oriented either towards or away from the reward zone, and that varied substantially in slope and onset of activation (Fig. 2c). Fixed-point analysis[56] showed that trained RNNs used action-dependent attractor dynamics, with a point attractor located ahead of the represented position during movement, and a line attractor that preserved location during unrewarded stops (Extended Data Fig. 3d; Supplementary Fig. 7).

These analyses are consistent with the view that low dimensional dynamics in RNNs provide a substrate for computations[2,9]. However, they are correlative and it's unclear whether the low dimensional representations are sufficient to explain the underlying computations[57]. To address this we implemented methods to more directly test computational roles of the putative latent variables.

**Manipulating low-dimensional latent variables predictably directs behaviour**

To evaluate causal roles of low-dimensional representations we developed a manipulation that we refer to as a low-dimensional activity clamp (LDAC), which uses the invertibility of PCA[58] to project a vector of PC values back into the original high-dimensional space (Fig. 2d). By clamping higher PCs we found that the first 4 PCs are sufficient to track location but with impaired behavioural performance (Fig. 2e) and mild degradation in decoding accuracy (Fig. 2f). Full task performance (>99% cued & uncued reward) required the first 15 PCs (Fig. 2e; Extended Data Fig. 4). When tested on novel speed distributions, > 35 PCs were required for >99% reward (Extended Data Fig. 4). By clamping individual PCs, we found that PCs 1, 2, and 4 were required for task performance and location decoding (Fig. 2g), whereas PC3 only appeared important when the movement speed distribution differed from the training distribution (Fig. 2g,h; Supplementary Fig. 8). Clamping higher PCs minimally affected location decoding (Fig. 2h), but could impact other aspects of computations that generate correct behaviour. For example, the 6th PC explains 4.1% of variance yet clamping it reduces uncued performance by 93%. This was not a trivial consequence of driving the network outside its normal operating regime, as the activity trajectory remained confined to the original manifold and single-neuron firing rates were within the same range as the original firing rates, whereas non-specifically increasing all recurrent weights by 1.1x impaired behaviour (12.8% reward) and drove activity far outside the manifold (Supplementary Fig. 9). The consequences of LDAC manipulations were similar for models with different seeds and hyperparameters (Extended Data Fig. 4).

To directly test whether spatial representation by PC1 and PC2 is used to direct behaviour, we asked if clamping them at values corresponding to particular regions of the manifold would produce behaviours appropriate to the corresponding track location (Fig. 2i). We found that on simulated trials in which PC1 and 2 were clamped to manifold locations that map to track locations far from the reward zone, the agent did not stop anywhere on the track (Fig. 2j i-iii, k). By contrast, when PC1 and 2 were clamped to manifold positions corresponding to track locations just in advance of or at the reward zone, the agent stopped at all track locations (Fig. 2j iv). Surprisingly, clamping the manifold impacted performance on cued as well as uncued trials (Fig. 2j-k), with cue-dependent stopping occurring only when the manifold was clamped at locations close to the reward zone (Fig. 2k), indicating that the cue signal does not fully override the influence of the internally generated location representation.

Together, these results establish the LDAC as a way to test roles for low-dimensional dynamics and establish that the leading PCs causally directed stopping behaviour. However, although the manifold captured task-relevant activity, analyses confined to this representation missed key aspects of the underlying computation. Neuron-level manipulations showed that PC loadings alone were insufficient to identify influential neurons (Extended Data Fig. 5; Extended Data Fig. 6; Extended Data Fig. 7; Supplementary Fig. 10, Supplementary Fig. 11; Supplementary Fig. 12; Supplementary Fig. 13). Moreover manipulating small (<2%) neural populations selected according to their effects on behaviour predictably shifted the agent's stopping location, whereas larger populations were required when neurons were selected by PC loading alone (Extended Data Fig. 5; Extended Data Fig. 6; Supplementary Fig. 14; Supplementary Fig. 15; Supplementary Fig. 16; Supplementary Fig. 17; Supplementary Fig. 18). Thus, although the manifold captured task-relevant activity, analyses confined to this representation missed key aspects of the underlying computation, implying that manifold-level explanations are incomplete.

**High-rank connectivity is necessary for generalisation to novel speed distributions**

Given analyses of other task-optimised RNNs[9–11,14,59,60], we hypothesised that low-rank structure would account for learned signal integration and generalisation capabilities of the trained RNNs. We used singular value decomposition to write the recurrent connectivity matrix of trained networks as the sum of unit-rank matrices (referred to as modes) ordered by decreasing singular value[60,61]. Throughout we refer to modes with larger singular values as 'leading' modes and those with smaller singular values as 'later' modes. Although the recurrent weights in trained networks were full rank, the distribution of singular values was highly skewed, with the first two singular values being on average 5.2-fold larger than the mean of the remaining values (Fig. 3b). The leading modes revealed strong connectivity between groups of neurons with similar displacement scores, consistent with these modes having captured functional organisation of the circuit (Fig. 3a; Extended Data Fig. 8; Supplementary Fig. 19; Supplementary Fig. 20; Supplementary Fig. 21). We obtained a similar structure when intra-mode connectivity was evaluated relative to the location correlation of neural activity or by contributions to PC2 (Supplementary Fig. 21; Supplementary Fig. 22; Supplementary Fig. 23).

If computations are implemented through the low rank structure then leading modes should encode task-relevant information. Projecting network activity onto each mode's left singular vector, which defines the axis along which that mode can drive activity, yielded a scalar value that we refer to as the network's activity in that mode (Supplementary Fig. 24). Activity in the first two modes was strongly correlated with location, but contained almost no speed information (Fig. 3c,d; Supplementary Fig. 25). When we computed the similarity between each mode's left singular vector and the PC axes of population activity, the top two left singular vectors closely matched the first two PC axes, suggesting that activity in PCs 1 and 2 is largely generated by connectivity modes 1 and 2 (Fig. 3e). By contrast, for modes ranked >2, location information was lower while speed information could be substantially higher (Fig. 3c, d; Supplementary Fig. 26). Thus, modes 1 and 2 provide core connectivity supporting spatial representation, while modes >2 support encoding of varying degrees of location and speed information.

To test whether the leading modes alone are sufficient for network computations, we evaluated network approximations in which the recurrent weight matrix was reconstructed from the first k modes. For networks simulated with the distribution of speed inputs used during training, a k = 5 approximation was sufficient to obtain >75% reward (Fig. 3f), for decoding of location from PC1 and PC2 with $R^2$ >0.65 (Fig. 3g), and for the agent's first stops on uncued trials to localise to the reward zone (Fig. 3h, Extended Data Fig. 9). The k = 5 network nevertheless showed excessive stopping after the reward zone (Extended Data Fig. 9), along with reduced reward and somewhat impaired location encoding, and k ≥224 was necessary for >99% reward. When we evaluated performance on trials with novel speed distributions, low rank networks continued to perform well on cued trials, but were substantially impaired on uncued trials. For example, when the speed distribution was 0.5 x the standard distribution, networks with k < 128 no longer obtained rewards on uncued trials (Fig. 3f), had impaired representation of location by PC1 and PC2 (Fig. 3g), and either didn't stop at all or tended to overshoot the reward zone (Fig. 3h). When the speed distribution was 2 x the standard distribution, behavioural performance of k = 8 networks on uncued trials appeared better relative to the standard speed distribution condition. However, lower rank networks in the 2 x speed condition nevertheless systematically under-estimated the first stop location, requiring rank > 128 for first stop accuracy comparable to full rank networks (Fig. 3h; Extended Data Fig. 9; Supplementary Fig. 27).

Together, these analyses show that while low-rank (k = 5) connectivity is sufficient for path integration under the speed distribution used for training, which is consistent with our initial hypothesis, their performance was nevertheless noisy and high-rank structure was required for accurate behavioural localisation (k >224) and generalisation to speed input distributions differing from that experienced during training (k >128).

**Later weak connectivity modes correct systematic path integration errors**

Because the impairment in performance of low compared to full rank networks was greatest for movement speeds outside the training distribution, we hypothesised that later modes provide corrective signals that reduce speed-dependent errors when inputs deviate from the training mean. To test this we simulated agent movement at a constant speed and then measured how activity in modes 1 and 2 changed in response to a single-step perturbation in movement speed (Fig. 4a).

In contrast to the performance of an ideal path integrator, which should be robust to speed perturbations, when the speed signal changed in full rank networks the activity in modes 1 and 2 deviated from the values expected if the network was accurately representing location (Fig. 4b, see Supplementary Fig. 28 for mode 1 and additional locations). Qualitatively, this deviation manifested, depending on the track location, as either an overshoot or undershoot in the location representation by activity in modes 1 and 2 (Fig. 4b). While this instantaneous error could be expected to rapidly impair performance of the agent, for full rank networks the error was corrected on the subsequent timestep, with the activity in modes 1 and 2 returning to their expected values (Fig. 4c,d; Extended Data Fig. 10). Thus, the recurrent network generates instantaneous path integration errors but then corrects for them on subsequent time steps.

We reasoned that the correction mechanisms must involve later connectivity modes. To test this we repeated simulations but with the network rank reduced before the simulation step on which the correction occurred. With k = 2, activity in modes 1 and 2 did not return to the expected trajectory (Fig. 4b; Extended Data Fig. 10; Supplementary Fig. 28). Instead, representations either overshot or undershot relative to the expected values for accurate correction, with the direction and degree of the error depending on track location (Fig. 4c; Extended Data Fig. 10). The extent of the recovery error was reduced for networks built from the first 4 connectivity modes, while > 224 modes were required to obtain performance comparable to full rank networks (Extended Data Fig. 10). Thus, while the leading modes implement a low-dimensional location tracking manifold, the later weaker modes provide speed and location modulated signals that correct speed-dependent errors in the low dimensional location estimate. The correction mechanism becomes especially important when successful behaviour requires generalisation to distributions of speed inputs not experienced during training.

**Mode to mode connectivity underlying rescue of systematic location error**

This correction mechanism implies a relationship between the leading modes and the later modes, with later modes being able to influence the update of the leading modes at the next timestep using information about recent speed and current location. To investigate this, we examined the recurrent computation in terms of the modes rather than neurons, by expressing the recurrent weights in the coordinate system defined by the left singular vectors (Fig. 5a). In this coordinate system, the recurrent update can be expressed as interactions between modes and their activity levels (Fig. 5b), yielding a mode-to-mode connectivity matrix that summarises how activity in one mode contributes to changes in other modes.

Two features of the mode-to-mode connectivity were aligned with the correction mechanism. First, strong self-recurrence was present in the first two modes, consistent with these modes having a location-tracking role. Second, inputs from later modes onto these two modes were weak and distributed, consistent with a distributed aggregate corrective influence from the later modes. To test whether these connections were causally meaningful, we selectively perturbed them in mode space and transformed the weights into neuron space for simulation. Removing self-recurrence within mode 1 or 2 reduced uncued reward to 0% (Fig. 5c), confirming that self-recurrence of the leading modes was required. By contrast, removing self-recurrence from later modes 4 onwards had no effect on performance. Finally, removing inputs from later modes onto mode 1 or 2 rapidly reduced uncued reward to 0% (Fig. 5d; Supplementary Fig. 29), whereas removing inputs onto later modes had minimal impact (uncued reward > 99%). Together, these results show that accurate path integration requires strong self-recurrence within the location-tracking modes, and requires distributed inputs from later modes to modes 1 and 2, consistent with later modes providing distributed corrective signals required to correct location updates.

**High-rank connectivity is required for precision and generalisation in a 2D task**

To test whether the importance of higher-rank connectivity generalised beyond the linear-track task, we trained agents on a context-dependent 2D spatial memory task (Figs. 1d, 6). On each trial, agents navigated from a fixed start location, stopped to obtain reward, and then moved to an end location. With training, diffuse trajectories and widespread stopping were replaced by direct routes and selective stopping at the reward zone (Fig. 6a). As in the 1D task, the RNN's activity was low-dimensional, with the first two PCs accounting for 66% of the variance and forming a manifold that encoded location (Fig. 6b). The learned recurrent connectivity was full rank but had a strongly skewed singular-value spectrum (Fig. 6c).

To determine how connectivity rank influenced navigation, we reconstructed the recurrent matrix using increasing numbers of leading modes. At rank 16, cued reward was already maximal, whereas uncued reward was only partially recovered; both reached maximal values by rank 64 (Fig. 6d). However, these networks did not reproduce the full navigation computation. At rank 64, location decoding remained below an $R^2$ of 0.7 and navigation efficiency was substantially reduced, particularly on uncued trials, with both measures continuing to improve as later modes were restored (Fig. 6d). Thus, relatively low-rank approximations could support reward acquisition, but accurate spatial representation and efficient memory-guided navigation depended on higher-rank connectivity.

To test whether later modes provided the same error-correcting function identified in the 1D task, we repeated the single-step speed perturbation analysis. In the full-rank network, mode 1 activity at the following timestep remained close to the value expected for each location, resulting in relatively small errors (< 0.2 activity units) across movement speeds (Fig. 6e,f; Supplementary Fig. 30). When recurrent connectivity was reduced to rank 2 before this update, the mode 1 projection deviated from the expected value and produced large systematic location- and speed-dependent errors (> 1.0 activity units). Together, these results extend the mechanism identified in the linear-track task, with leading modes supporting a low-dimensional location estimate, while distributed later modes correct its update and enable accurate navigation in a more complex environment.

**Discussion**

Low-rank connectivity has provided a compelling framework for linking recurrent connectivity to low-dimensional neural dynamics and task computation[9–15]. Here we show that, for 1D and 2D continuous integration tasks that demand precision and generalisation, this framework is not by itself sufficient. In trained RNNs, a low-rank recurrent core generated the dominant task manifold, but accurate localisation and robustness to novel input statistics required additional higher-rank connectivity modes carrying distributed correction signals. This core/scaffold decomposition at the level of connectivity modes assigns a distinct computational role to higher-rank recurrent structure and argues against treating it as residual complexity around an otherwise interpretable low-rank mechanism[9,13,15,61–63]. More generally, our results indicate that the circuit features that generate dominant latent dynamics need not be identical to those that mediate accuracy and adaptability, providing a route to reconcile low-dimensional population-level explanations with the heterogeneous connectivity of real neural circuits.

Our results also clarify relationships between low-dimensional population dynamics and heterogeneous single-neuron influence. The dominant latent dynamics were causally important, as shown by low-dimensional activity clamp (Fig. 2), yet neurons with similar firing-rate profiles or similar contributions to the leading principal components could differ markedly in their effects on behaviour (Extended Data Fig. 5). Small targeted neuronal perturbations were sufficient to predictably alter localisation, indicating concentrated causal leverage within an otherwise distributed code. Identifying the dominant low-dimensional manifold was therefore necessary but not sufficient for identifying the circuit elements that most strongly influenced performance. In this respect, our study extends prior low-rank accounts of neural computation from simpler tasks and constrained architectures to unconstrained reinforcement-learning-trained networks performing continual integration and action selection[9–13,61,64–67].

For machine learning, our results show that interpretable low-dimensional computation and robust generalisation can be explained by different connectivity modes coexisting within the same recurrent network (Figs. 3-5). The low-rank core and high-rank scaffold are implemented by overlapping neurons and synapses, but are distinct in the mode space obtained from the singular-value decomposition of the recurrent connectivity. This division of labour implies that analyses based only on dominant latent activity or low-rank connectivity truncations may capture the main computation while missing the mechanisms that support robustness under novel input statistics. More generally, the findings suggest that higher-order connectivity structure in recurrent networks (e.g. [68–71]) may be functionally important even when it is distributed over the same neurons and connections that implement the dominant low-dimensional dynamics.

For neuroscience, these findings suggest a way to reconcile low-dimensional population dynamics with the dense and heterogeneous connectivity of real circuits[72–78]. In areas such as medial entorhinal cortex, behaviourally relevant activity can often be described by low-dimensional latent structure[47,79–85], yet the underlying circuitry appears unlikely to be organised as a simple low-rank network. Our results raise the possibility that these computational roles are separated in connectivity mode space rather than in anatomically distinct cell groups. In our networks computational roles were superimposed across the same population, so the same neurons and synapses could participate both in the dominant low-dimensional dynamics and in higher-order corrective interactions. This view implies that low-dimensional activity does not necessarily reflect a low-complexity circuit, and suggests that experimentally observed neural heterogeneity may contribute to robustness even when it does not strongly alter the dominant latent dynamics.

Several limitations and open questions remain. Although the consistency of the results across 1D and 2D tasks, and across model parameters, argues that the core/scaffold decomposition is not specific to a single task, it will be important to determine whether similar mechanisms emerge in other continuous estimation problems[29–38] and how they manifest in more biologically constrained network models[86–91]. It also remains unclear why reinforcement learning converges to solutions with superimposed low- and high-rank structure. Finally, establishing whether analogous principles operate in biological circuits will require experiments that move beyond observational latent-space analyses to selectively perturb recurrent interactions contributing to dominant and corrective connectivity modes[92–98]. Such experiments will be important for determining whether higher-rank recurrent structure contributes to behavioural precision and generalisation in biological circuits as it does in trained networks.

## Main figures

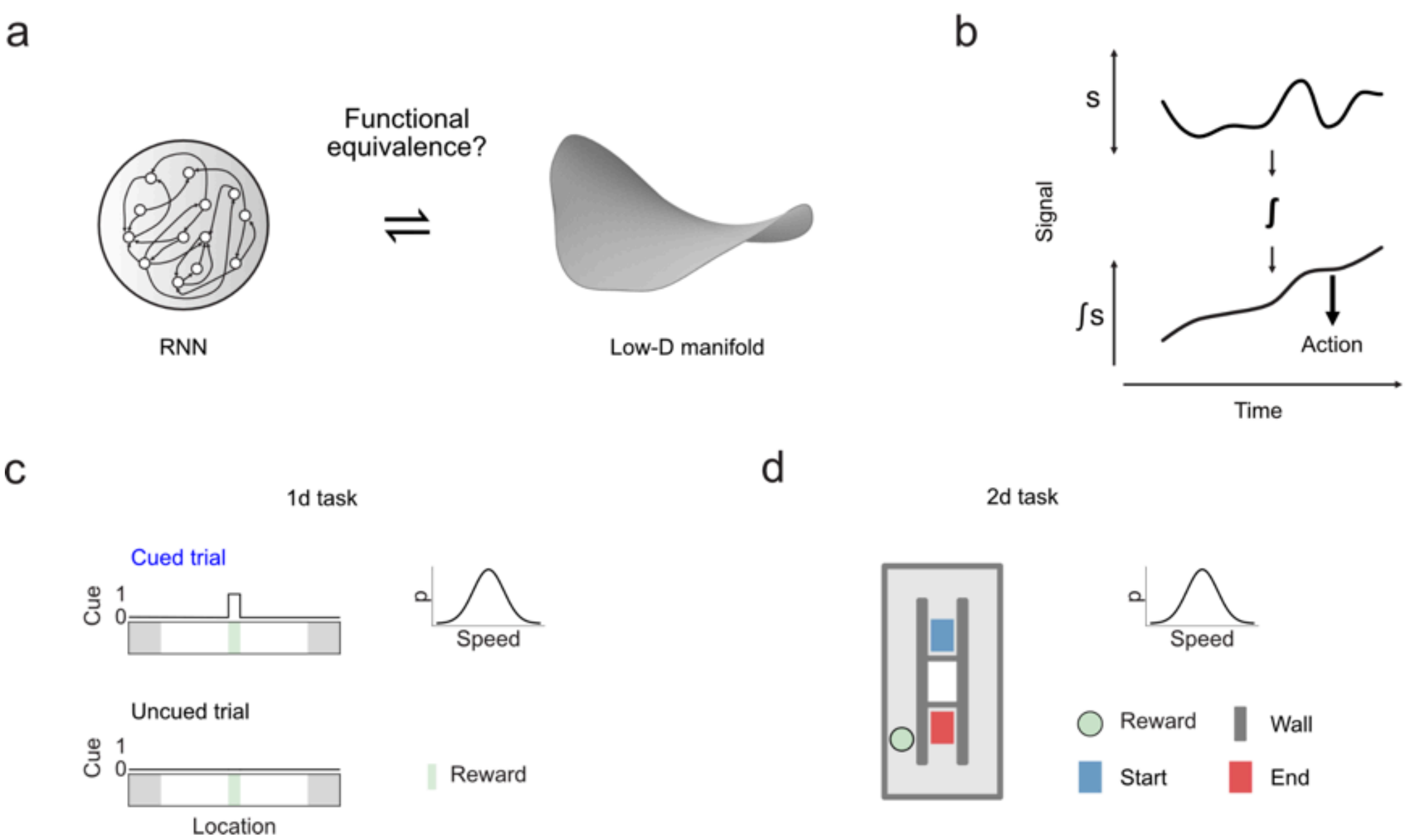

**Fig. 1: Framework for investigating path integration-dependent location estimation**. **a**, Relationship between an RNN and its low-dimensional activity manifold. **b,** Actions are selected when the integral ('∫s') of a continuously varying signal ('s') reaches a learned value ('Action'). **c**, A spatial memory task in which agents obtain rewards for stopping within a target zone on a linear track. On cued trials, the cue signal is set to '1' when the agent is in the reward zone. On uncued trials the cue signal is always 0. Movement speed is drawn from a Gaussian distribution. **d**, Two dimensional variant of the spatial memory task. Agents can navigate by choosing one of eight compass directions.

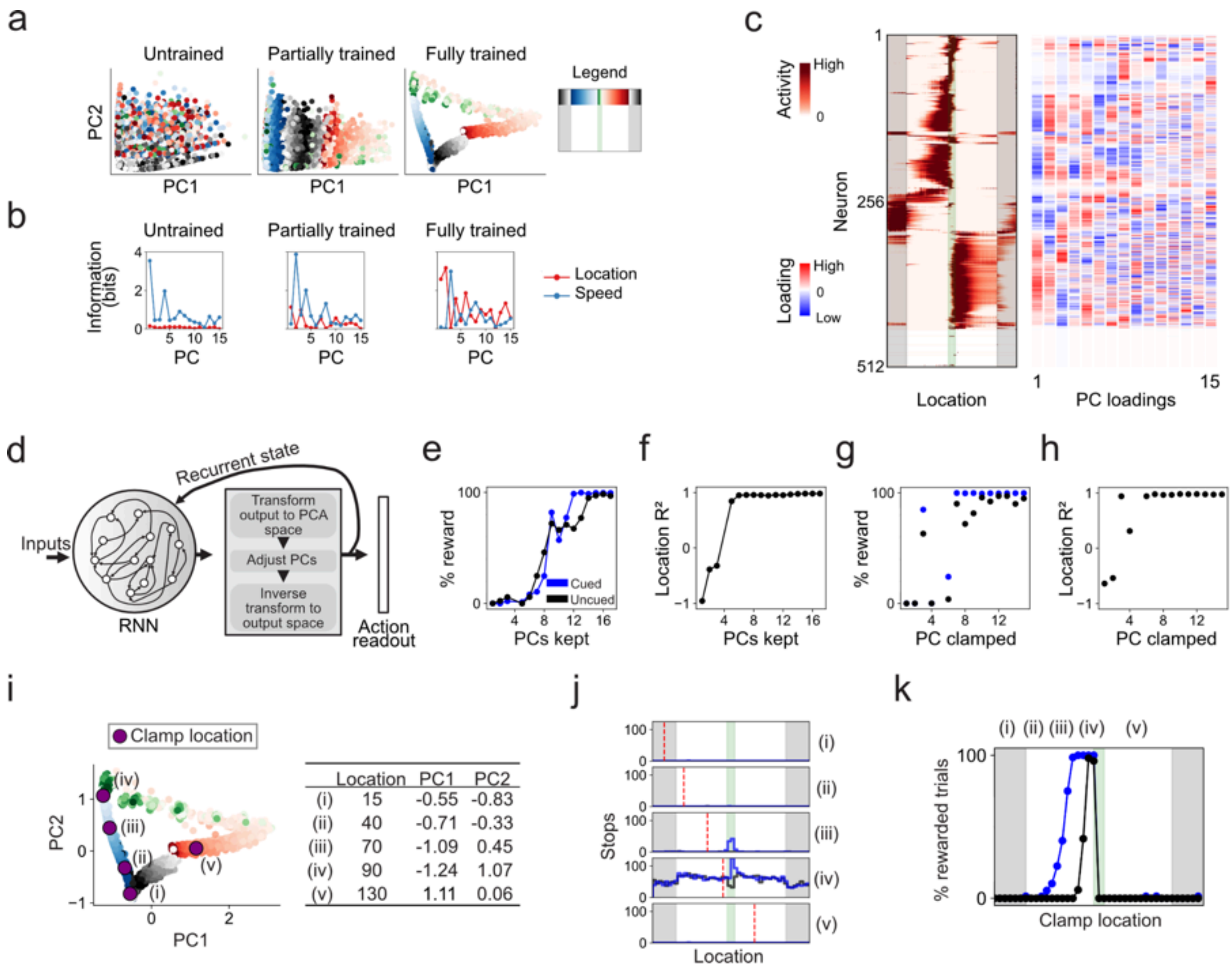

**Fig. 2: Clamping the first two principal components predictably directs agent behaviour. a,** RNN activity at three training stages plotted in PCA space and coloured by track location (PCA fit separately for each phase). **b**, Location and speed information content of first 15 PCs across training. **c**, Heatmap of trial-averaged neuron activity ordered by Rastermap fit (left) [99], and the corresponding PC loadings for each neuron (right). **d**, Low Dimensional Activity Clamp procedure. At each timestep the RNN activity is projected into PC space, the PCs are modified and then the PCs are transformed back into the RNN space by transposing the loadings matrix and undoing the mean subtraction. **e-f**, Reward (e) and location decoding performance (f) as a function of the number of PCs retained ('PCs kept'). Higher PCs were clamped to 0. **g-h** Reward (g) and location decoding performance (h) when individual PCs were clamped to zero. Zeroing component 5 caused the simulation to fail and so this datapoint is not shown. **i**, Average PC1 and PC2 values at manifold locations corresponding to selected regions on the track (i-v). The manifold-location of selected regions is indicated by purple dots. **j**, Stop histograms when PC1 and PC2 were clamped at values corresponding to the regions labelled in (i) indicated here by the dotted red lines. **k**, Reward received as a function of clamp location for 100 locations.

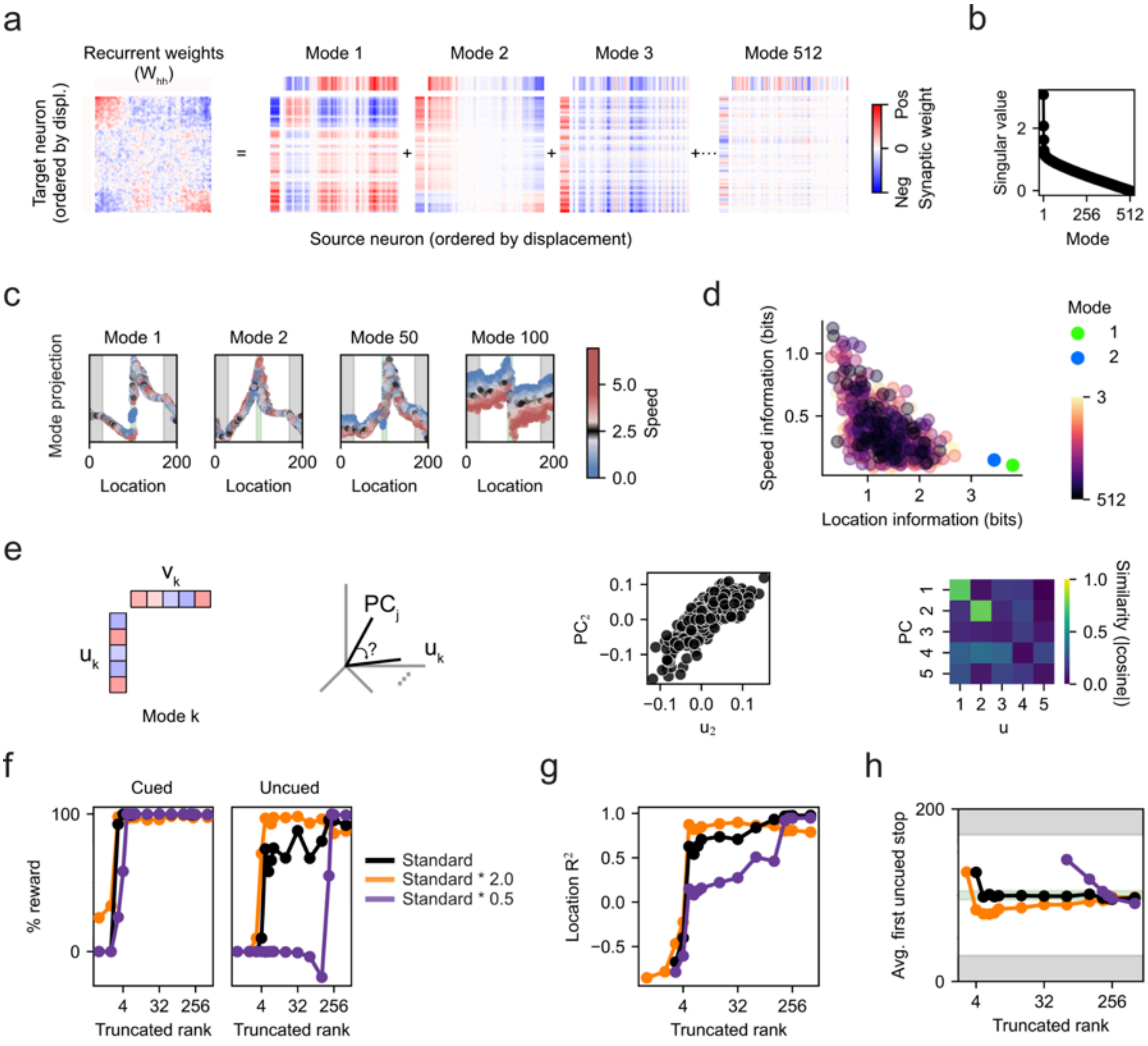

**Fig. 3: High-rank weights are required for generalisation to novel input distributions. a**, Singular value decomposition of the full-rank recurrent weights matrix into a sum of unit-rank modes equal in number to the number of neurons. Each mode is the outer product of the i-th left and right singular vectors, scaled by the i-th singular value. Selected mode matrices and their corresponding left and right singular vectors are shown. Neuron entries in the vectors or matrices are ordered by their displacement score. **b**, Singular values for each mode. **c**, Activity computed by projecting a control simulation's neural activity along the u-vectors for the selected mode and plotting the resulting scalar value at each timestep. **d**, Mutual information of location and speed for each mode. **e**, Similarity between left singular vectors of the recurrent weights (u) and activity PC vectors. Example neuron-wise coefficients for $u_2$ and $PC_2$ are shown in the scatterplot. Cosine similarity between the first 5 PCs and first 5 u-vectors is shown in the heatmap. **f**, Uncued and cued reward as a function of the number of modes retained. **g-h**, Location decoding accuracy from PCs 1 and 2 (g) and location of the first stop on uncued trials (h) as a function of low-rank truncation.

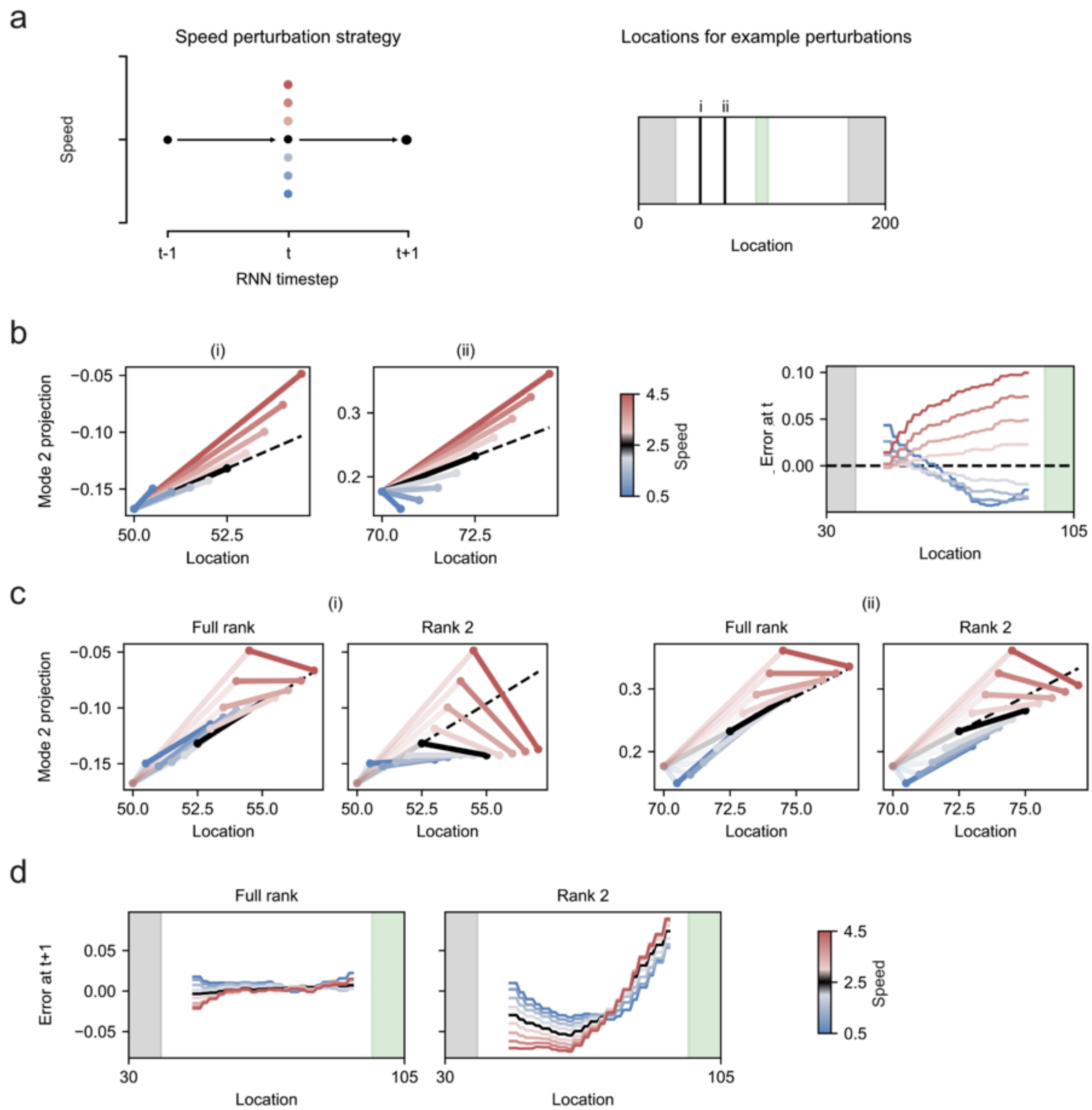

**Fig. 4: Higher rank connectivity modes correct instantaneous location estimation error. a,** Framework for investigating location estimation error. **b**, The activity change in mode 2 resulting from applying speed inputs from a shared initial state corresponding to the network activity at locations i and ii (indicated in a) (left). The difference between the expected projection in the absence of error (extrapolated from the average speed) and the actual projection as a function of the track location on which the manipulation was applied. **c**, The projected activity value at t+1 for mode 2 as a function of track location for the full rank network and for networks that at time t was reconfigured to rank 2. **d**,

Error (calculated as in b) as a function of track location for the full rank network (left) and network that was configured to rank 2 at t.

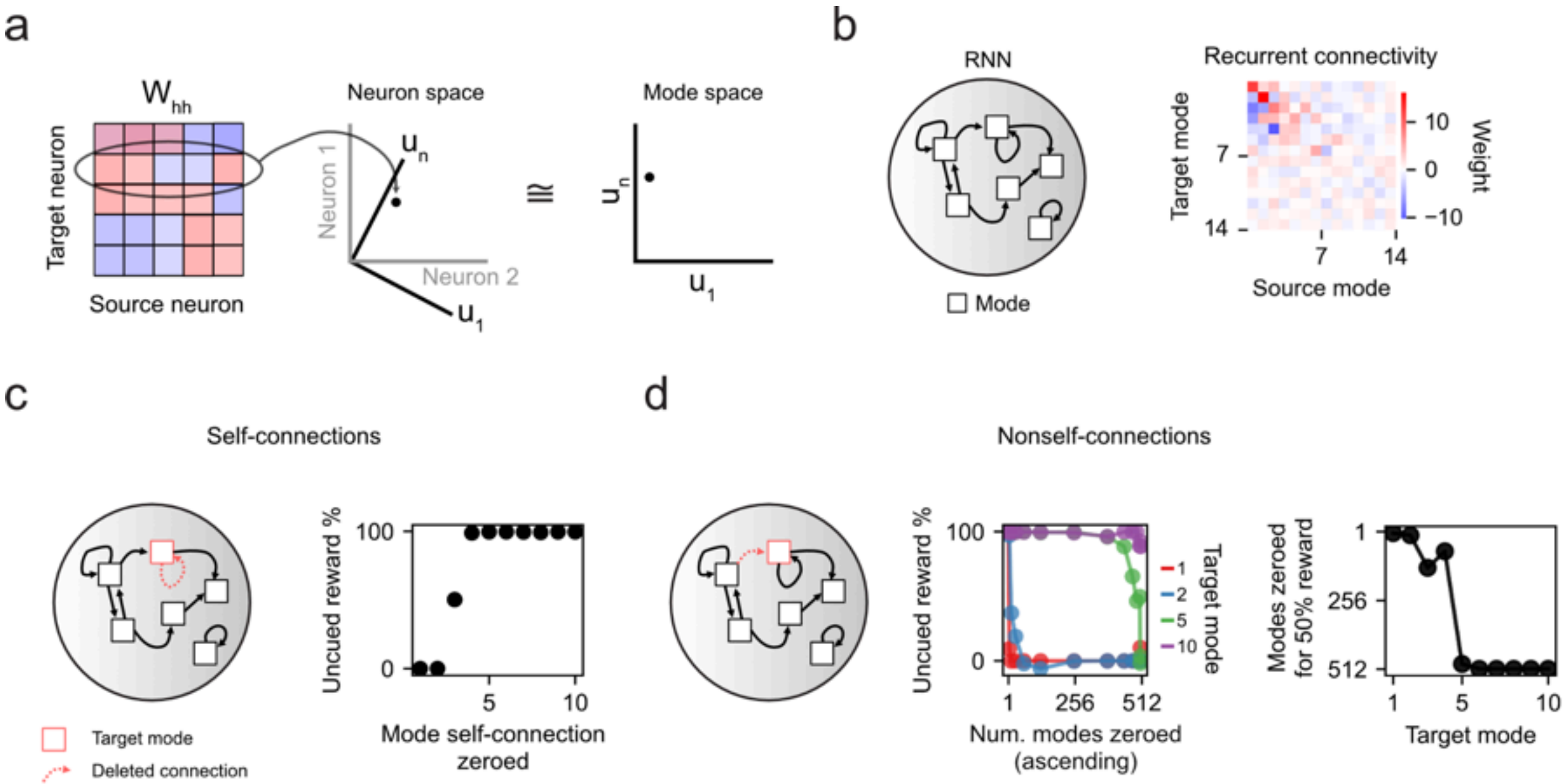

**Fig. 5: Mode to mode connectivity underlies correction of systematic location error. a**, Transformation of the recurrent weight matrix ($W_{hh}$) into mode space. All u-vectors were stacked into a matrix (M). The matrix change of basis operation $M^T$ $W_{hh}$ M produces a matrix whose rows and columns are in mode-space rather than neuron-space. **b**, Schematic of recurrent connectivity in mode space (left) and weights for inter-mode connections (right). **c**, Reward on uncued trials with the Standard * 0.5 speed distribution when the self-connection (diagonal entries in b) for each of the top 10 modes was set to zero. **d**, Reward on uncued trials (centre) under the Standard * 0.5 speed distribution as a function of the number of targeted mode's inputs that are removed (sorted by ascending singular value). The number of mode inputs zeroed to reduce the uncued reward to 50% for different target modes (right).

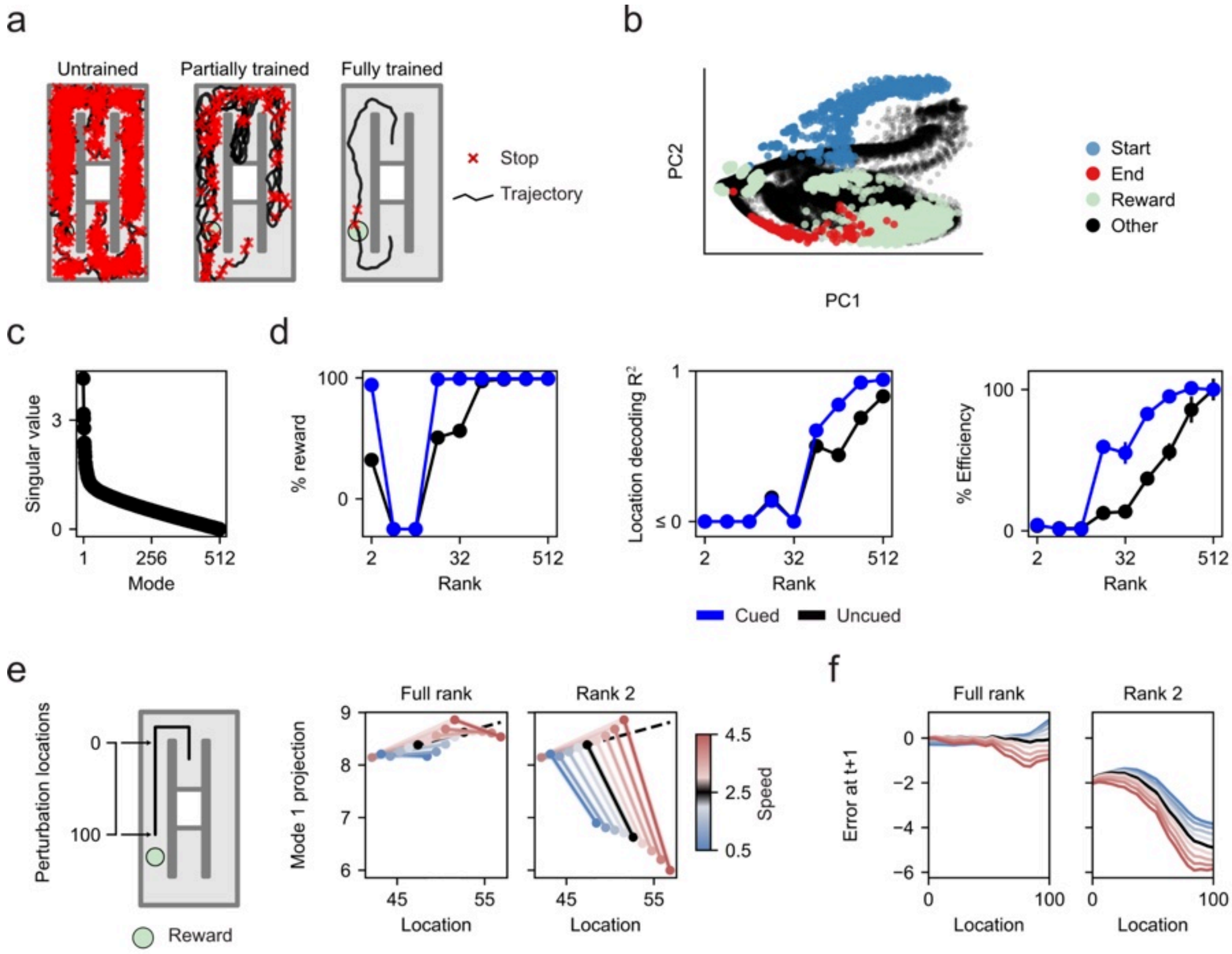

**Fig. 6: Higher rank connectivity improves navigation in a 2D spatial task by correcting errors in location representation.** **a**, Agent behaviour across training stages. **b**, RNN activity in top two PCs coloured by track location. **c**, Singular values for each mode. **d**, (left) uncued and cued reward as a function of the number of modes retained, (middle) location decoding accuracy, and (right) navigation efficiency metric computed by normalising the number of trials performed in 10,000 frames to that achieved with full rank. **e**, Projected and actual values at t+1 for mode 1 as in Fig. 4c. **f**, Error as a function of track location as in Fig. 4d.

## Extended Data Figures

### Extended data related to Fig. 1

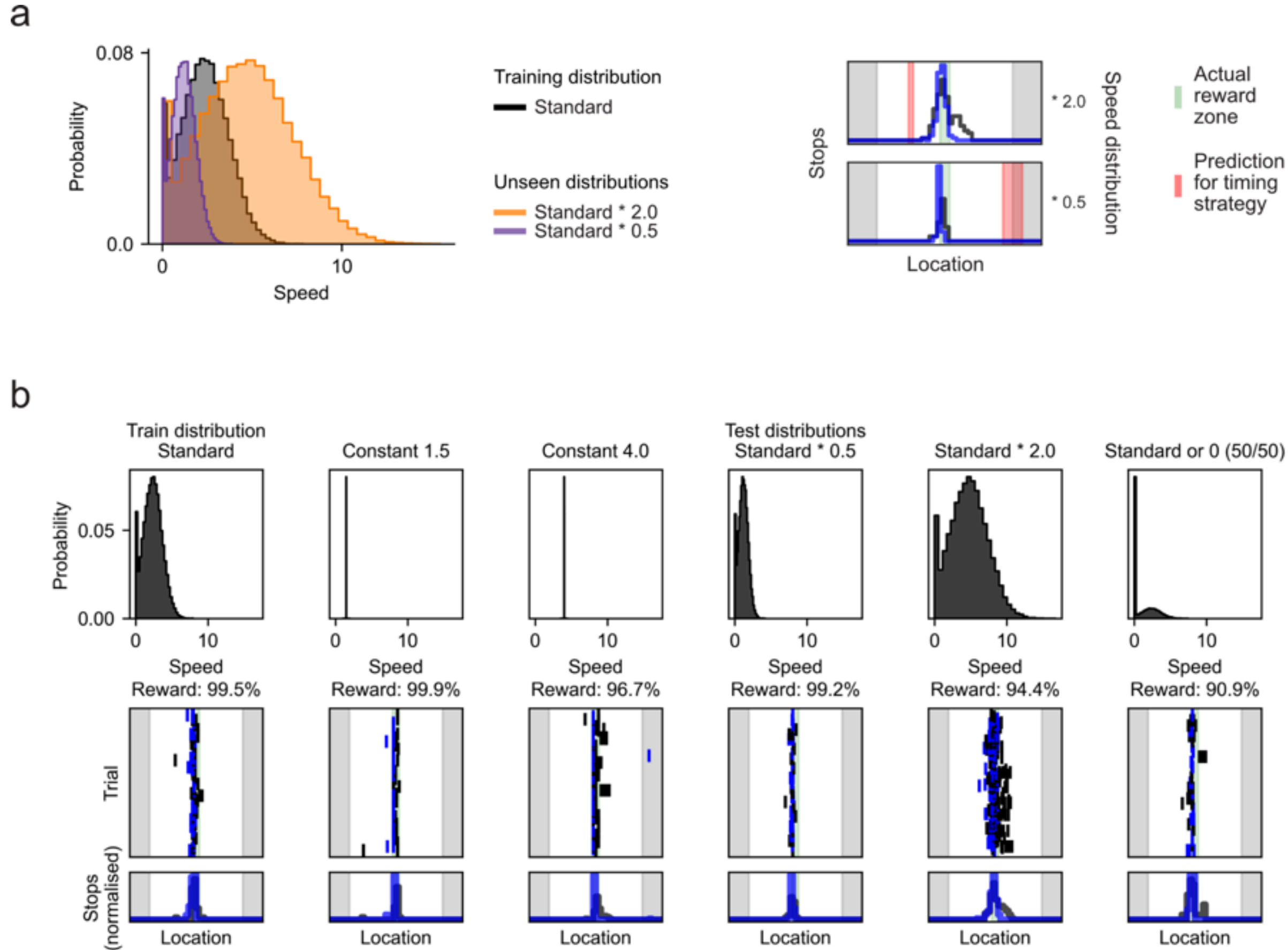

**Extended Data Fig. 1**: **Agents rely on an integration computation rather than a timing computation.** **a**, Left: Speed input distributions. *Standard* is the distribution used for training. Right: Stopping locations for trials with the Standard x 0.5 and Standard x 2 speed distributions. Red band indicates the location of the anticipated stopping location if the agent were using a timing rather than a path integration strategy. **b,** Top row: Histograms showing speed input distributions used for training and standard test (left column) and for additional testing to evaluate generalisation (right columns). Bottom row: Rasters of stops (upper panels) and normalised stop probability (lower panel) as a function of location during test trials using speed distributions in the corresponding upper panels. Blue rasters/lines indicate cued trials and black rasters/lines indicate uncued trials.

**Extended data related to Fig. 2**

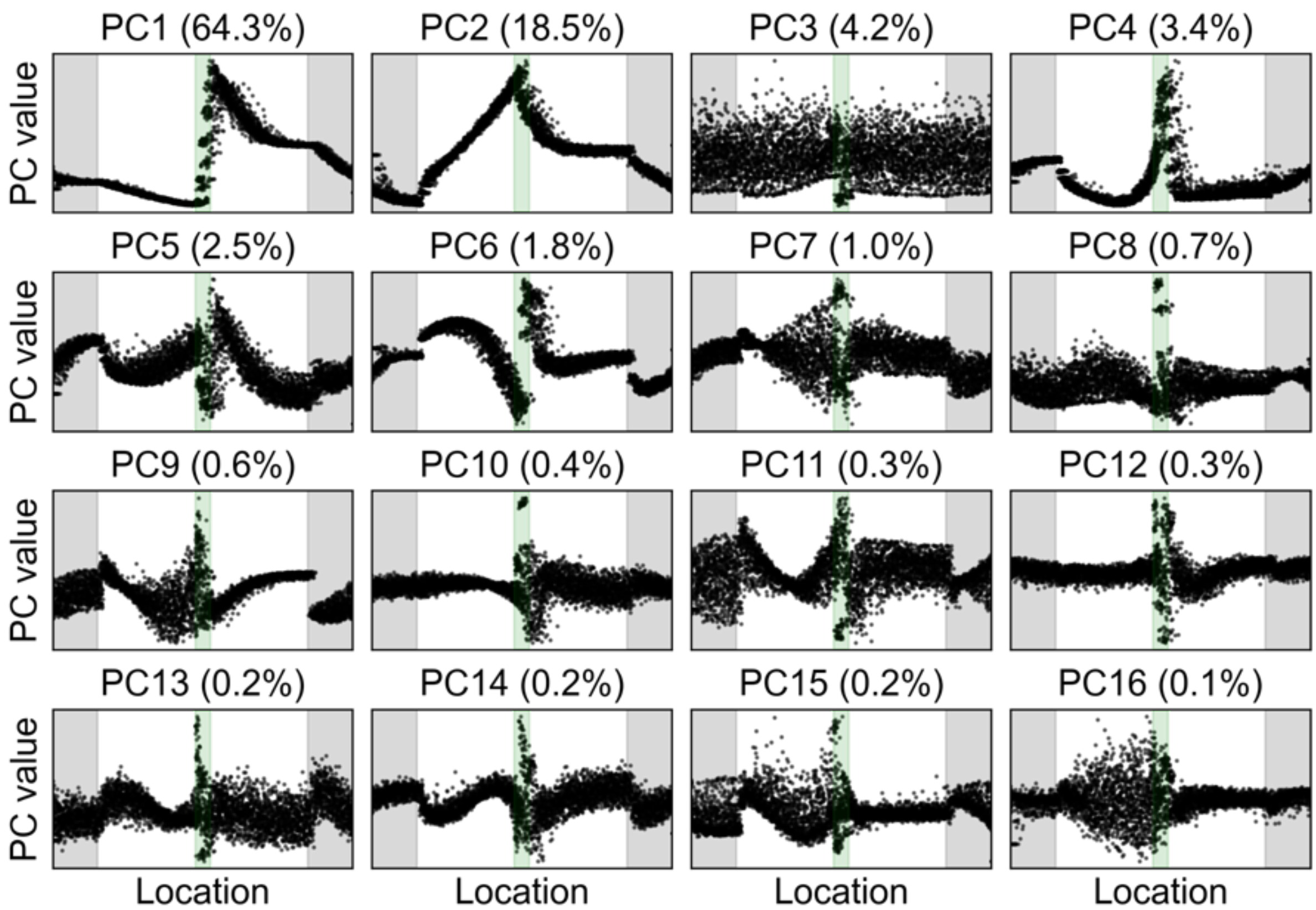

**Extended Data Fig. 2**: **RNN activity projected into the top principal component axes.** Value of each of the first 16 PCs of RNN activity as a function of track location during simulation of the trained exemplar network under the standard speed distribution. Percentages indicate the variance explained per PC.

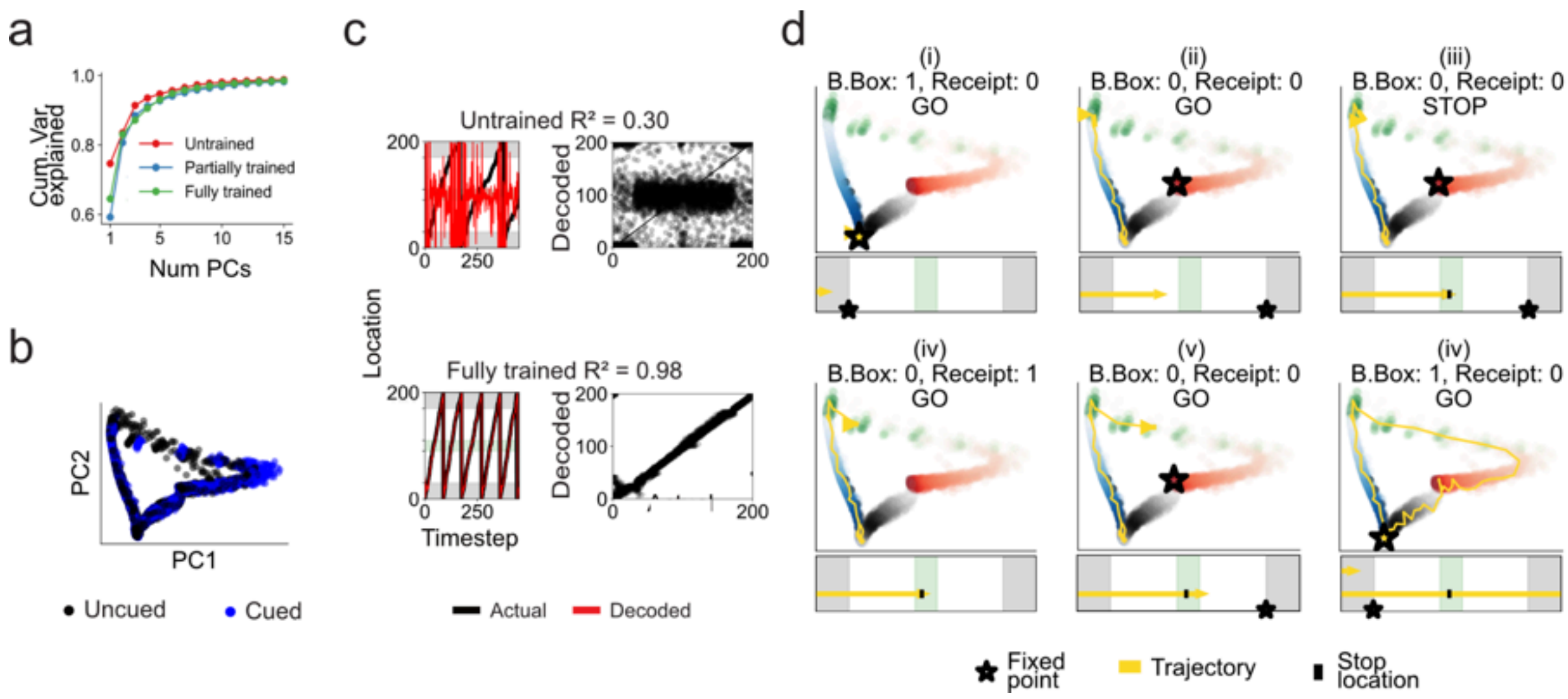

**Extended Data Fig. 3: Fixed point structure of the task. a**, Cumulative explained variance across principal components. **b**, Fully trained manifold coloured by trial type. **c**, Decoding of location from the first two components using a k-nearest neighbours decoder. Left plot shows raw decoding of track location across time, and rightmost plots show the corresponding aggregate decoding performance, with each point being a simulation timestep. Decoders were fit individually to the training stages and were evaluated on unseen simulation data (which is shown here). **d**, Progression of the agent on the track (lower, yellow line) and corresponding progression on the manifold (upper) across key phases of the trial. Inputs and actions are shown above the manifold plots. Stars indicate the position of the stable fixed point (if present) on the manifold and corresponding location on the track.

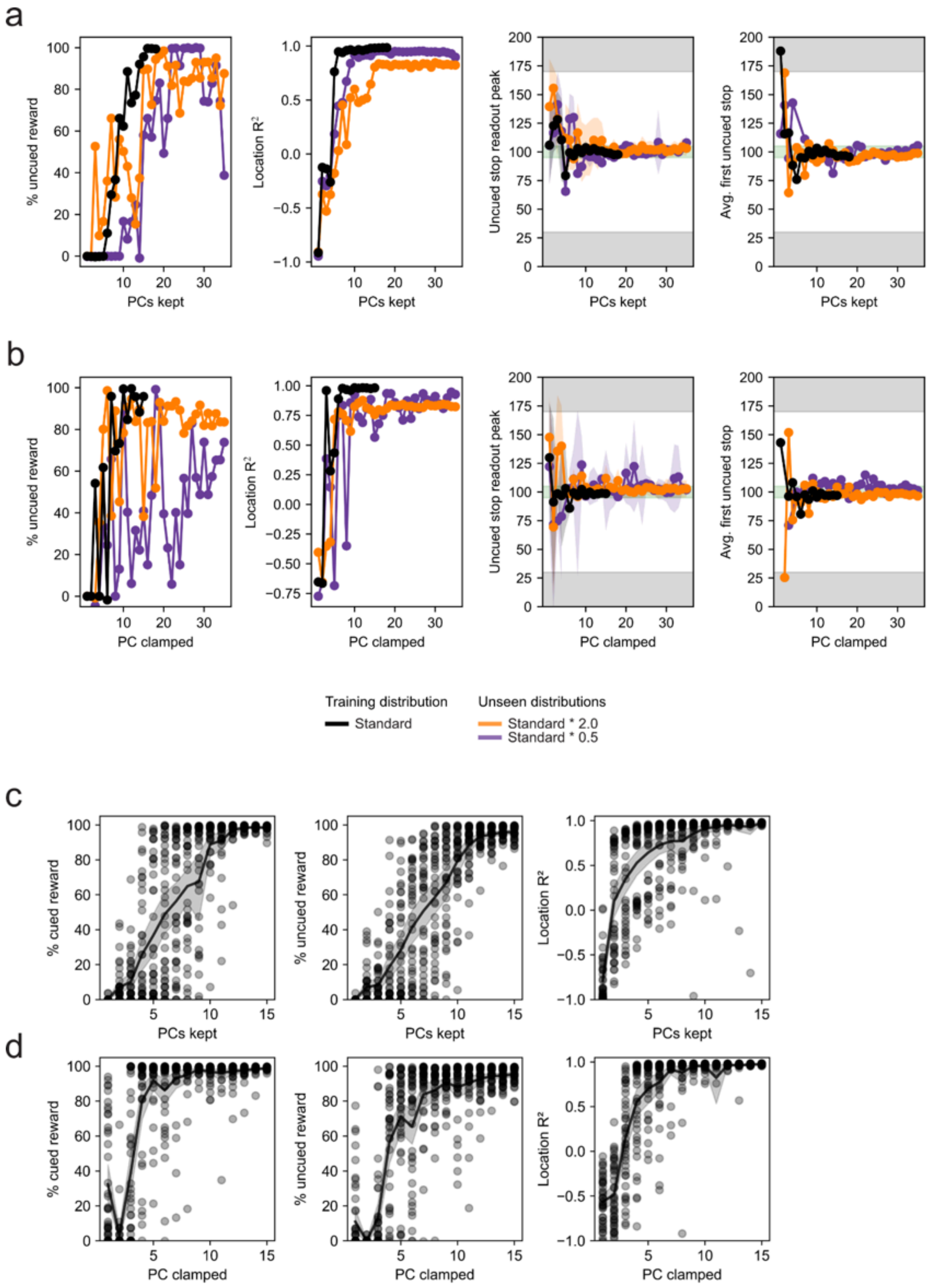

**Extended Data Fig. 4: Application of LDAC manipulations to models with different speed distributions, architectures, and random seeds. a**, Reward, decoding accuracy, track location with highest stop probability, and location of the first stop, each plotted as a function of the ascending

number of PCs that are kept in the model and colour coded according to the speed distribution used for testing. ’Uncued stop readout peak’ is the average location of where the peak of the stop action readout is, with shaded regions showing 1 standard deviation from the mean. **b**, as in **a** but clamping different individual PCs. **c-d,** Evaluations similar to **a-b** but across 92 successfully trained models. The dark line shows the mean and the shaded region shows the 95% CI across multiple model seeds/architectures.

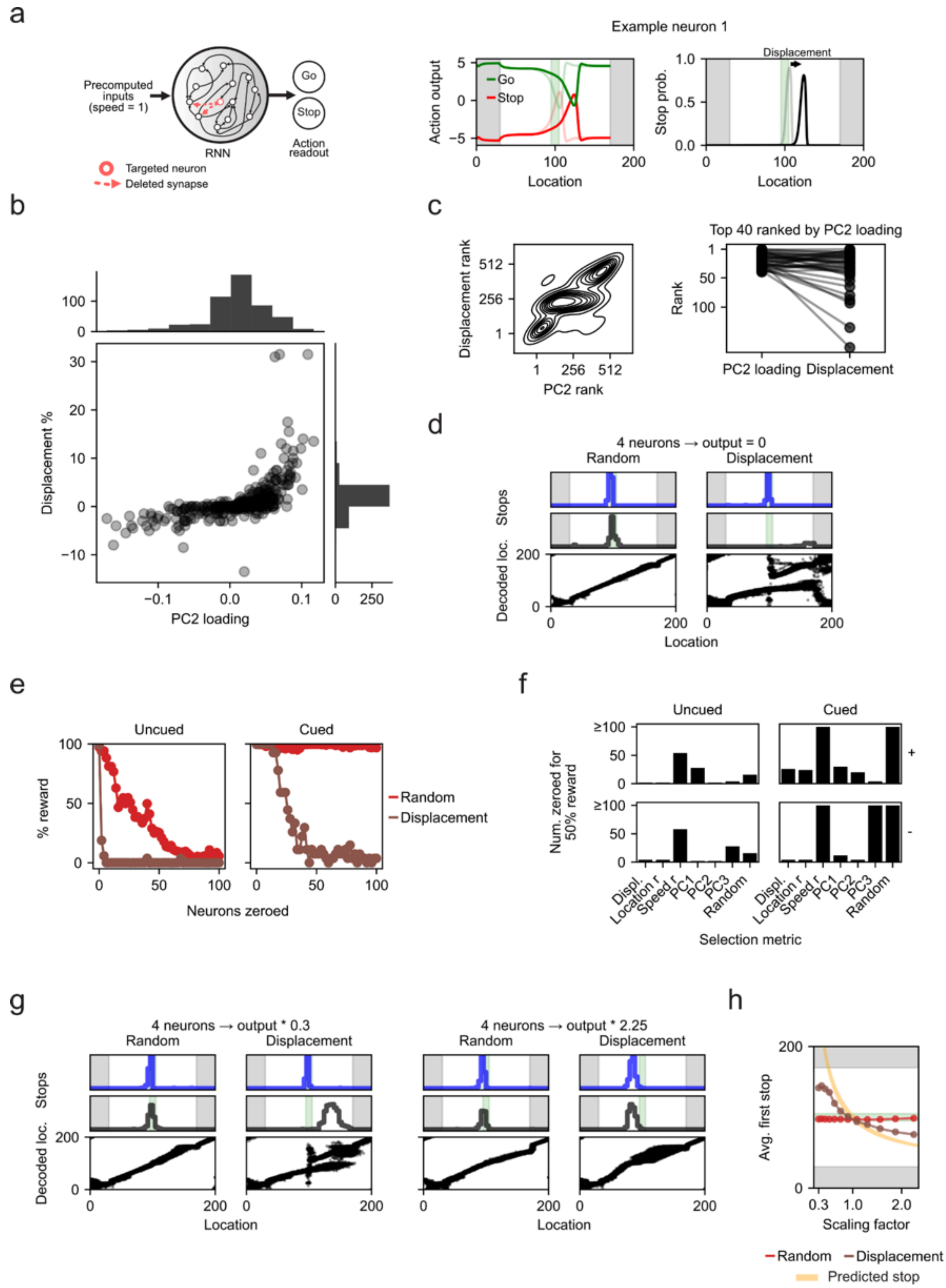

**Extended Data Fig. 5: Small numbers of neurons can direct localisation behaviour. a**,

Displacement scores were calculated by evaluating the effect of inactivating target neurons on the location with maximal stopping probability, calculated from the activity of stop/go neurons. **b,** Displacement scores for all neurons as a function of their PC2 loading. **c**, Rank relationship between PC2 loading and displacement scores for all neurons (left) and for the top 40 neurons by PC2 loading (right). **d,** Effect on behaviour and location representation when zeroing the top 4 displacement ranked neurons or randomly selected neurons. **e**, Reward on cued (left) and uncued (right) trials as a function of number of inactivated neurons, when targeting for inactivation was random or ordered according to displacement score. **f**, Summary of the number of neurons perturbed to reduce the reward by 50% with different ordering strategies. **g**, as in **d** but scaling the output of the targeted neurons by 0.3 (left) or 2.25 (right). **h**, The average first stop location as a function of the factor used to scale neuronal output in (g). Orange region shows the location of the reward zone scaled by the same factor.

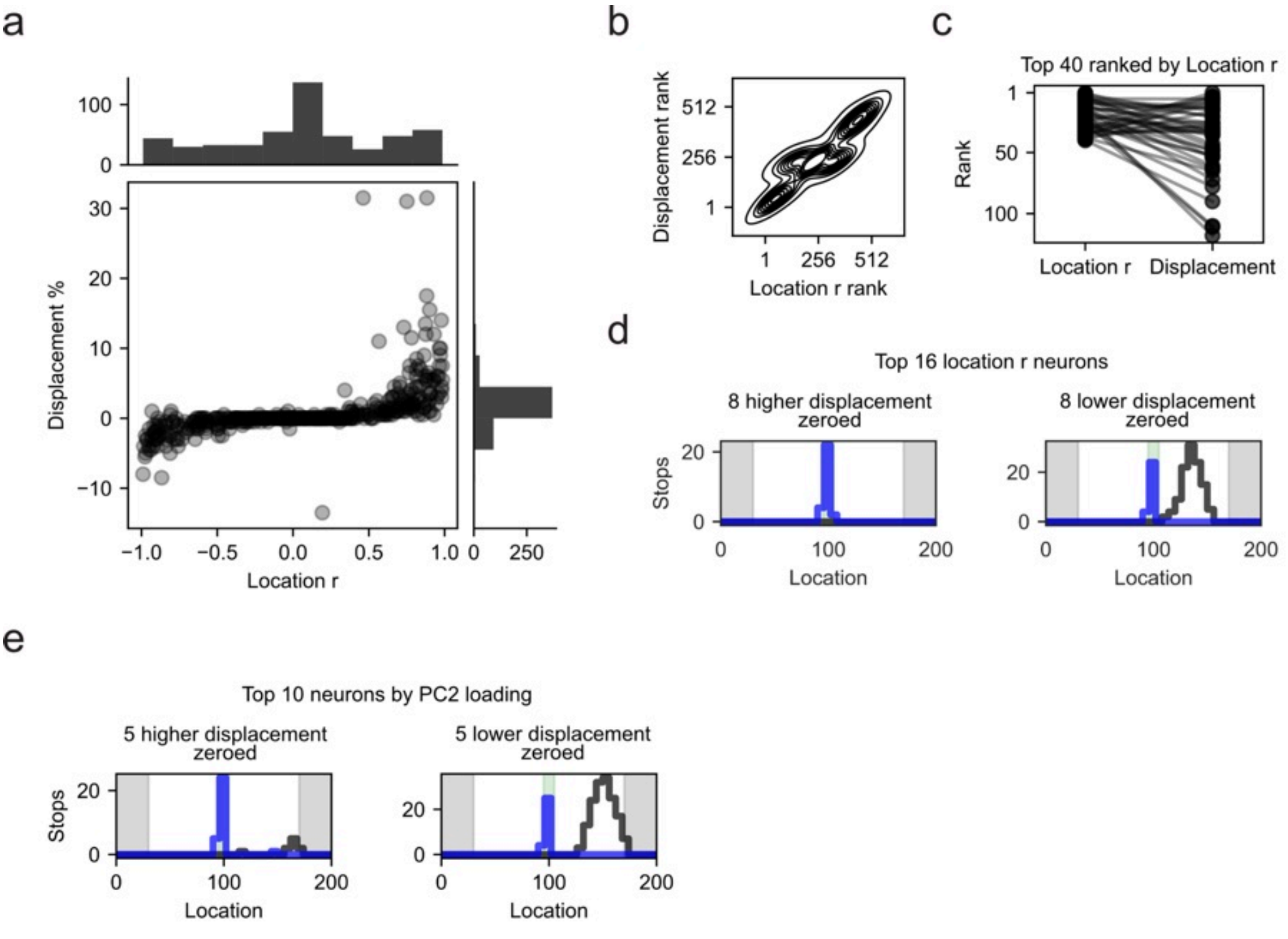

**Extended Data Fig. 6: Single neuron influence on localisation behaviour diverges from spatial representation.** **a**, Displacement scores for all neurons as a function of location correlation. **b**, Relationship between location and displacement scores expressed as ranks. **c**, Ranks of the top 40 location correlated neurons and corresponding displacement ranking. **d**, Effect of zeroing 8 neurons with higher or lower displacement scores within the top 16 location correlated neurons. **e**, Effect of zeroing 5 neurons with higher or lower displacement scores within the top 10 neurons by PC2 loading.

a

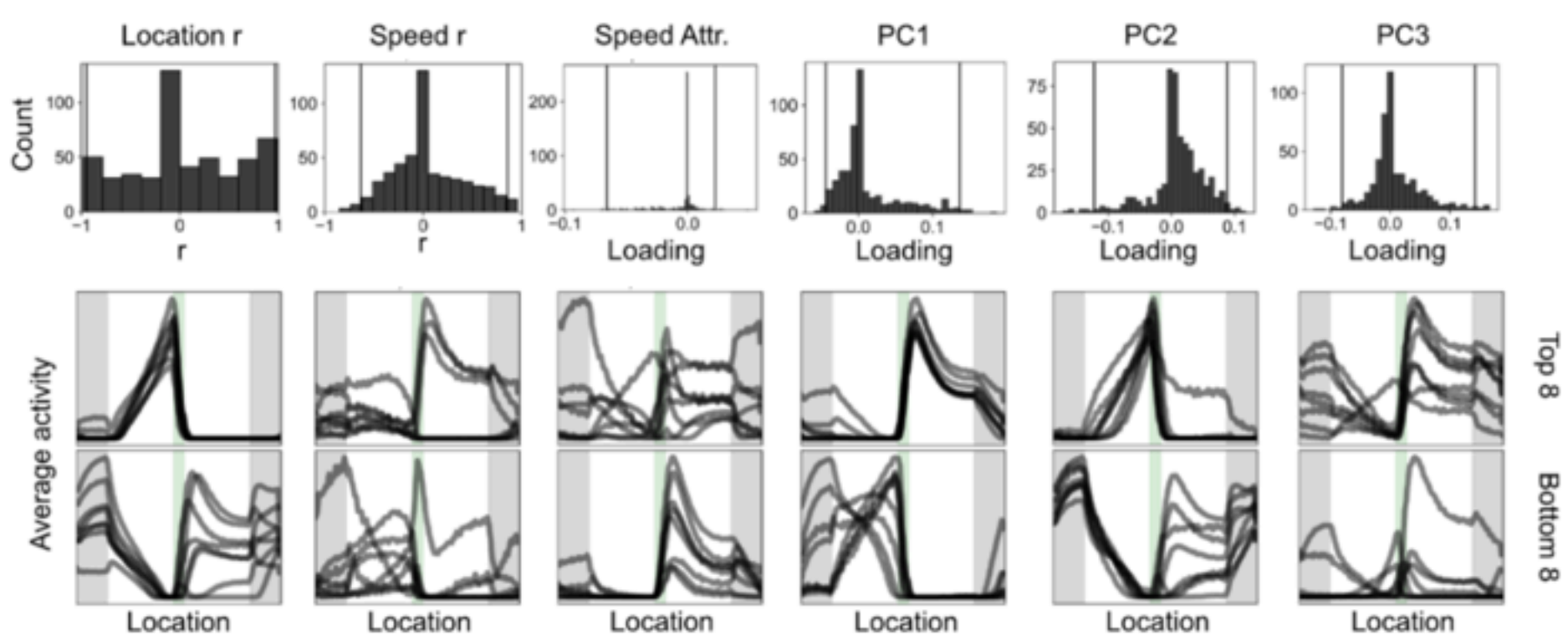

Location r
Speed r
Speed Attr.
PC1
PC2
PC3
Count
Loading
Average activity
Location
Top 8
Bottom 8

b

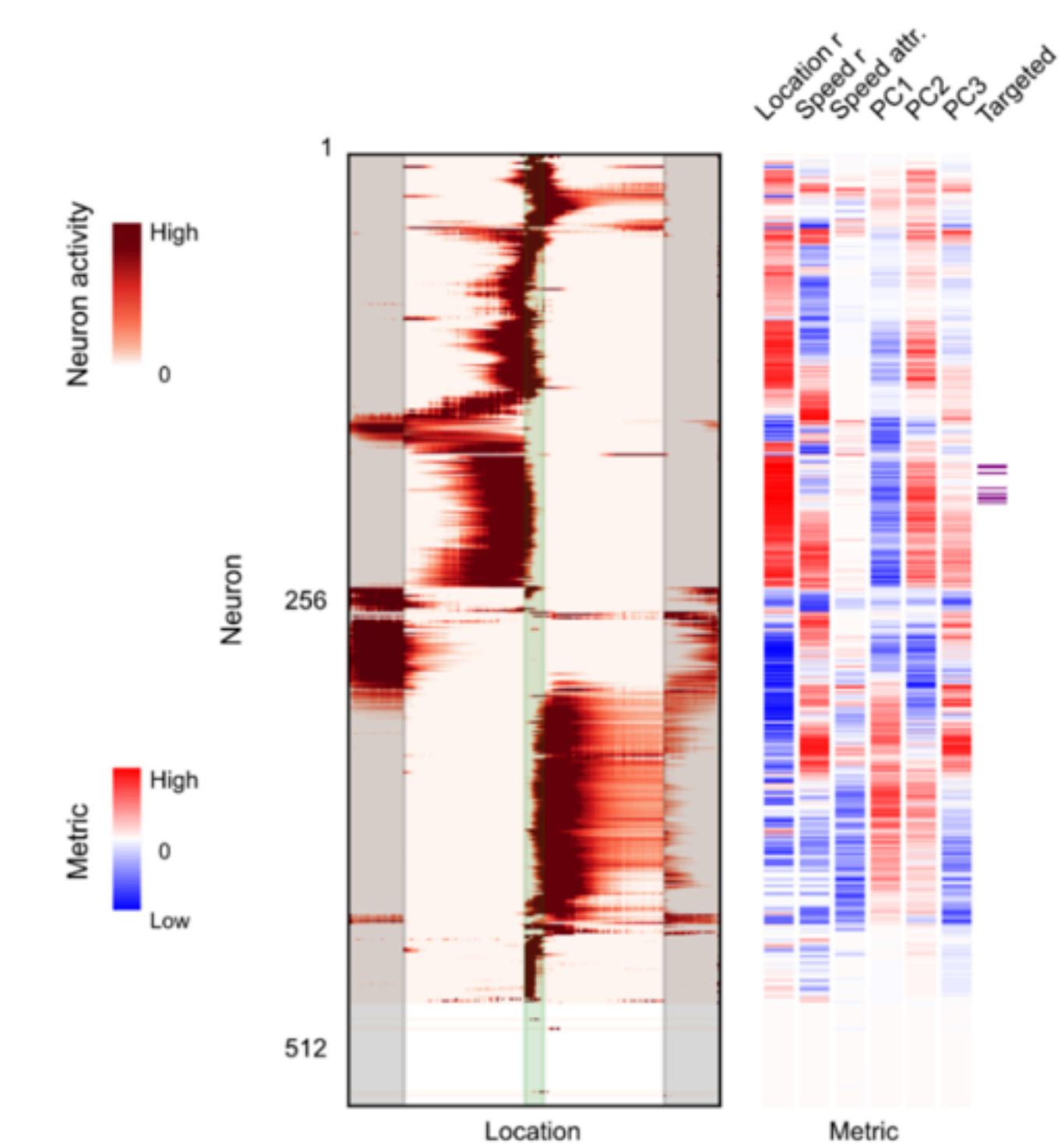

Neuron activity
High
0
Metric
Low
Neuron
Location
Metric
Location r
Speed r
Speed attr.
PC1
PC2
PC3
Targeted

c

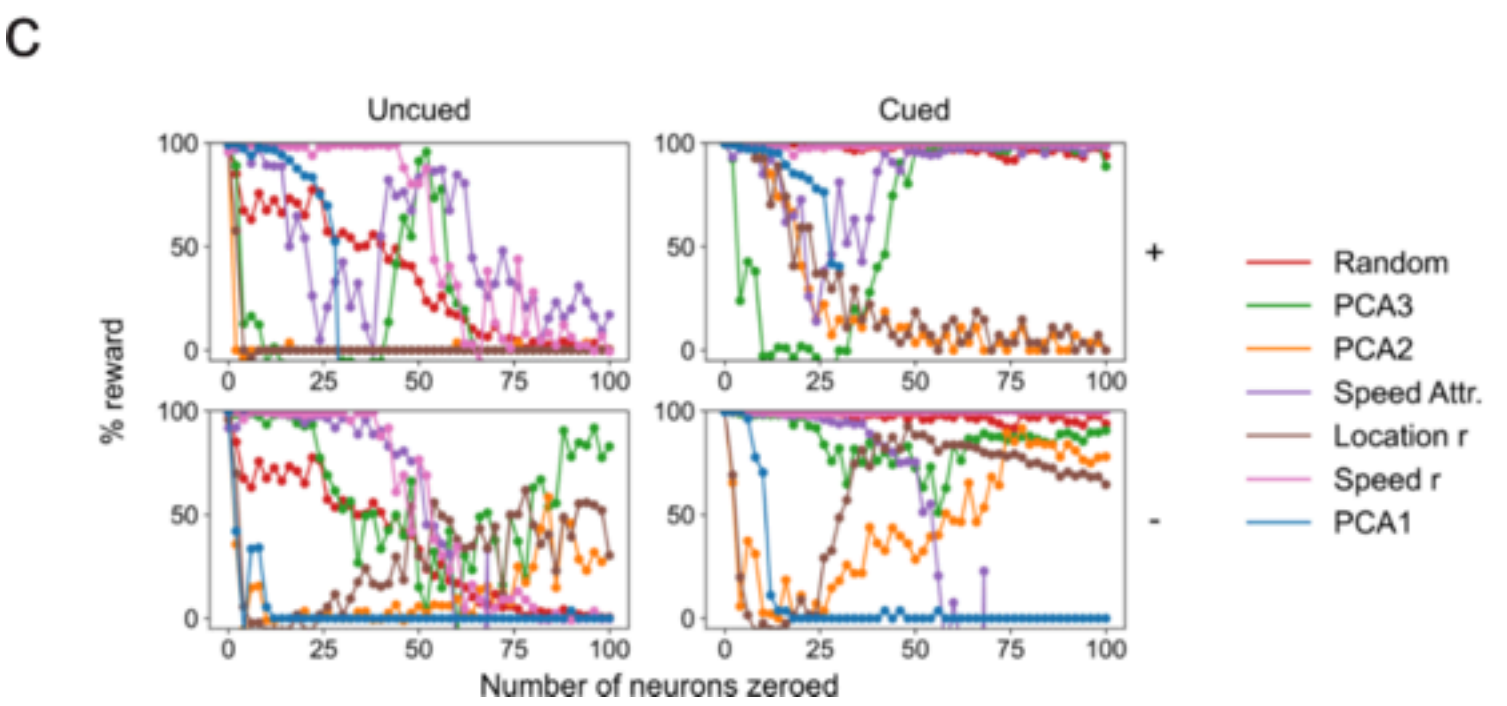

Uncued
Cued
% reward
Number of neurons zeroed
Random
PCA3
PCA2
Speed Attr.
Location r
Speed r
PCA1

d

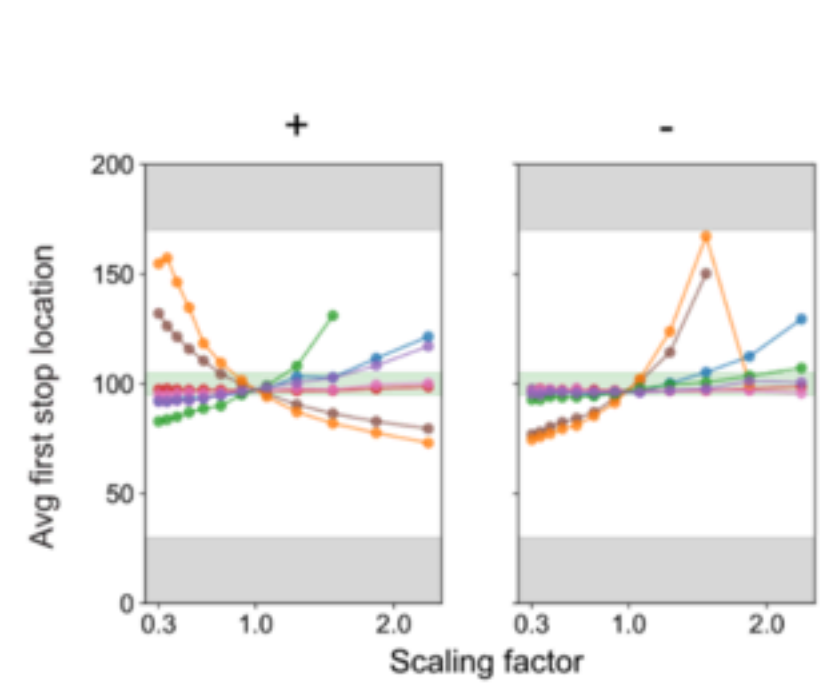

Avg first stop location
Scaling factor

**Extended Data Fig. 7. Identification and manipulation of single neurons**. **a**, Top: Distributions of metrics used to select neurons. Vertical bars indicate the position of the bottom 8 and top 8 ranking values. Bottom two rows show the average activity of the top and bottom 8 neurons. **b**, Heatmap of trial-averaged neuron activity ordered by Rastermap fit as in Fig. 2 but with single neuron metrics in **a** plotted as heatmaps on the right. Values are aligned to the neurons on the left. **c**, Effect on uncued (left) and cued (right) performance when increasing numbers of neurons are zeroed ordered by various criteria (coloured lines). Top row (+) indicates selecting neurons by descending values of the criteria, and bottom row (-) indicates selecting neurons by ascending values. **d**, Effect on average first stop location when the outputs of 8 neurons selected by various criteria are modified by different scaling factors.

## Extended data related to Fig. 3

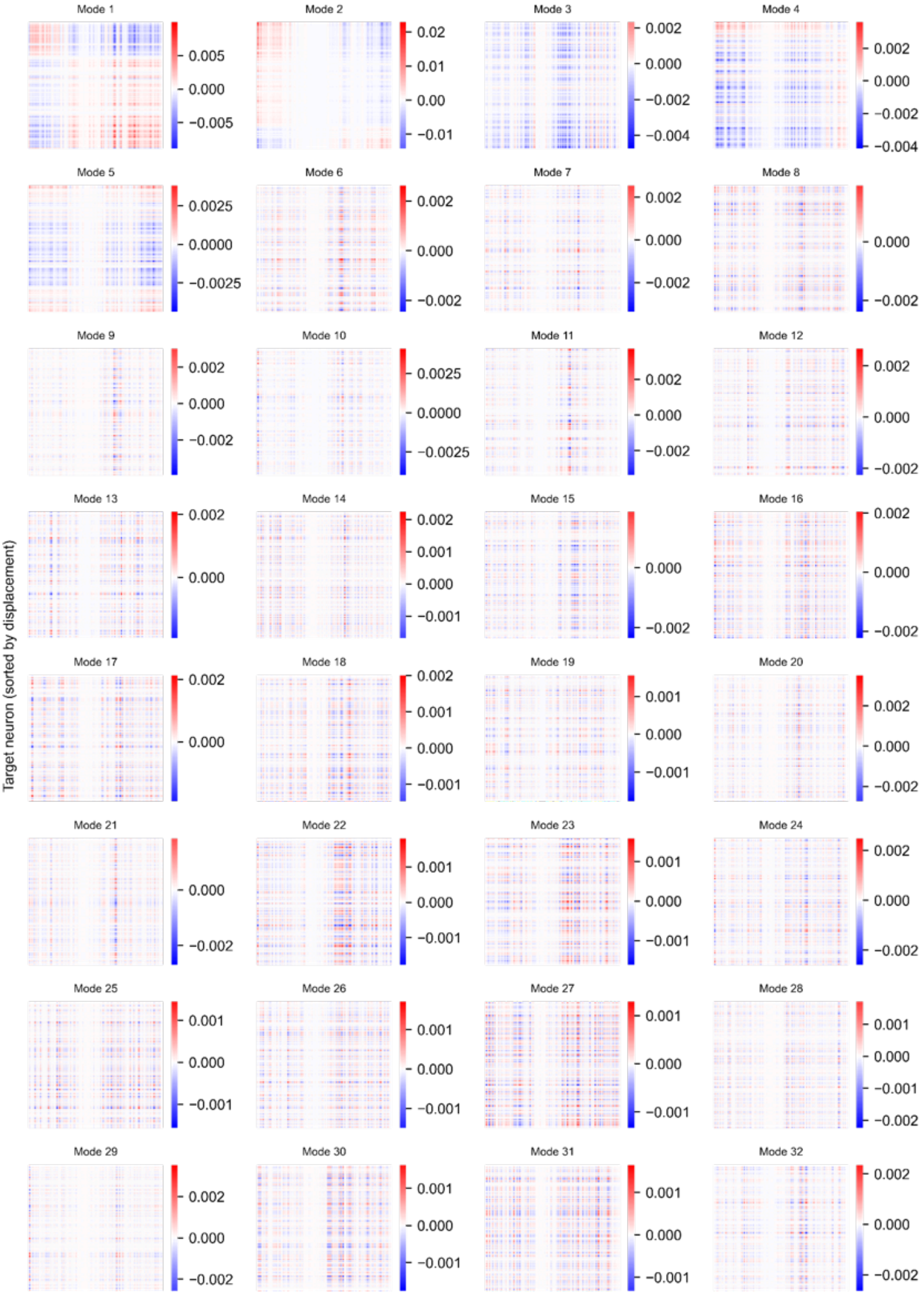

**Extended Data Fig. 8: Singular modes of the recurrent connectivity matrix of the exemplar model.** Modes were computed by performing singular value decomposition on the recurrent weight matrix, then computing the outer product of the i-th left and right singular vectors and multiplying those by the i-th singular value. Rows and columns for each mode were sorted according to each neuron's displacement score.

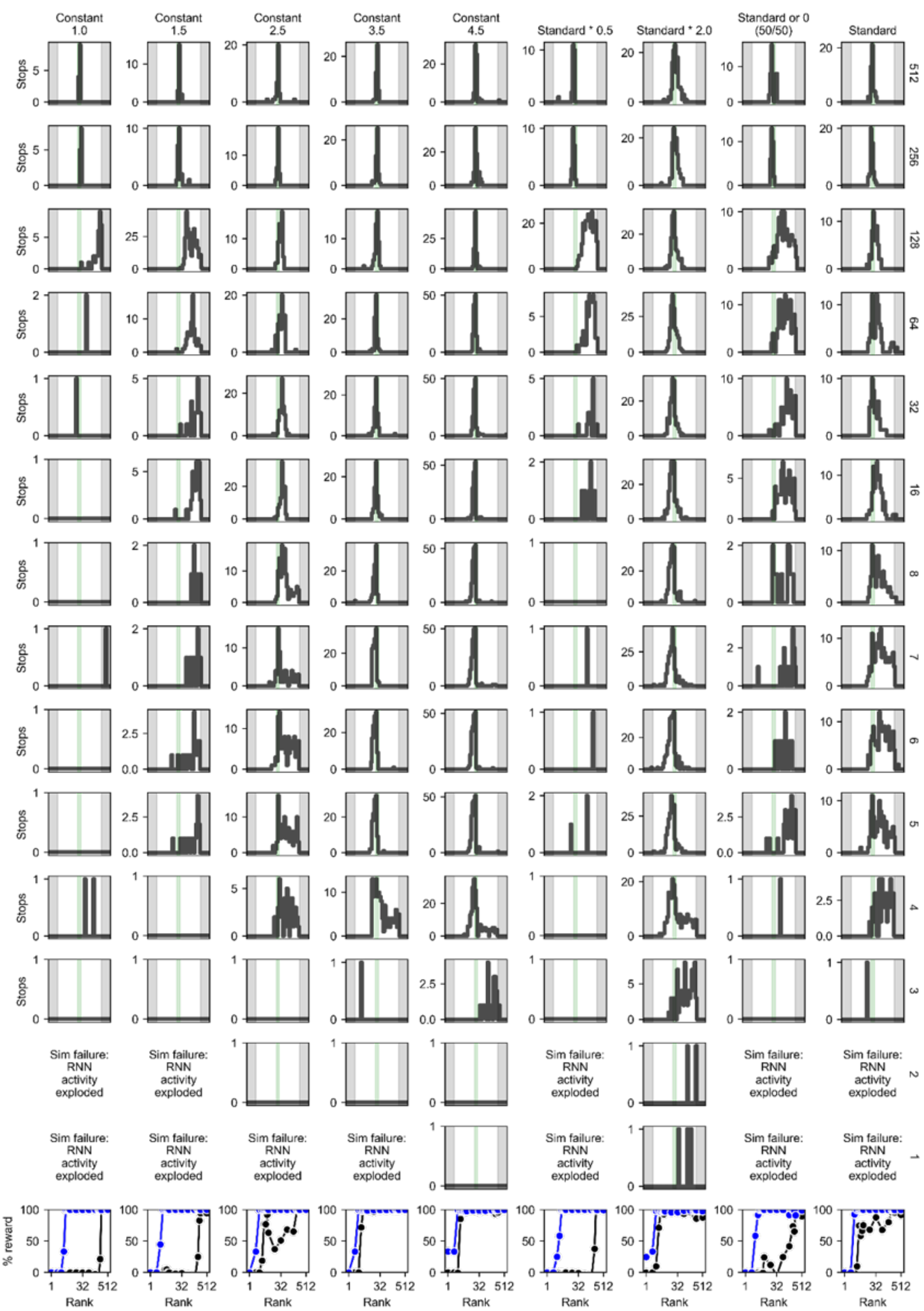

Constant 1.0
Constant 1.5
Constant 2.5
Constant 3.5
Constant 4.5
Standard * 0.5
Standard * 2.0
Standard or 0 (50/50)
Standard
Stops
512
256
128
64
32
16
8
7
6
5
4
3
2
1
Sim failure: RNN activity exploded
% reward
Rank

**Extended Data Fig. 9: Low rank approximations.** Stop probability as a function of track location for models of different rank each tested with different speed distributions. Each column shows results for a different speed distribution as in Extended Data Fig. 1. Each row shows the rank of the recurrent connectivity matrix of the tested model. Bottom row shows the cued (blue) and uncued (black) reward as a function of rank for each speed distribution.

**Extended data related to Fig. 4**

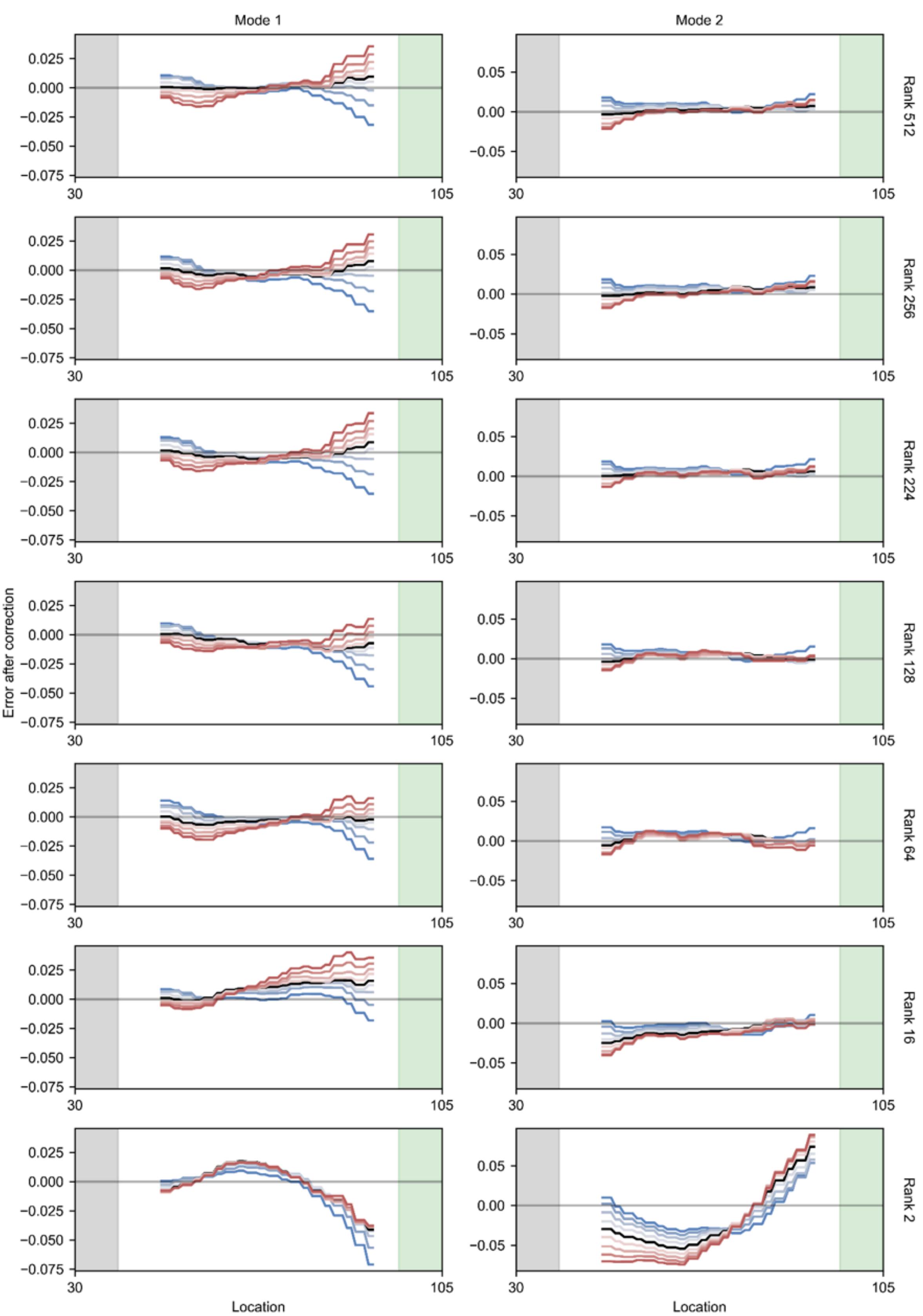

**Extended Data Fig. 10: Location estimation error compensation across low-rank approximations.** Location estimation error was computed as the difference between the expected activity indicated by the dashed line in Fig. 4c and the final network activity. Columns show errors across modes 1 and 2, and each corresponds to a rank-k approximation of the recurrent connectivity matrix. Colour indicates speed, with brighter red values corresponding to higher speed values and blue values corresponding to lower speeds.

## Methods

### Task

We implemented a linear navigation task[22,46] as an OpenAI gym[100]. The task was implemented as a linear track (200 units long), with zones corresponding to a previous experimental configuration [22], consisting of a black box (at locations 0-30 and 170-200) and track segment (30-170) containing a reward zone (95-105). Upon reaching the end of the track, the agent was teleported back to position 0. The goal of the task was to obtain rewards, which requires movement to the reward zone and stopping once. A new trial is initiated by then continuing to the end of the second black box region. On each timestep the agent selected one of two actions: stop or go. A 'go' choice caused the agent to move forwards with a speed drawn from a rectified Gaussian distribution to simulate behavioural variability. For evaluations on unseen speed distributions, speeds were rescaled by factors of 0.5 or 2.0. The agent received 4 observations: a speed input (0-6), a reward receipt (0 or 1), a reward zone cue (0 or 1), and a black box input (0 or 1). The speed input corresponded to the speed value assigned after the choice on the previous timestep. The reward receipt, which acts as an artificial analogue of the soy milk dispensed in the previous experimental task, was set to 1 for a single timestep after the first stop choice in the reward zone on each trial and was zero otherwise. On cued trials the reward cue was set to 1 when the agent was in the reward zone and was zero otherwise. On uncued trials the reward zone cue was set to zero throughout. The blackbox input was set to 1 when the agent was in the blackbox region and was zero otherwise.

To test whether our findings extended to richer spatial behaviour, we trained recurrent agents on a 2D navigation task in which one of five context cues specified a distinct reward zone within a maze. The agent received observations encoding speed, head direction, nearby boundaries, context, and reward receipt. On cued trials, a beacon was available when the agent was in the reward zone. The agent selected one of 8 compass movement directions or to stop. Successful trials required leaving a common start location, navigating to and stopping within the context-appropriate reward zone, then proceeding to a common end location. A correct stop and subsequent arrival at the end location yielded a +100 reward, while each time step, including an unsuccessful stop, yielded -0.01, encouraging efficient trajectories. For evaluation we focused on a single context and its associated reward zone. The environment was designed using RatInABox[101].

**Training**

To train the agent to perform the 1D task we used the actor-critic proximal policy optimisation (PPO) algorithm[54]. We used the Adam stochastic gradient descent method to update the network weights in the opposite direction of the PPO-clip loss function[102]. We distributed training across the Eddie high-performance computing cluster provided by the Edinburgh Compute and Data Facility (ECDF) and used the Sample Factory implementation of PPO[103]. Gradient L2 norm clipping with a value of 4.0 was used to facilitate training with vanilla RNNs. Weight decay of 0.0001 was also used.

During training, a scalar reward value was available to the reinforcement learning algorithm. The reward structure involved a total possible reward of 100 per episode. Stopping in the correct reward zones yielded a reward of 20. Stopping unnecessarily was penalised by a reward of -0.2, and moving forwards was neutral, with a reward of 0. These rewards are distinct from the 'reward receipt' input, which acts as a stimulus without any associated valence. Five trials were in a training episode. During training, the agents were simulated with the standard speed distribution (visualised by sampling the distribution 50,000 times in Extended Data Fig. 1).

~50% of the trained networks reached a final mean reward of >95% (varying with hyperparameter combination) and were kept for analysis (Supplementary Fig. 2). The moderate training success likely stems from the high task difficulty, nonstationary training data sampling, and the vanishing/exploding gradient problems associated with training vanilla RNNs[104,105]. While gated RNNs such as LSTMs and GRUs could be used instead, they rely on biologically less plausible connectivities and computations[106]. Learning was successful when all weights were plastic, but not if only input, output, or recurrent weights were plastic (Supplementary Fig. 1). Because RNN architecture may influence emergent representations and computational solutions[107], we evaluated key results with 54 network hyperparameter configurations, each with 3 random number seeds and including agents with variable numbers of neurons, feedforward input and output layers and different activation functions (Supplementary Table 1; Supplementary Fig. 2). We focus here on a representative agent composed of an RNN containing 512 neurons with ReLU activation functions, each receiving input signals directly, and with outputs mediated via 4 feedforward decoder layers converging on neurons signaling stop and go actions. For the 2D task we train an agent with the same architecture as the representative 1D agent, but the RNN bias was disabled as this improved training stability.

**Model architecture**

The exemplar agent consisted of a vanilla RNN ($h_t = f(W_{hh}\, h_{t-1} + W_{ih}\, x_t)$), where the hidden state ($h_t$) is updated from the previous state ($h_{t-1}$) and current input ($x_t$) through recurrent ($W_{hh}$) and input ($W_{ih}$) weights followed by a ReLU nonlinearity (f). There were 4 decoder layers (ELU nonlinearity) followed by a final value decoder and action decoder layer. We also trained additional networks with different hyperparameters, consisting of 0/1/4 encoder layers, 0/1/4 decoder layers, and a Tanh or ReLU RNN nonlinearity. Training used an actor-critic configuration with the networks learning both a value function and a behavioural policy. During experiments the value function output was ignored. The default weight and bias initialisation distributions were used from PyTorch 2.0.0. Actions are chosen by converting the raw values of the stop/go neurons to probabilities (by applying the softmax function) and taking the action associated with the highest value.

**Dimensionality reduction**

We used PCA to reduce the dimensionality of the neural data. Dimensionality reduction was performed on the neural activity obtained during the task, which was 5000 timesteps of simulation.

Each feature was a neuron and each sample was a timestep. The results were not averaged and reflect the raw RNN activity before it was passed to any decoder layers.

**Location decoding**

To decode locations we used PCA to reduce the dimensionality, then trained a K-nearest neighbour classifier on the reduced neural data and location labels. PCA and the decoder were evaluated on a 'test' simulation dataset created by rerunning the simulation (which has randomness due to the movement and action selection).

**Fixed point identification**

To find fixed points, we used an optimisation-based procedure[56]. We randomly sampled 1000 initial starting states during normal behaviour, then ran the procedure from those points. The procedure optimised the initial states such that the norm between the input state and the next state (where next_state = rnn(inputs, state)) was minimised. Fixed points were then classified as such if the norm (akin to speed) was orders of magnitude below the speeds experienced during normal behaviour of the network. We ran non-batched fixed point finding[108]. Fixed points were classified as stable or unstable by assessing the eigenvalues of the recurrent Jacobian. Linearisation was performed after the RNN nonlinearity[109].

**Low-dimensional activity clamp (LDAC)**

LDAC was implemented as in Fig. 2d, where the PCs of the RNN states were altered by transforming the RNN outputs to low-dimensional PC space, modifying the appropriate PC values given the experimental goal, then inverse transforming the states back to the initial high-dimensional space. The transformed neural activity was used by both the downstream decoder layers and as part of the RNN's next timestep. The scikit-learn implementation of PCA inverse transform was used[58], where the transpose of the loadings matrix is multiplied by a vector of PC values and the means are added back. Clamping a PC's activity to 0 was equivalent to setting its coefficients to zero in the loadings matrix.

**Single neuron scores**

Speed correlation was calculated by taking the Pearson correlation of neuron output vs speed (for the region from the end of the first black box to the start of the reward zone), and output vs location for the location correlation score. Contributions to PC axes were quantified using the corresponding loading vector, whose entries indicate the contribution of each neuron's activity to that PC. Speed attribution was calculated using the integrated gradients method (50 steps, speed input: 1) implemented in Captum[110,111].

**Single neuron perturbations**

Perturbations of single neurons as in Extended Data Fig. 5 were performed by zeroing the elements of $h_t$ corresponding to targeted neurons. For control experiments with randomly selected neurons, results were averaged over five random seeds. Scaling perturbations were implemented by multiplying the corresponding elements of $h_t$ by a scaling factor.

**Displacement score**

To test the influence of individual neurons we simulated full trials with the RNN fed precomputed inputs with speed = 1, removing the influence of speed noise and behavioural variability. Throughout the trial one neuron's recurrent outputs were set to zero while leaving its connections to the output layer intact. We collected activities of the stop/go neurons across the precomputed track locations. These raw values were converted to a stop probability, and the displacement between the peak of the unperturbed condition and the perturbed peak was measured. We repeated this procedure for each neuron in the network.

**Low rank approximations and perturbations**

Low-rank approximations of the (unconstrained, full-rank) recurrent weight matrix were obtained by computing its singular value decomposition[60,61]. Each mode was defined as the unit-rank outer product of a left and a right singular vector, scaled by its singular value. A low-rank approximation was formed by summing the largest modes, ordered by singular value. To perturb a specific mode, we multiplied that mode's singular value by a constant factor and then recomposed the weight matrix by summing over all modes.

To analyse interactions between recurrent connectivity modes, we expressed the recurrent weight matrix in the basis defined by the left singular vectors of the recurrent weight matrix. In this mode-space representation, diagonal entries reflected the self-recurrence between modes and off-diagonals represented the mode-to-mode couplings. To test the causal roles of these interactions, we set selected entries to zero and transformed the perturbed matrix back into neuron space (because the left singular vectors form an orthonormal basis allowing a change of basis without loss of information). The network was then simulated with the modified recurrent weights.

### Code availability

All analysis code will be made available from the Nolan Lab GitHub page (https://github.com/orgs/MattNolanLab/).

### Data availability

All data will be made available from the Nolan Lab repository on the University of Edinburgh's DataShare site (https://datashare.is.ed.ac.uk/handle/10283/777).

### Acknowledgements

We thank Chris Halcrow, Wolf de Wulf, Harry Clark, Marino Pagan, Angus Chadwick, Matthias Hennig, David Price, Tara Spires-Jones, Richard Morris, Mark Humphries, and members of the Nolan Lab for discussions and comments on the manuscript. This work was supported by the Simons Initiative for the Developing Brain, by grants to MN from the Wellcome Trust (200855/Z/16/Z) and by the Wellcome Trust (108890/Z/15/Z) Translational Neuroscience PhD programme to IH. IH is supported by the UKRI Biotechnology and Biological Sciences Research Council (BBSRC) (BB/X01861X/1). This work made use of resources provided by the Edinburgh Compute and Data Facility.

### Author contributions

IH and MFN conceptualised the study. IH performed simulations, developed code and performed analyses. MFN and IH wrote the manuscript. MFN obtained funding and supervised the project.

### Competing interest

The authors declare no competing interest.

## Supplementary information

### Supplementary information related to Fig. 1

**Supplementary Table 1**

| Hyperparameter | Values |
|---|---|
| RNN nonlinearity | [ReLU, tanh] |
| Encoder layers | [0, 1, 4] |
| Decoder layers | [0, 1, 4] |
| RNN neurons | [64, 256, 512] |
| Seed | [1, 2, 3] |

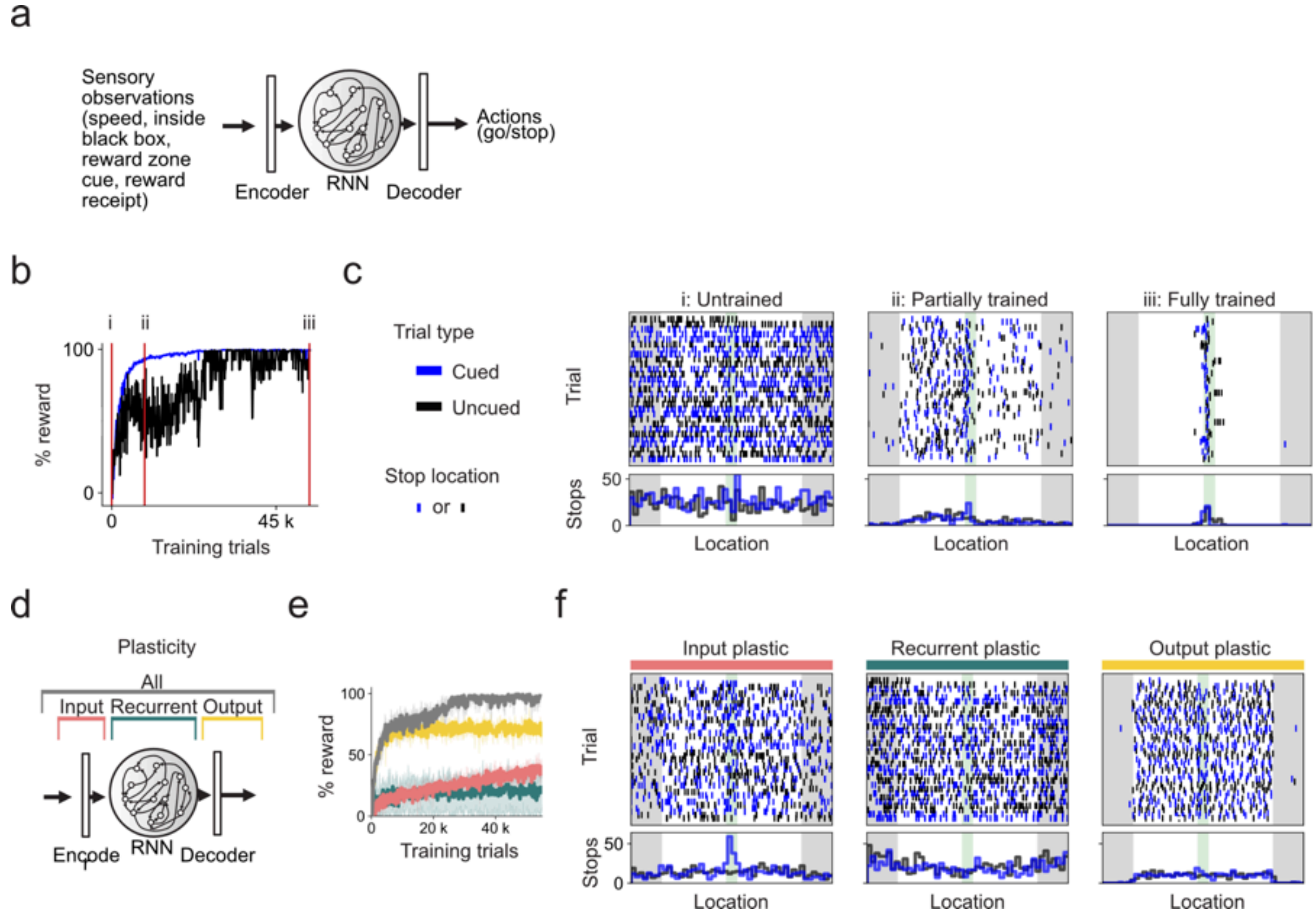

**Supplementary Fig. 1: Successful training requires plasticity at input, recurrent and output connections**. **a**, Schematic of the agent network. The inputs pass through 0-4 encoder layers, an RNN, then 0-4 decoder layers. **b**, Training curve showing % of total reward across cued (blue) and uncued (black) training trials. **c**, Stopping behaviour across three stages of training (corresponding to the vertical red lines in b). Panels show stop rasters coloured according to trial type (upper) and histograms of stops as a function of track position (lower). **d**, Schematic of neuron populations in the network. **e,** Reward obtained as a function of training trials when plasticity is confined to different network populations in the exemplar model configuration used in the main text. Shading represents the 95% CI and dark lines indicate the mean of 5 random seeds. **f**, Stopping rasters (upper panels) and stopping probability as a function of track position at the end of training for three models.

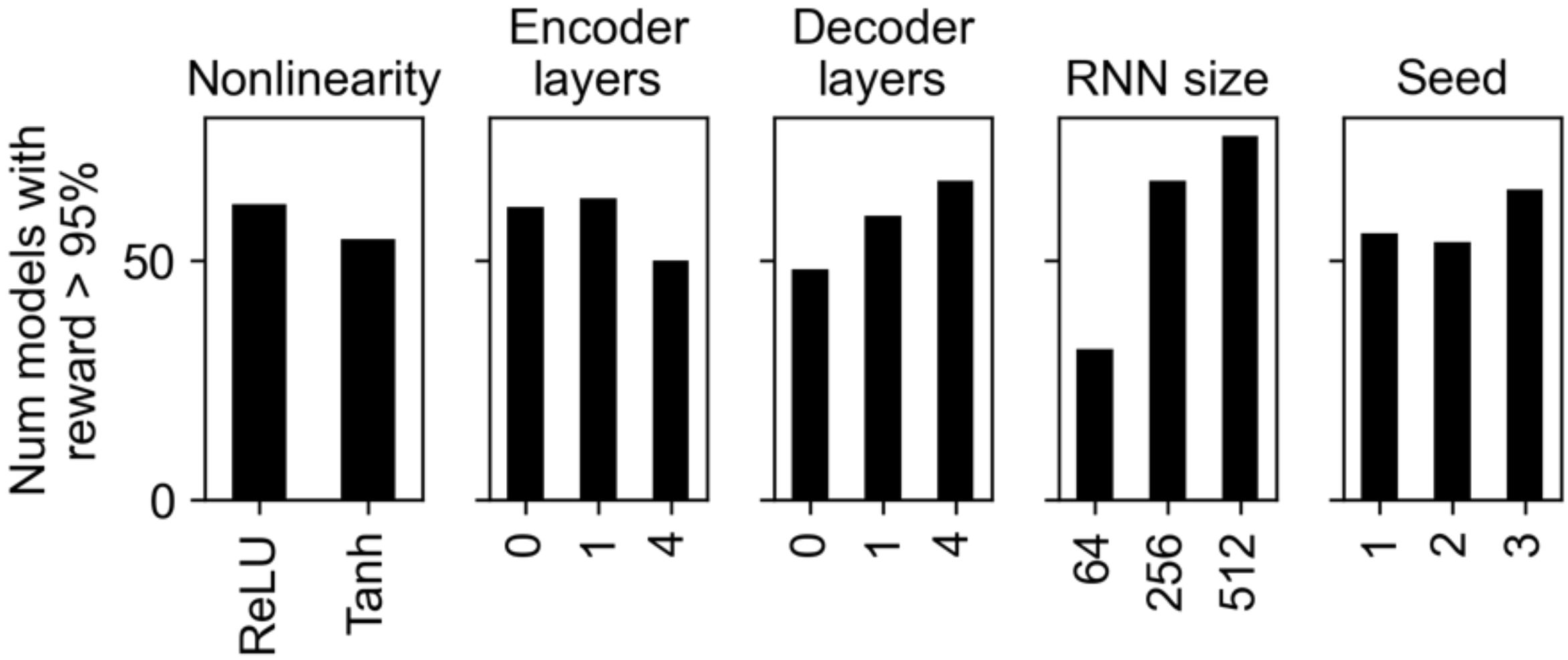

**Supplementary Fig. 2: Influence of hyperparameters on training success**. Training success across the hyperparameter sweep, defined as mean reward > 95%. Each plot groups models by hyperparameter, with all other tested hyperparameters combined. Each bar shows the percentage of models successfully trained for the indicated hyperparameter value.

**Supplementary information related to Fig. 2**

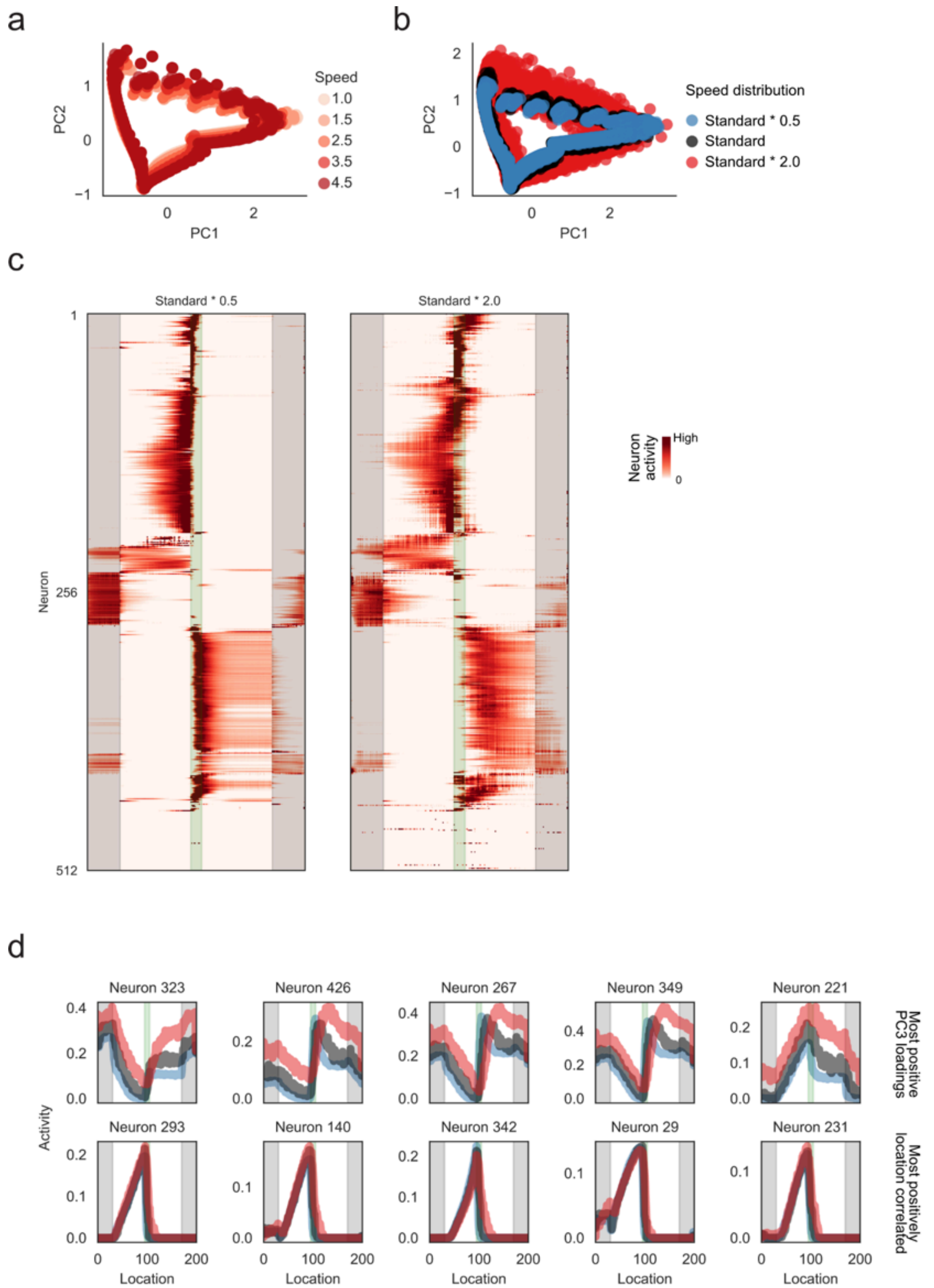

**Supplementary Fig. 3: Representations in networks evaluated with different speed distributions. a**, First two PCs for networks simulated with different constant speed inputs (as in Extended Data Fig. 1). **b**, As in **a** but for randomised speed distributions. **c**, Heatmap of location-averaged single neuron activity sorted by Rastermap ordering (from control simulation). **d**, Example single neuron firing profiles across speed distributions as used in **b.** Top row shows most positively contributing PC3 loadings (because PC3 carries the strongest speed-related activity), and bottom row shows most positively location-correlated neurons.

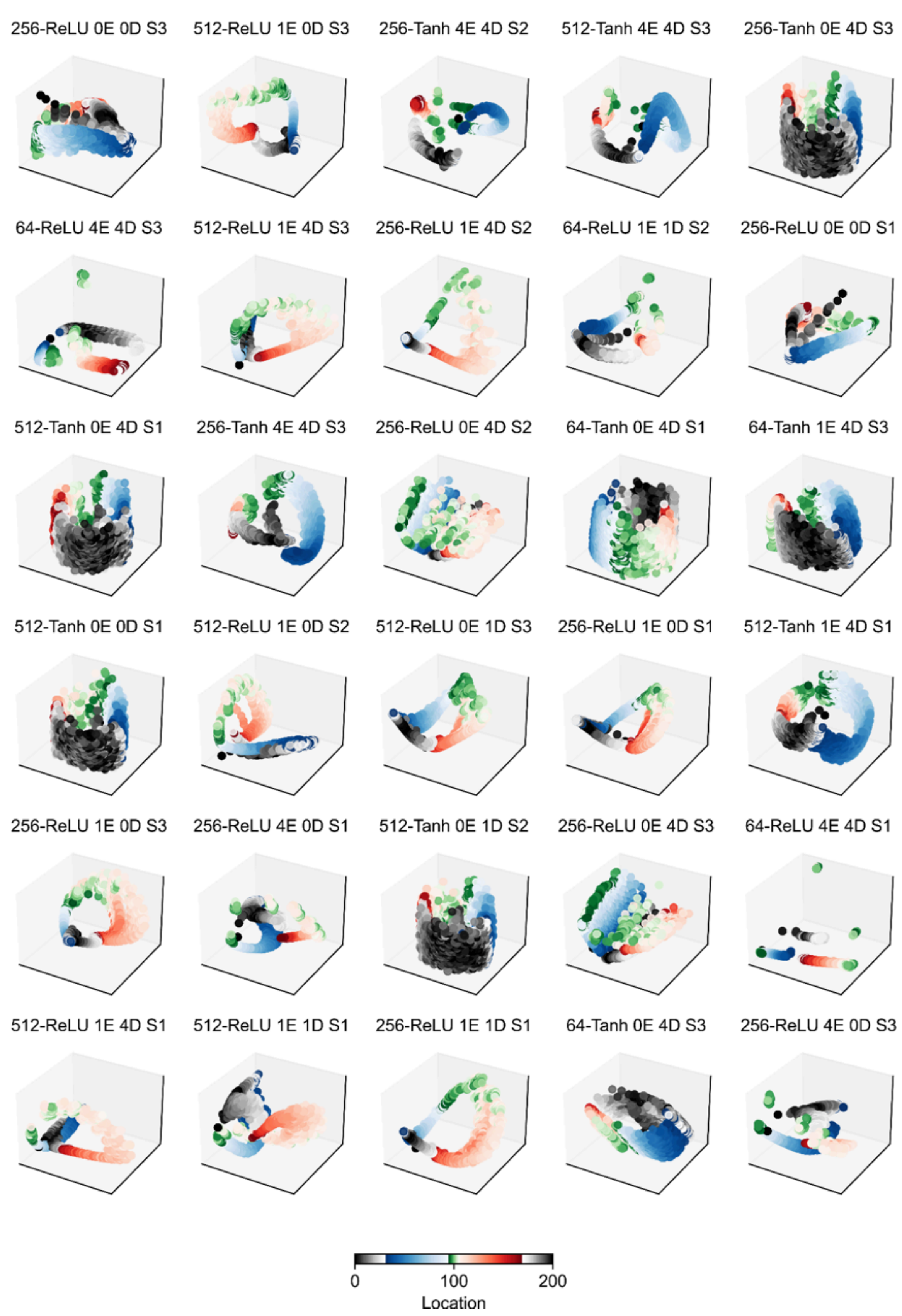

**Supplementary Fig. 4: Location representation in manifolds formed by PCs1-3 for different**

**model seeds and architectures.** Each plot shows representations of neural activity in the first 3 PCs during standard test simulations. The colour code indicates the agent's location on the track. The plot titles indicate the model configuration shown. Format: [n neurons]-[nonlinearity] [n encoder layers]E [n decoder layers]D S[seed]. Manifolds are coloured by the track.

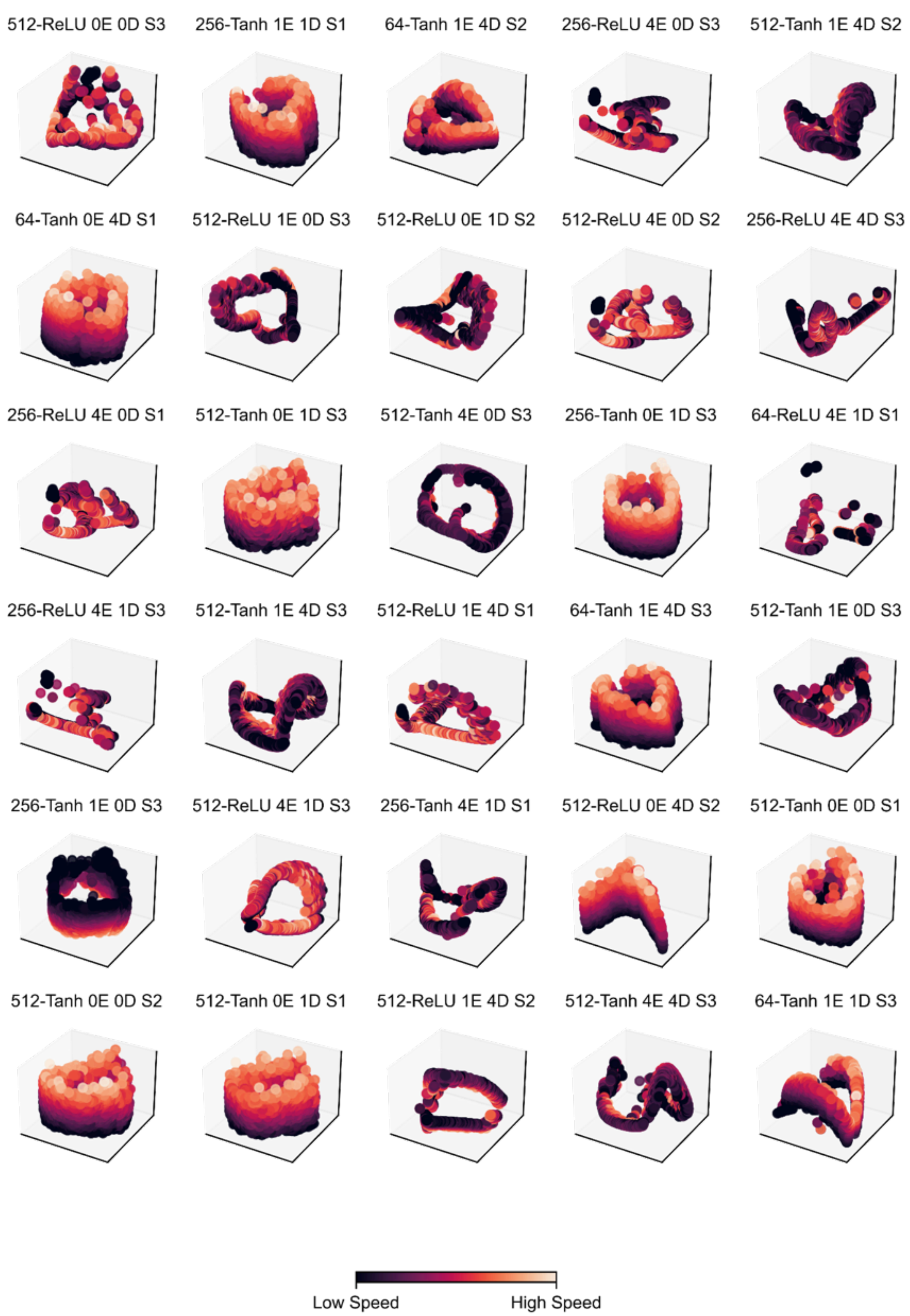

**Supplementary Fig. 5: Manifolds across different model seeds and architectures coloured by speed.** Activity in the first 3 PCs as in Supplementary Fig. 4 but the colour code indicates the agent's

speed. Title format: [n neurons]-[nonlinearity] [n encoder layers]E [n decoder layers]D S[seed]. Manifolds are coloured by speed.

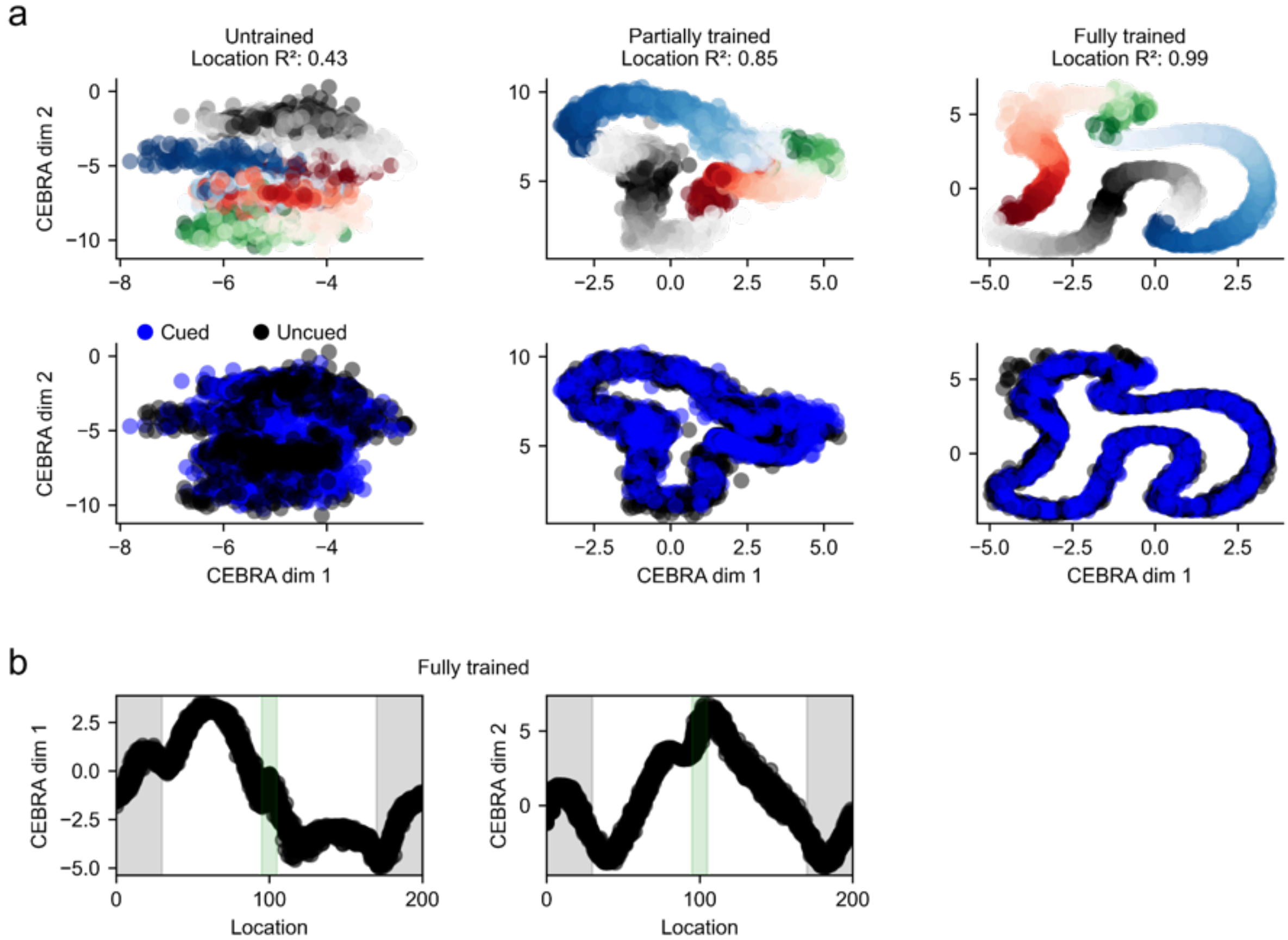

**Supplementary Fig. 6: Dimensionality reduction with CEBRA**. **a**, Non-linear dimensionality reduction with CEBRA[112] across training phases coloured by track location (first row) or trial type (second row). CEBRA models were fit separately for each training phase. **b**, CEBRA embedding coordinates plotted as a function of location on the track for the fully trained model.

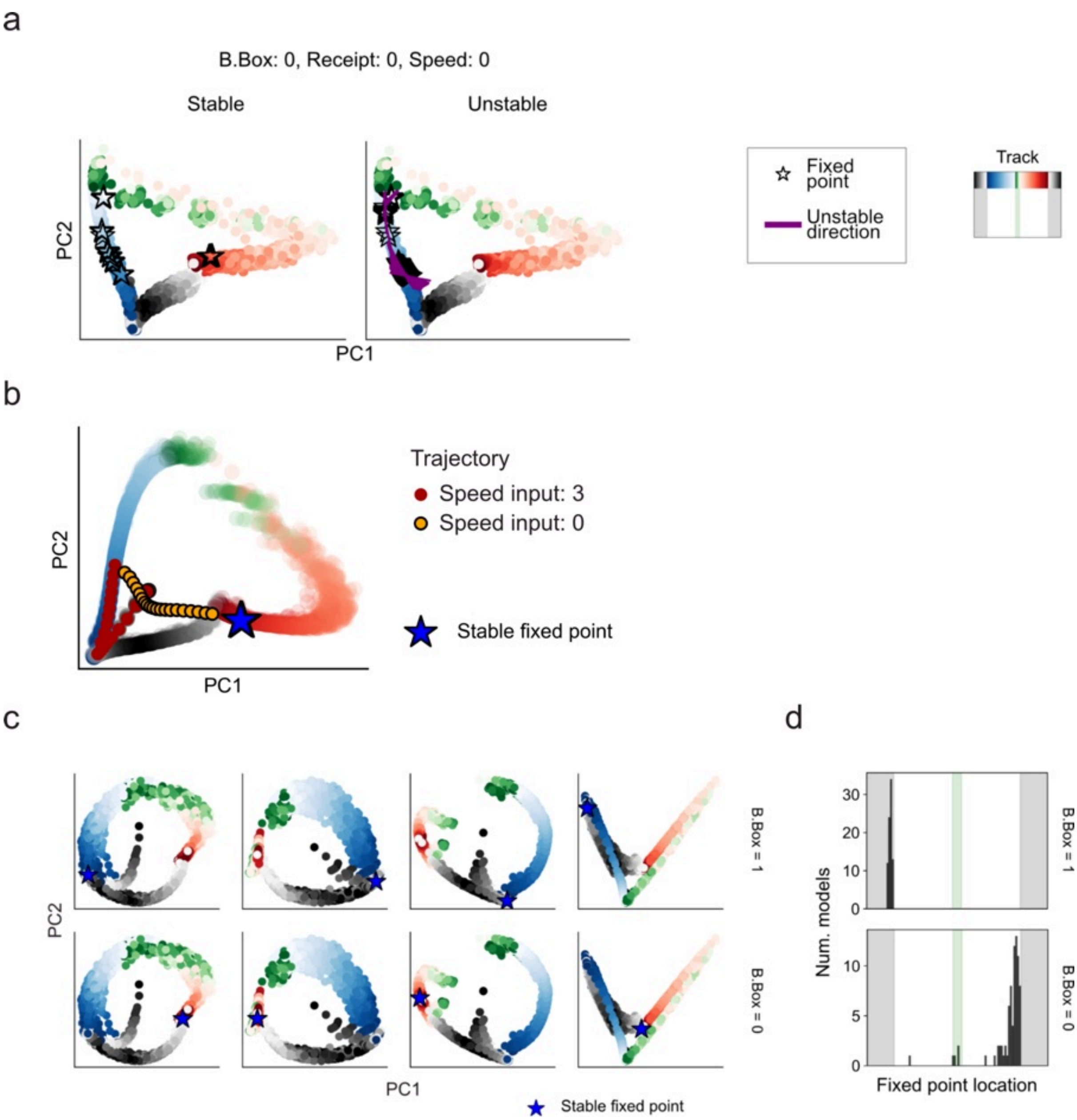

**Supplementary Fig. 7: Emergent attractor dynamics**. **a**, Plots of PC1 and PC2 colour coded according to track positions for simulations of the exemplar network when the agent speed is set to zero. Stable (left) and unstable (right) fixed points are superimposed on the plot. In the track region before the reward zone the points form an approximate line attractor enabling the agent to maintain a stable encoding of location during stops. **b**, Similar to **a** but for a model trained on a variant of the task where the speed was not allowed to be 0. When the agent was exposed to speed inputs of 0 after training, information about the current location was lost as the trajectory fell away from the manifold. **c**, First two PCs for selected models, showing stable fixed points when the black box input is either 1

(top) or 0 (bottom), corresponding to whether the agent is in the black box or in the middle track respectively. Additional example models were selected at random and varied in their nonlinearity, network size, and the numbers of encoder and decoder layers. **d**, Histograms of decoded location of stable fixed points across all models under the two input conditions.

**Supplementary information related to Fig. 2**

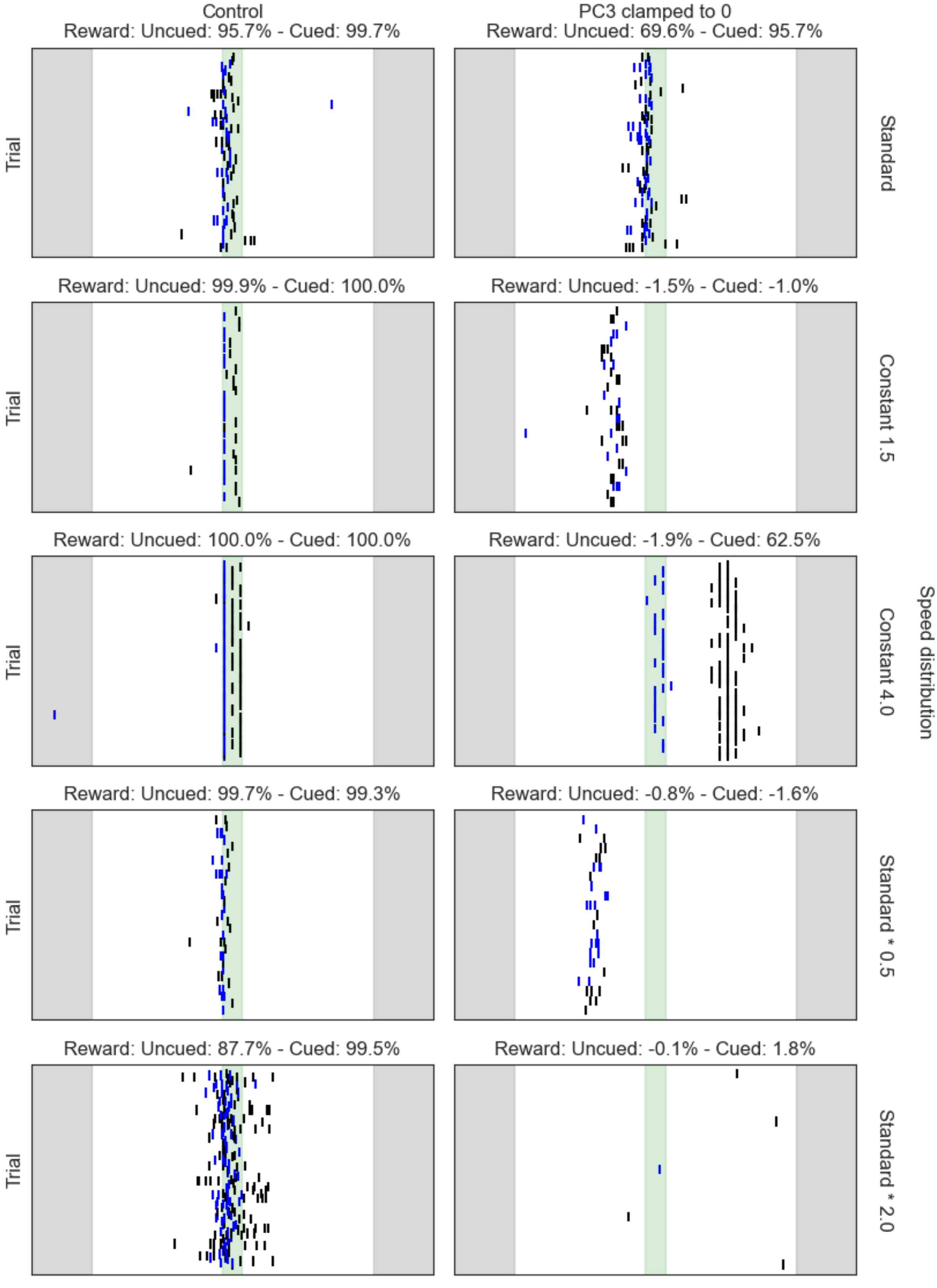
Control
Reward: Uncued: 95.7% - Cued: 99.7%
PC3 clamped to 0
Reward: Uncued: 69.6% - Cued: 95.7%
Standard
Reward: Uncued: 99.9% - Cued: 100.0%
Reward: Uncued: -1.5% - Cued: -1.0%
Constant 1.5
Reward: Uncued: 100.0% - Cued: 100.0%
Reward: Uncued: -1.9% - Cued: 62.5%
Constant 4.0
Speed distribution
Reward: Uncued: 99.7% - Cued: 99.3%
Reward: Uncued: -0.8% - Cued: -1.6%
Standard * 0.5
Reward: Uncued: 87.7% - Cued: 99.5%
Reward: Uncued: -0.1% - Cued: 1.8%
Standard * 2.0
Trial
Location

**Supplementary Fig. 8. LDAC manipulations of the speed PC.** Behavioural performance when PC3 (which represents speed) is clamped to 0 across different speed distributions (as introduced in Extended Data Fig. 1).

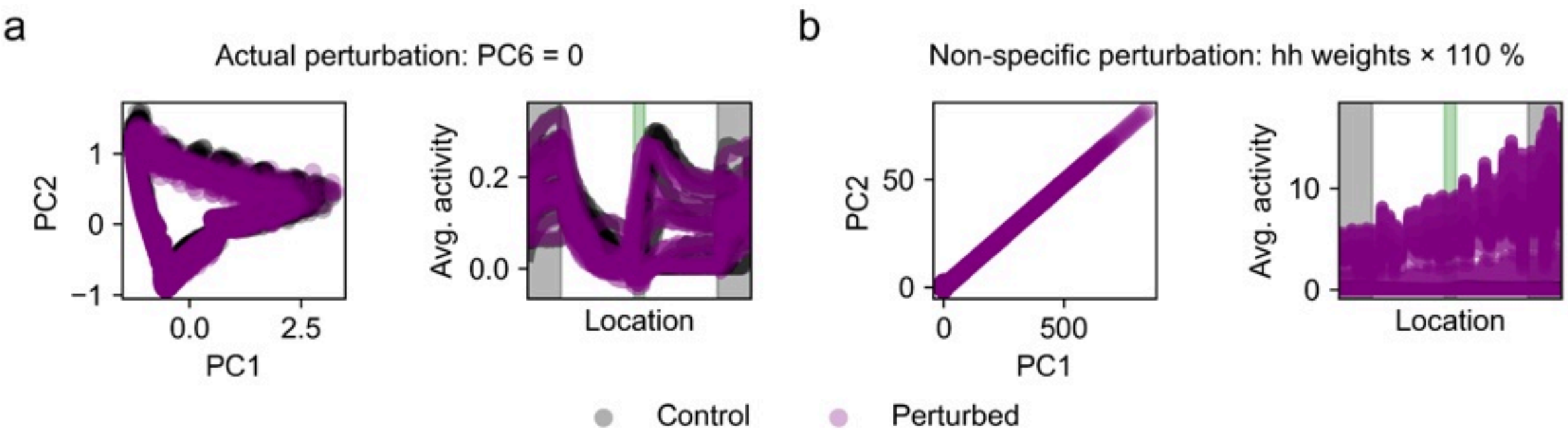

**Supplementary Fig. 9. LDAC perturbations keep the network within its normal operating regime. a,** Effect of clamping PC6 = 0 (as in Fig. 2g-h) on the manifold (purple) and single neuron activity, compared to unaltered network activity (black). **b,** Effect of a nonspecific perturbation obtained by multiplying all weights in the recurrent connectivity matrix by 110%.

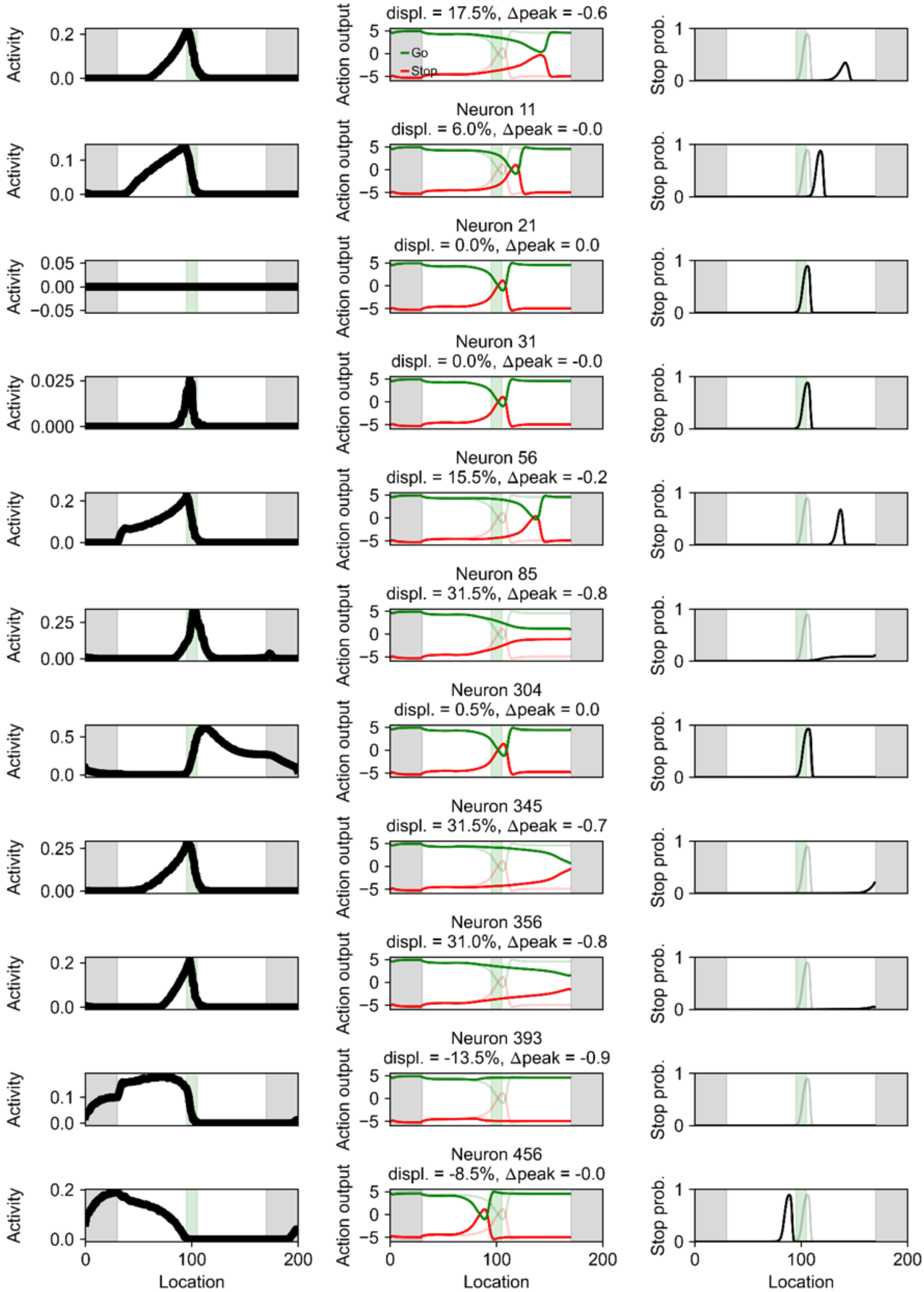

Neuron 1
displ. = 17.5%, Δpeak = -0.6
Go
Stop
Neuron 11
displ. = 6.0%, Δpeak = -0.0
Neuron 21
displ. = 0.0%, Δpeak = 0.0
Neuron 31
displ. = 0.0%, Δpeak = -0.0
Neuron 56
displ. = 15.5%, Δpeak = -0.2
Neuron 85
displ. = 31.5%, Δpeak = -0.8
Neuron 304
displ. = 0.5%, Δpeak = 0.0
Neuron 345
displ. = 31.5%, Δpeak = -0.7
Neuron 356
displ. = 31.0%, Δpeak = -0.8
Neuron 393
displ. = -13.5%, Δpeak = -0.9
Neuron 456
displ. = -8.5%, Δpeak = -0.0
Activity
Action output
Stop prob.
Location

**Supplementary Fig. 10: Stop-action readouts and stopping probabilities after perturbation of example neurons.** Left column shows the activity of selected example neurons during control simulations. Rows correspond to different neurons. Middle column shows the stop/go action readouts when the corresponding neuron was inactivated during the displacement score calculation. Right column shows the resulting stop probability. Text indicates the displacement score and the change in peak stop probability for each neuron. Example neurons were randomly selected.

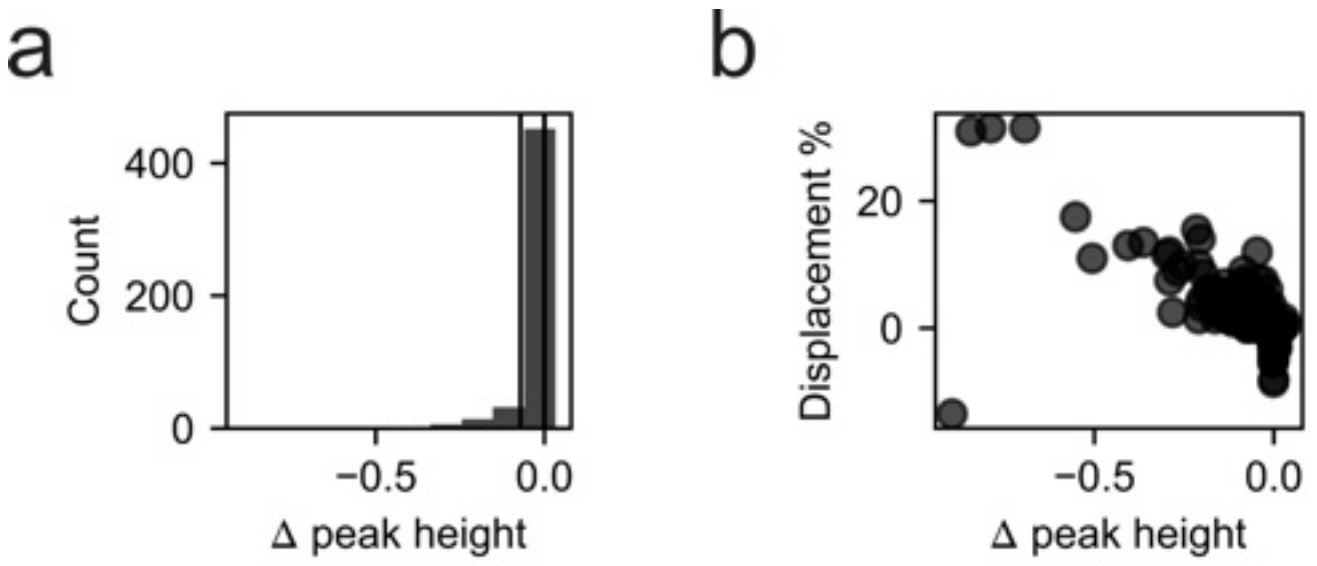

**Supplementary Fig. 11: Changes in stop-probability peak height after single-neuron inactivation. a,** Distribution of changes in peak stop probability after inactivation of individual neurons during displacement score calculation. Bars correspond to the 10th and 90th percentiles. **b**, Displacement as a function of peak height.

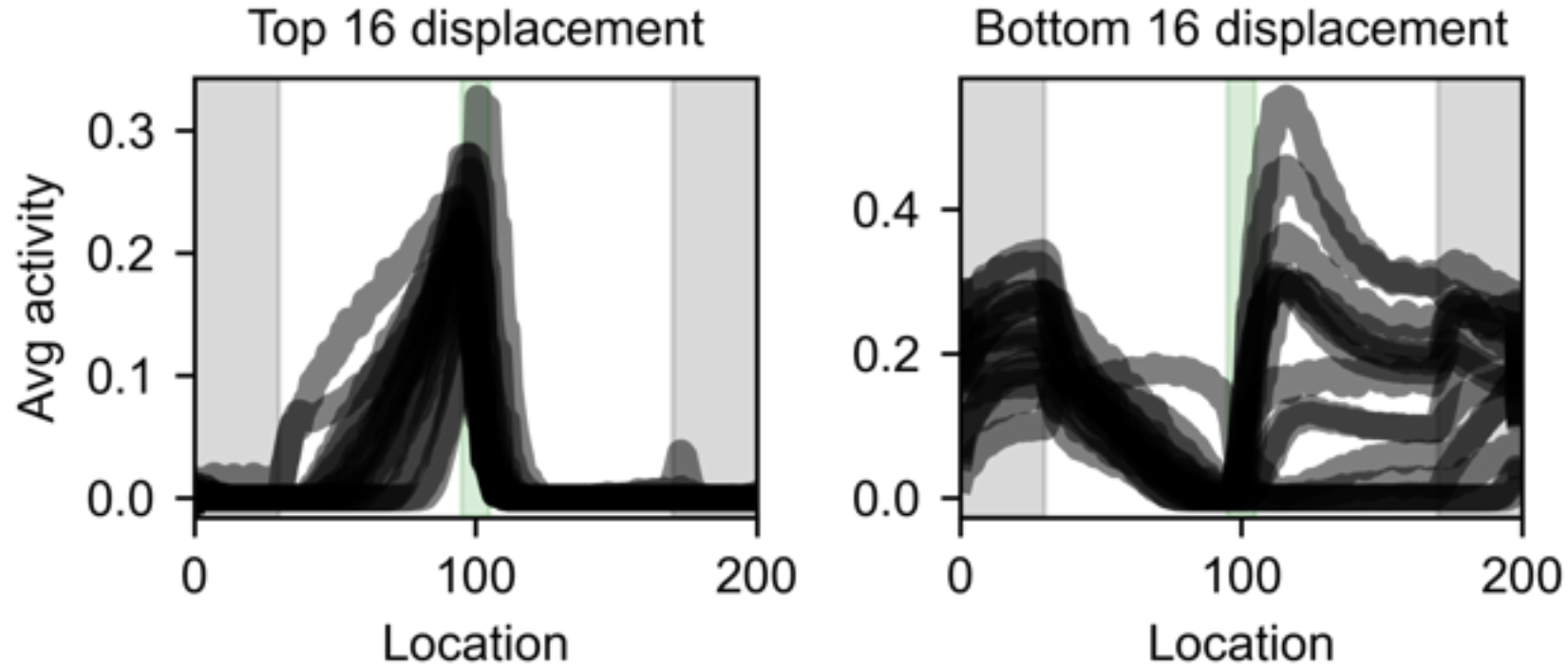

**Supplementary Fig. 12: Activity profiles of neurons with high or low displacement scores.** Trial-averaged activity profiles of neurons with the 16 highest or lowest displacement scores.

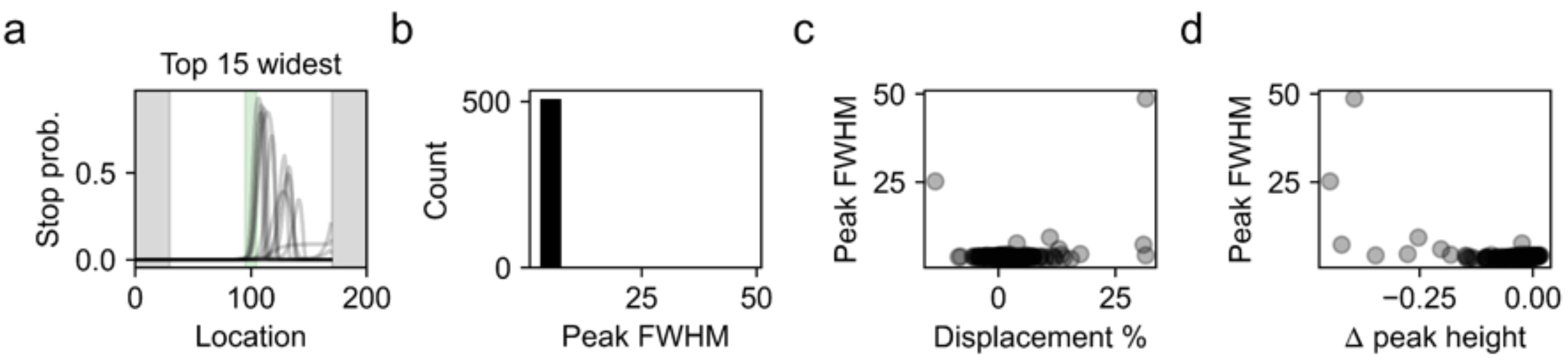

**Supplementary Fig. 13: Widths of stop-probability peaks after single-neuron inactivation. a,** Stop-probability profiles for the 15 neurons with the widest peaks. **b**, Distribution of peak widths, measured as full width at half maximum. **c**, Peak width as a function of displacement score. **d**, Peak width as a function of change in peak stop probability.

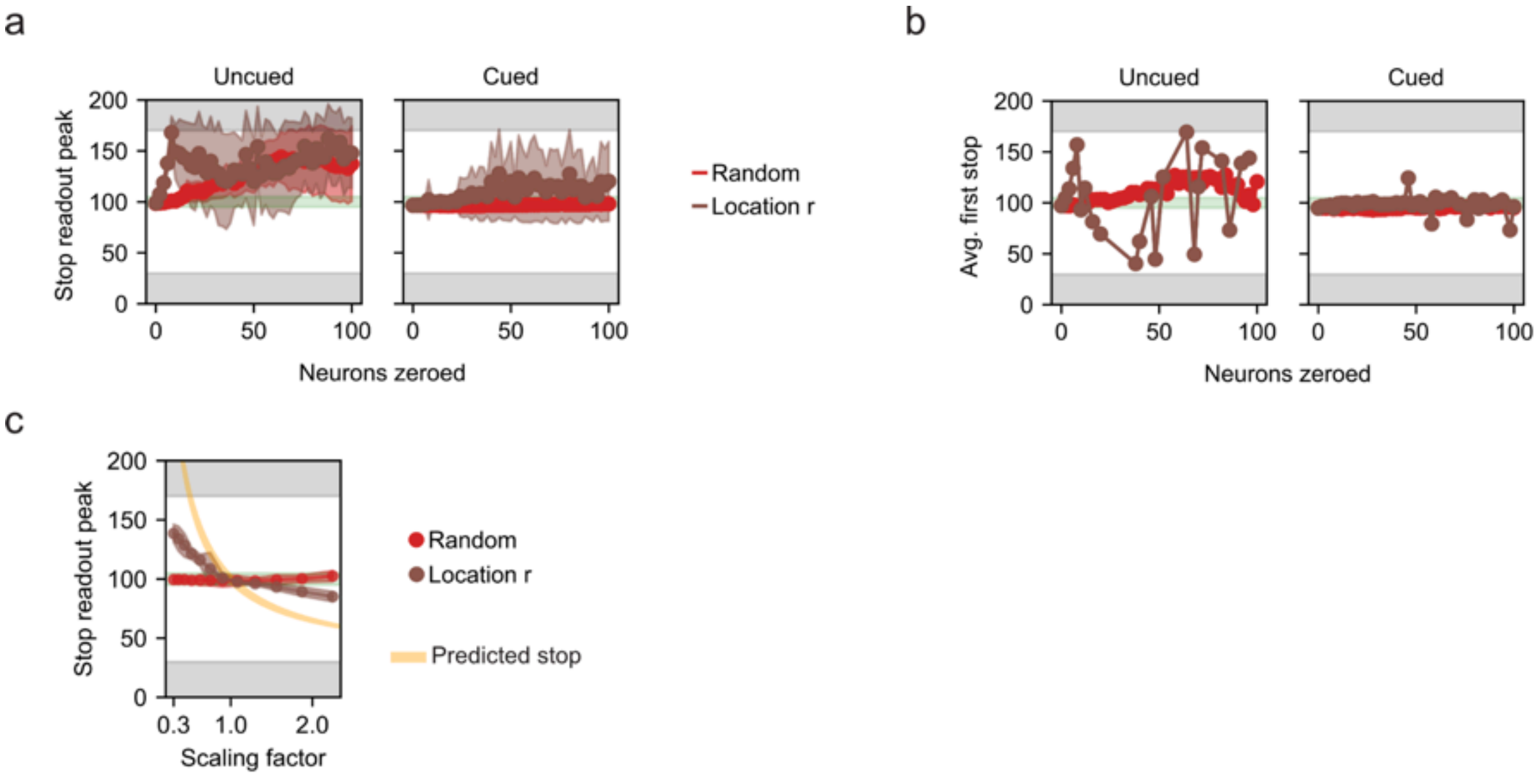

**Supplementary Fig. 14. Effects of perturbing small numbers of neurons on stopping-related metrics. a**, Average location of the peak of the stop-action readout as a function of the number of inactivated neurons. Shaded regions indicate 1 s.d. **b**, as in **a** but showing the average first stop-location. **c**, Average first-stop location after scaling the activity of the eight targeted neurons

across a range of scaling factors as in Extended Data Fig. 5h. Orange region shows the location of the reward zone scaled by the same factor.

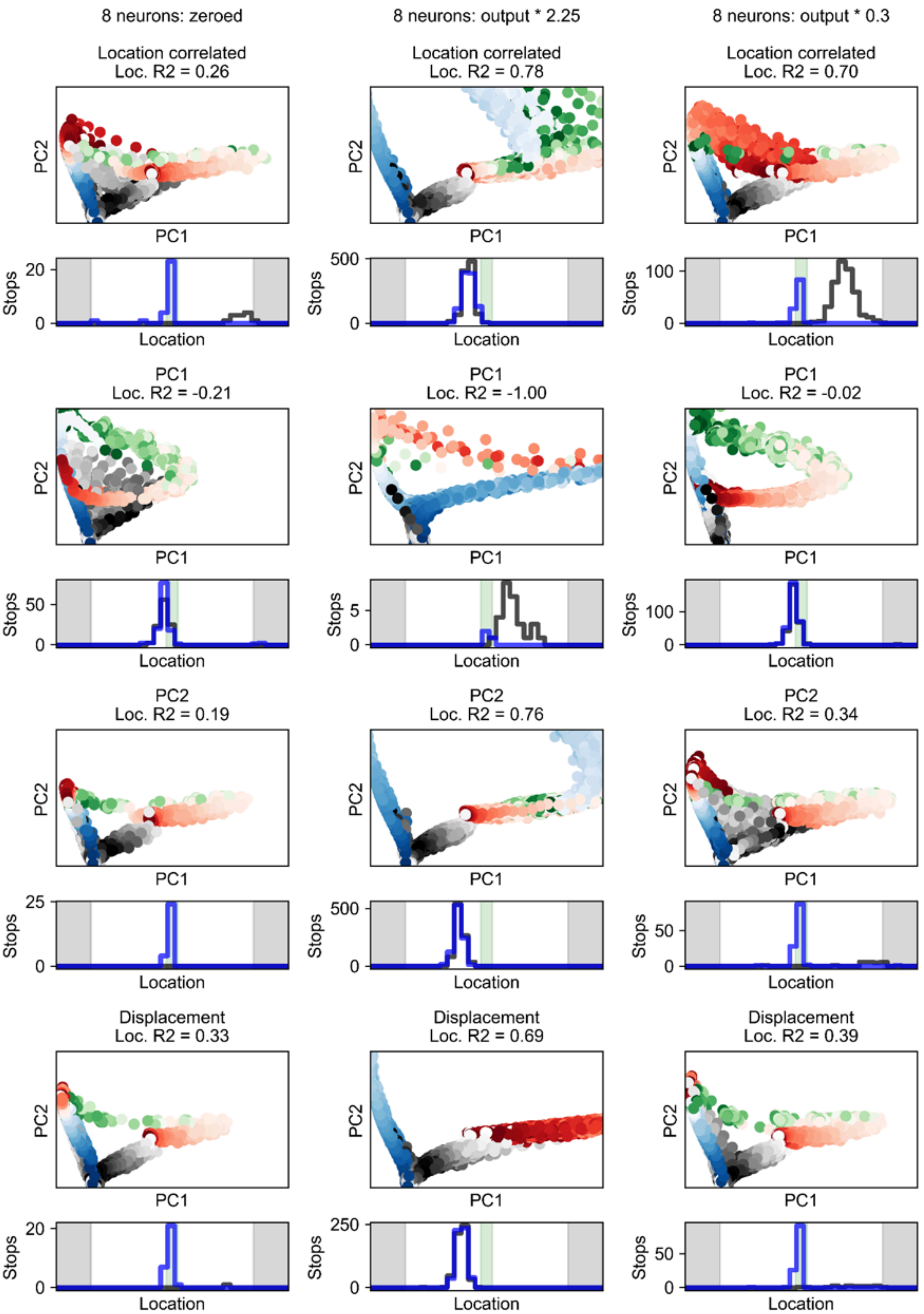

8 neurons: zeroed
8 neurons: output * 2.25
8 neurons: output * 0.3
Location correlated
Loc. R2 = 0.26
Location correlated
Loc. R2 = 0.78
Location correlated
Loc. R2 = 0.70
PC2
PC1
Stops
Location
PC1
Loc. R2 = -0.21
PC1
Loc. R2 = -1.00
PC1
Loc. R2 = -0.02
PC2
Loc. R2 = 0.19
PC2
Loc. R2 = 0.76
PC2
Loc. R2 = 0.34
Displacement
Loc. R2 = 0.33
Displacement
Loc. R2 = 0.69
Displacement
Loc. R2 = 0.39

**Supplementary Fig. 15: Effects of targeted neuronal perturbations on low-dimensional activity and behaviour.** Rows show perturbations of the 8 neurons ranked highest by location correlation, PC1 loading, PC2 loading, or displacement score. Columns show inactivation, output scaling by 2.25 or by 0.30. For each condition, the upper plot shows activity projected onto the control PC1/2 axes with location decoding R2 indicated, and the lower plot shows the corresponding stop distribution. PC axes are fixed to the control ranges.

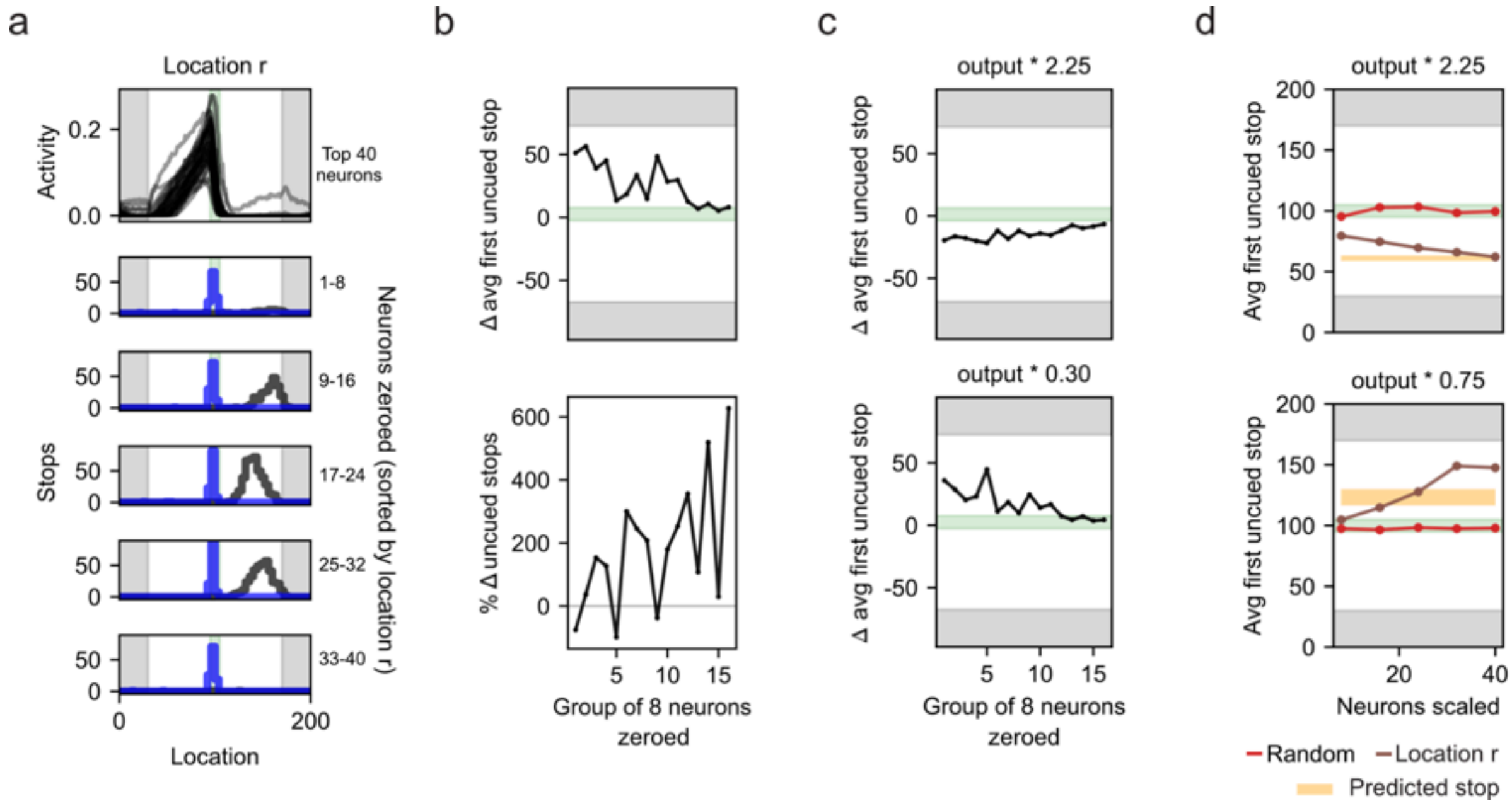

**Supplementary Fig. 16. Effects of manipulating groups of neurons selected by location correlation. a**, Effect of inactivating non-overlapping blocks of eight neurons ordered by descending location-correlation. **b**, Average first uncued stop location (top) and uncued stopping behaviour (bottom) after inactivation of each eight-neuron block. Blocks were constructed by either location-correlation or randomly as a control. **c**, As in **b** but with the activity of neurons in each block scaled up or down rather than inactivated. **d**, Effect of scaling cumulative numbers of neurons by location correlation (brown) or randomly (red). Orange region shows the location of the reward zone scaled by the same factor.

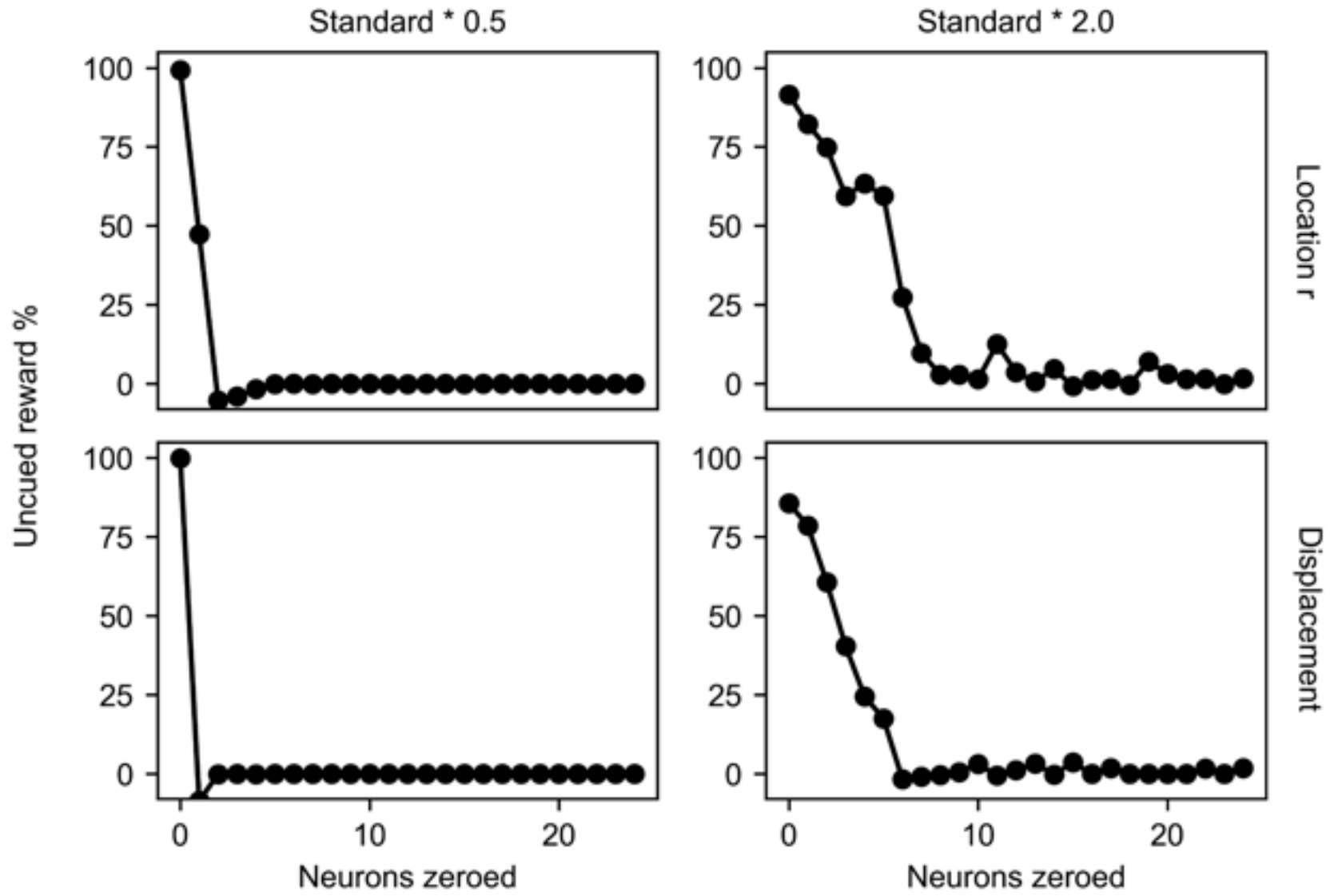

**Supplementary Fig. 17. Effects of targeted neuron inactivation across unseen speed distributions.** Reward as a function of the number of inactivated neurons when neurons were selected by descending positive location correlation score (top row) or descending displacement scores (bottom row). Columns show the speed distribution used for evaluation.

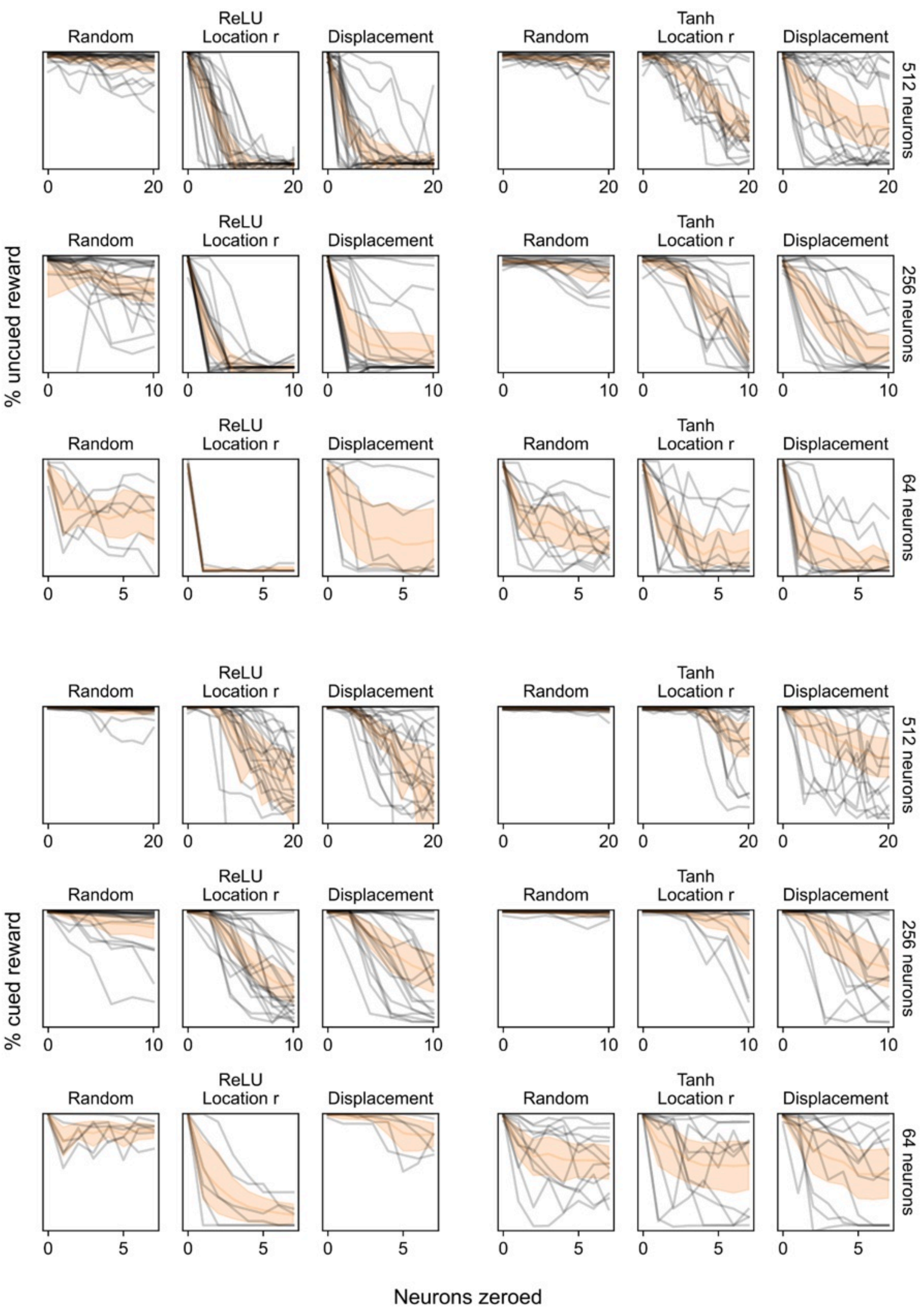

**Supplementary Fig. 18: Effects of inactivating neurons ordered by location, displacement, and**

**randomly, for different model seeds and architectures.** Results were split by nonlinearity and number of neurons. Top three rows shows uncued reward and bottom three shows cued reward.

**Supplementary information related to Fig. 3**

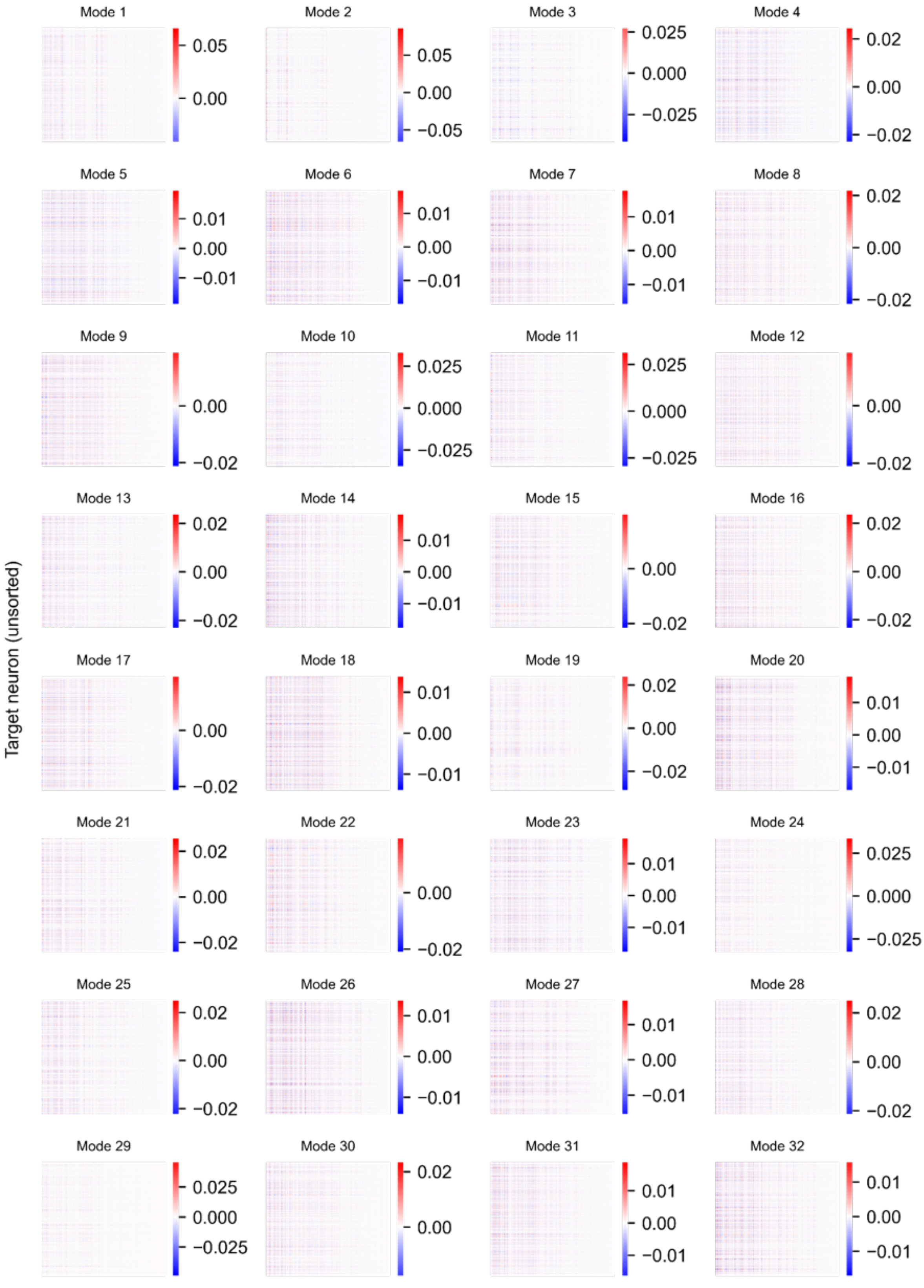

**Supplementary Fig. 19: Singular modes of the recurrent connectivity matrix without displacement-score sorting.** Top 32 singular modes of the recurrent connectivity matrix. Rows and columns show neurons ordered by their arbitrary index in the network rather than by displacement score.

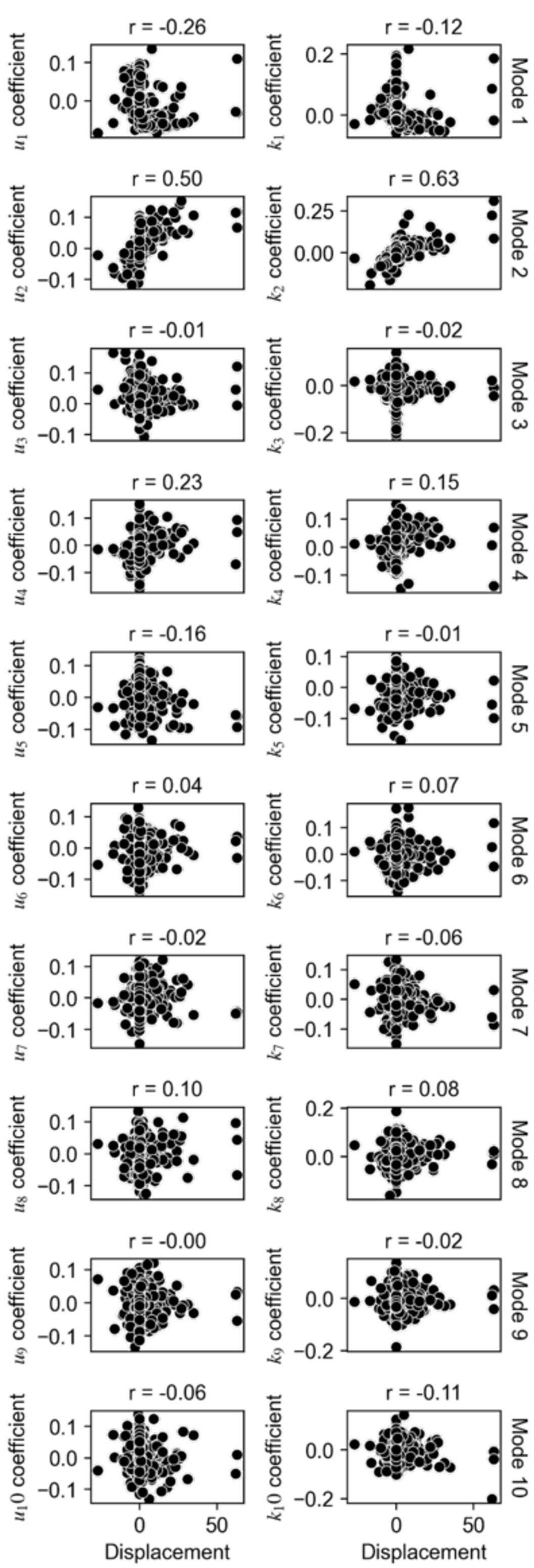

**Supplementary Fig. 20: Relationship between singular-vector coefficients and displacement.** Coefficients of the left and right singular vectors for each recurrent connectivity mode plotted against each neuron's displacement score.

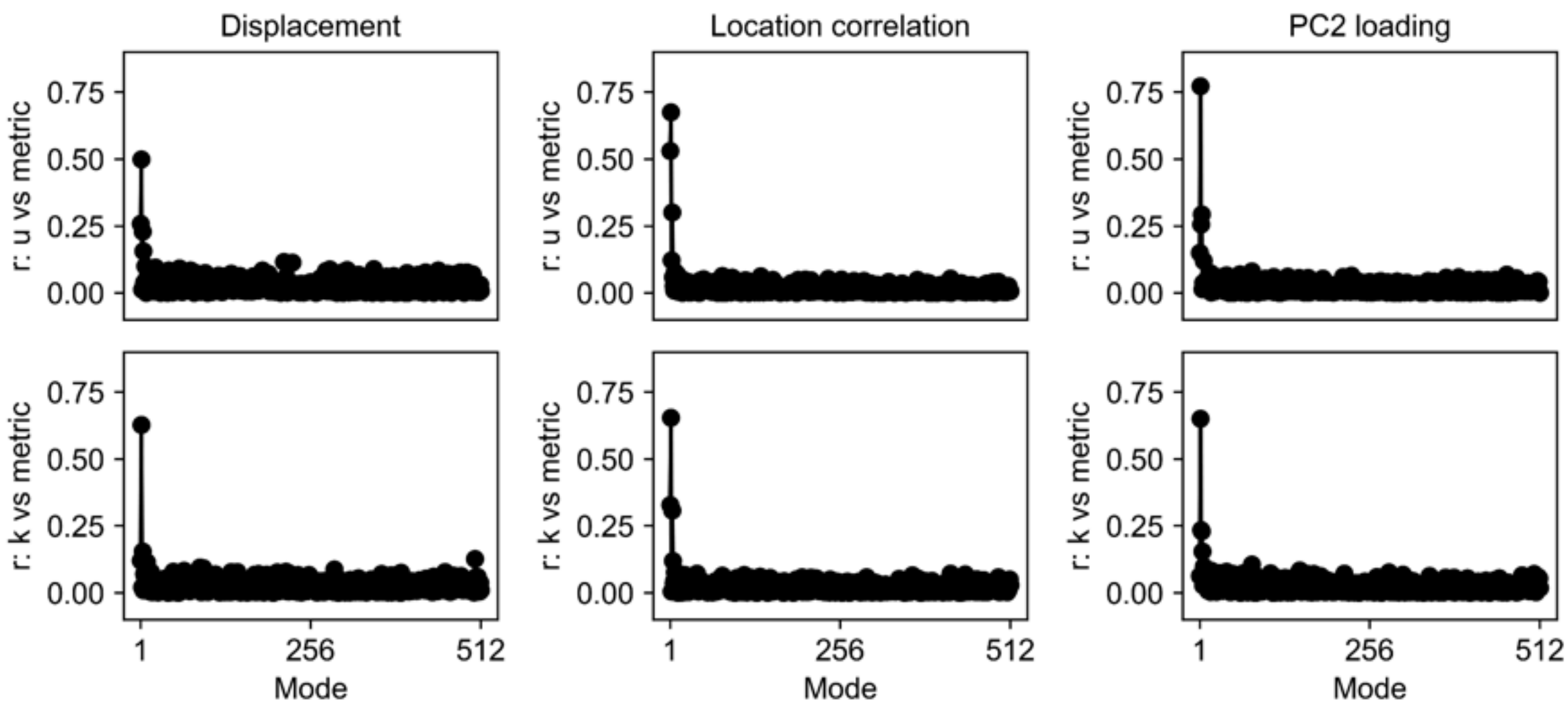

**Supplementary Fig. 21: Summary of relationships between singular-vector coefficients and single-neuron metrics.** Absolute Pearson correlations between left singular-vector coefficients (u, top row) or right singular-vector coefficients (v, bottom row) and displacement score, location-correlation score, or PC2 loading. Modes were ordered by descending singular value.

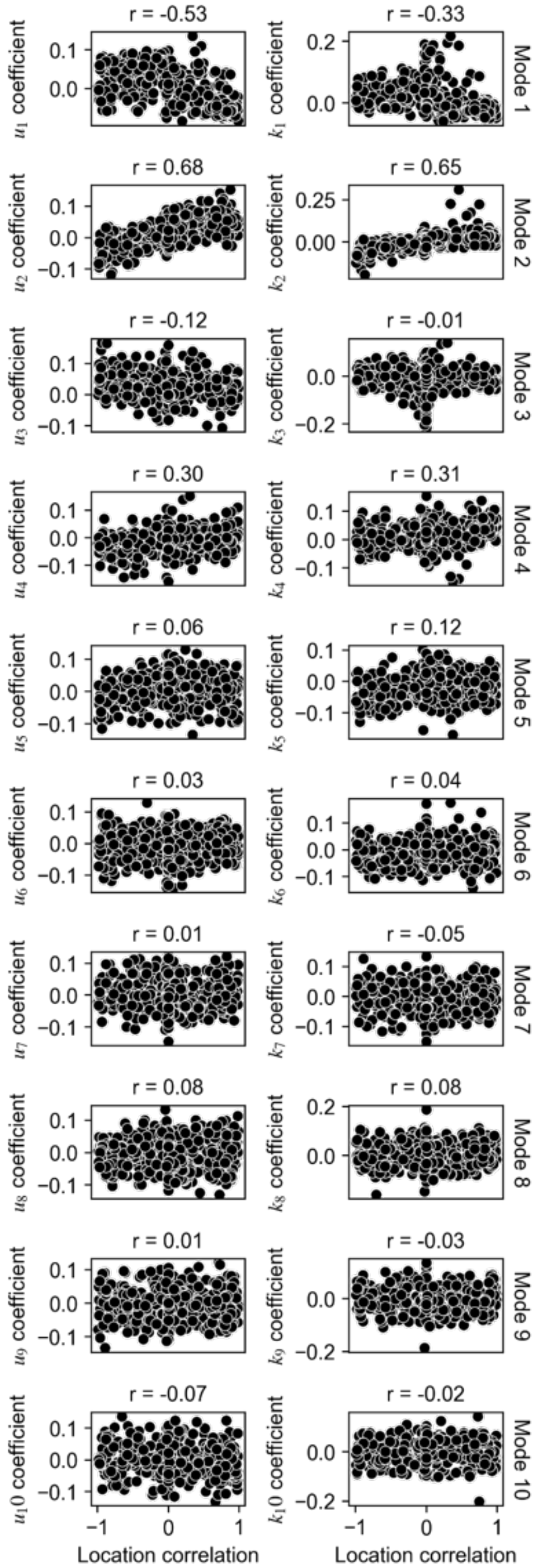

**Supplementary Fig. 22: Relationship between singular-vector coefficients and location correlation.** Coefficients of the left and right singular vectors for each recurrent connectivity mode plotted against each neuron's location-correlation score.

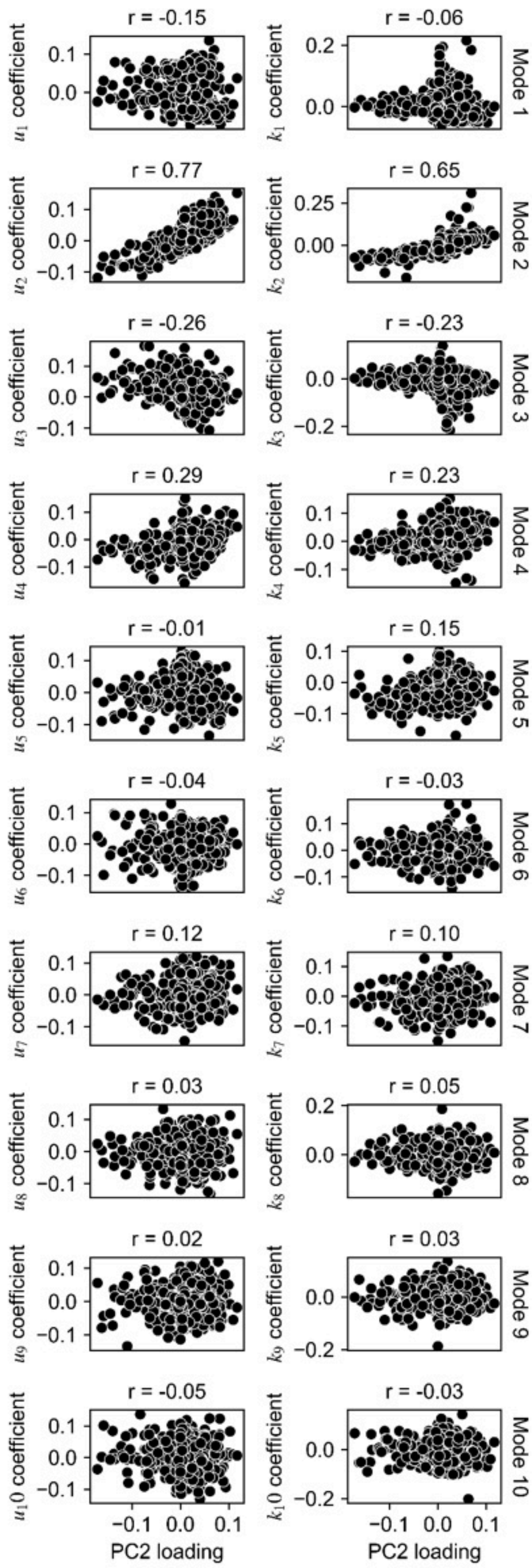

r = -0.15
r = -0.06
Mode 1
$u_1$ coefficient
$k_1$ coefficient
r = 0.77
r = 0.65
Mode 2
$u_2$ coefficient
$k_2$ coefficient
r = -0.26
r = -0.23
Mode 3
$u_3$ coefficient
$k_3$ coefficient
r = 0.29
r = 0.23
Mode 4
$u_4$ coefficient
$k_4$ coefficient
r = -0.01
r = 0.15
Mode 5
$u_5$ coefficient
$k_5$ coefficient
r = -0.04
r = -0.03
Mode 6
$u_6$ coefficient
$k_6$ coefficient
r = 0.12
r = 0.10
Mode 7
$u_7$ coefficient
$k_7$ coefficient
r = 0.03
r = 0.05
Mode 8
$u_8$ coefficient
$k_8$ coefficient
r = 0.02
r = 0.03
Mode 9
$u_9$ coefficient
$k_9$ coefficient
r = -0.05
r = -0.03
Mode 10
$u_10$ coefficient
$k_10$ coefficient
PC2 loading
PC2 loading

**Supplementary Fig. 23: Relationship between singular-vector coefficients and PC2 loading.** Coefficients of the left and right singular vectors for each recurrent connectivity mode plotted against each neuron's PC2 loading.

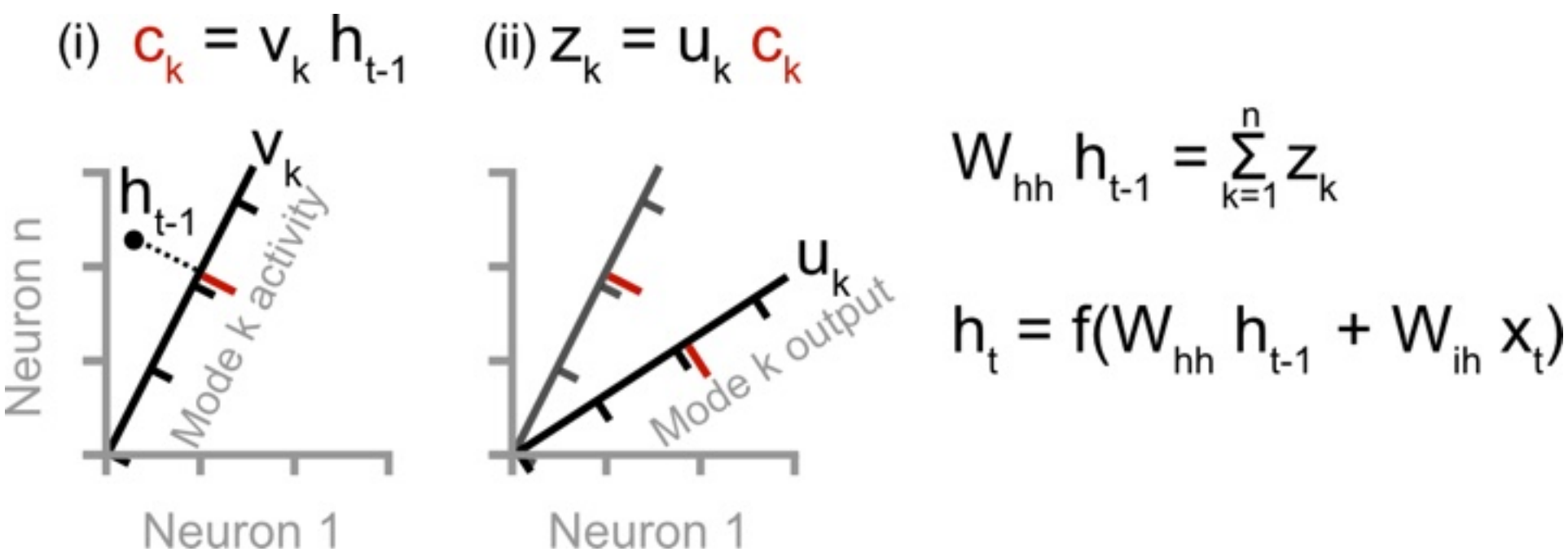

**Supplementary Fig. 24: Relationship between left and right singular vectors and recurrent computation.** Schematic showing how each unit-rank connectivity mode contributes to the RNN update. Each mode is defined by its left and right singular vector, $u_k$ and $v_k$, with $v_k$ scaled by the corresponding singular value. The internal activity of mode k, $c_k$ is computed as the dot product between the previous network state, $h_{t-1}$ and $v_k$. This scalar then scales the output vector $u_k$ generating the mode's contribution, $z_k$ in neuron space. Summing these contributions across all modes is equivalent to applying the recurrent weight matrix $W_{hh}$ $h_{t-1}$.

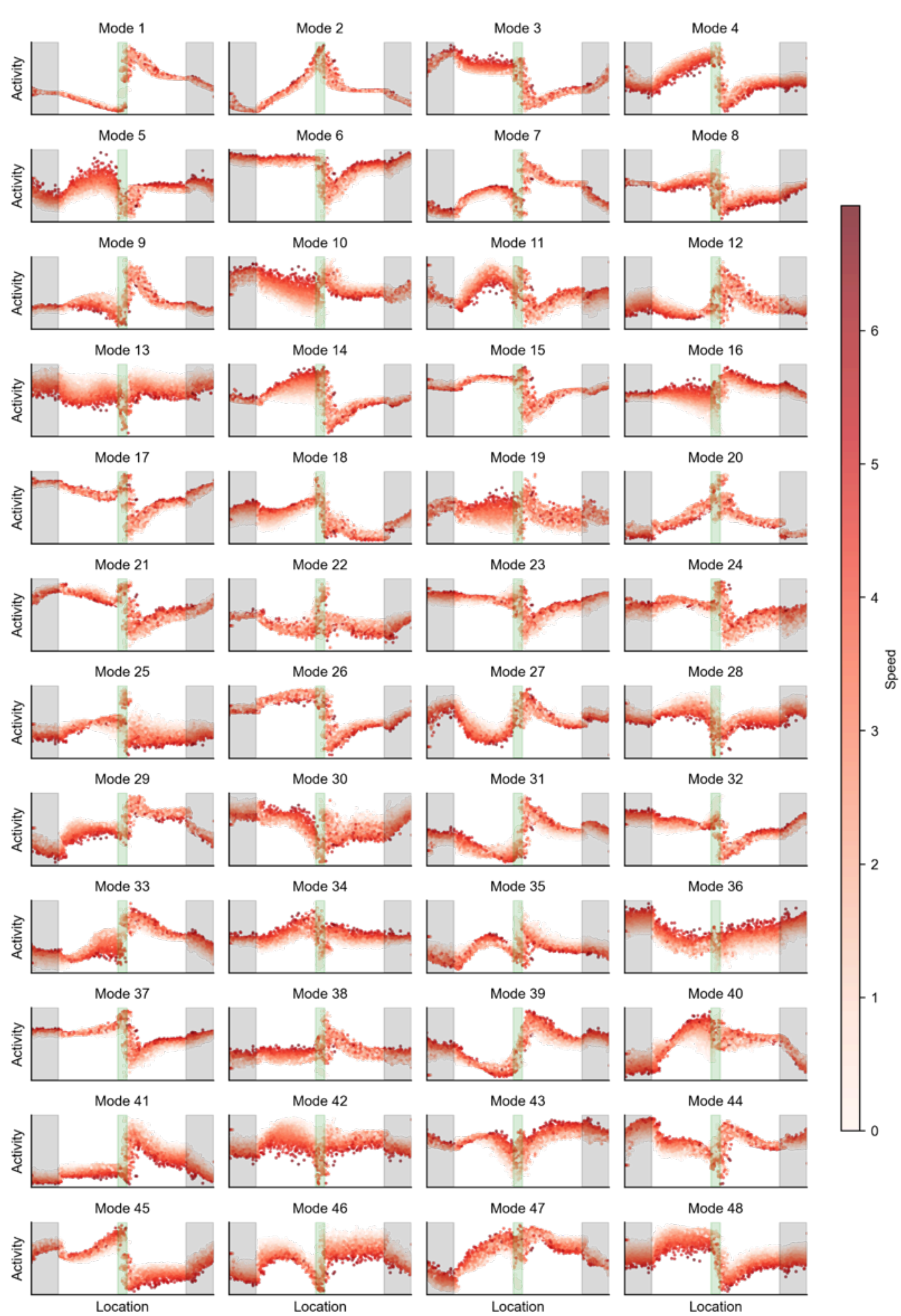

**Supplementary Fig. 25: Network activity projected onto the leading recurrent connectivity modes.** Network activity projected onto the left singular vectors of the top 48 recurrent connectivity modes. Points are coloured by movement speed.

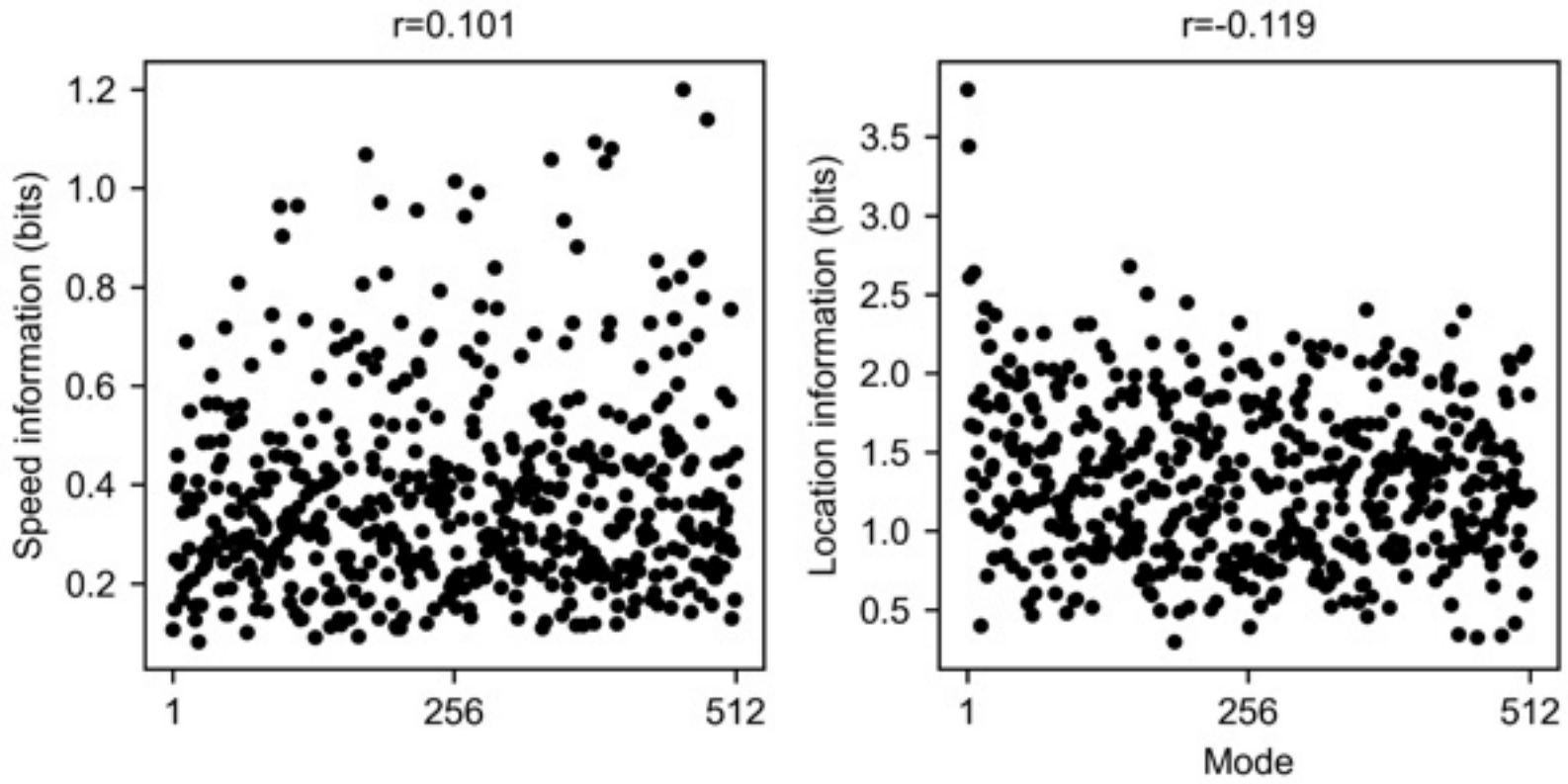

**Supplementary Fig. 26: Information content of activity in each recurrent connectivity mode.** Location and speed information in activity projected onto the left singular vector of each recurrent connectivity mode. Modes are ordered by descending singular value.

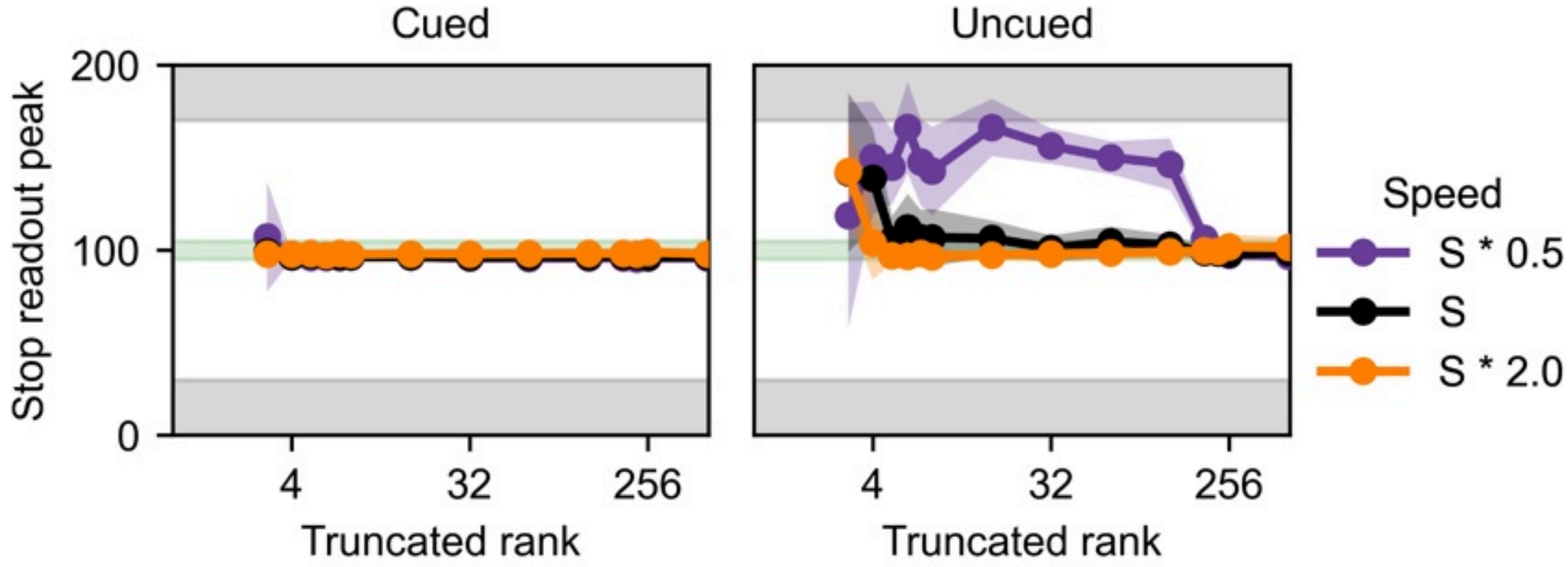

**Supplementary Fig. 27: Effect of low-rank approximations on stop-action readout location.** Location of the stop-action readout peak on cued and uncued trials as a function of recurrent connectivity rank. Colours indicate the speed distribution used for evaluation. Shading indicates 1 s.d. across trials.

**Supplementary information related to Fig. 4**

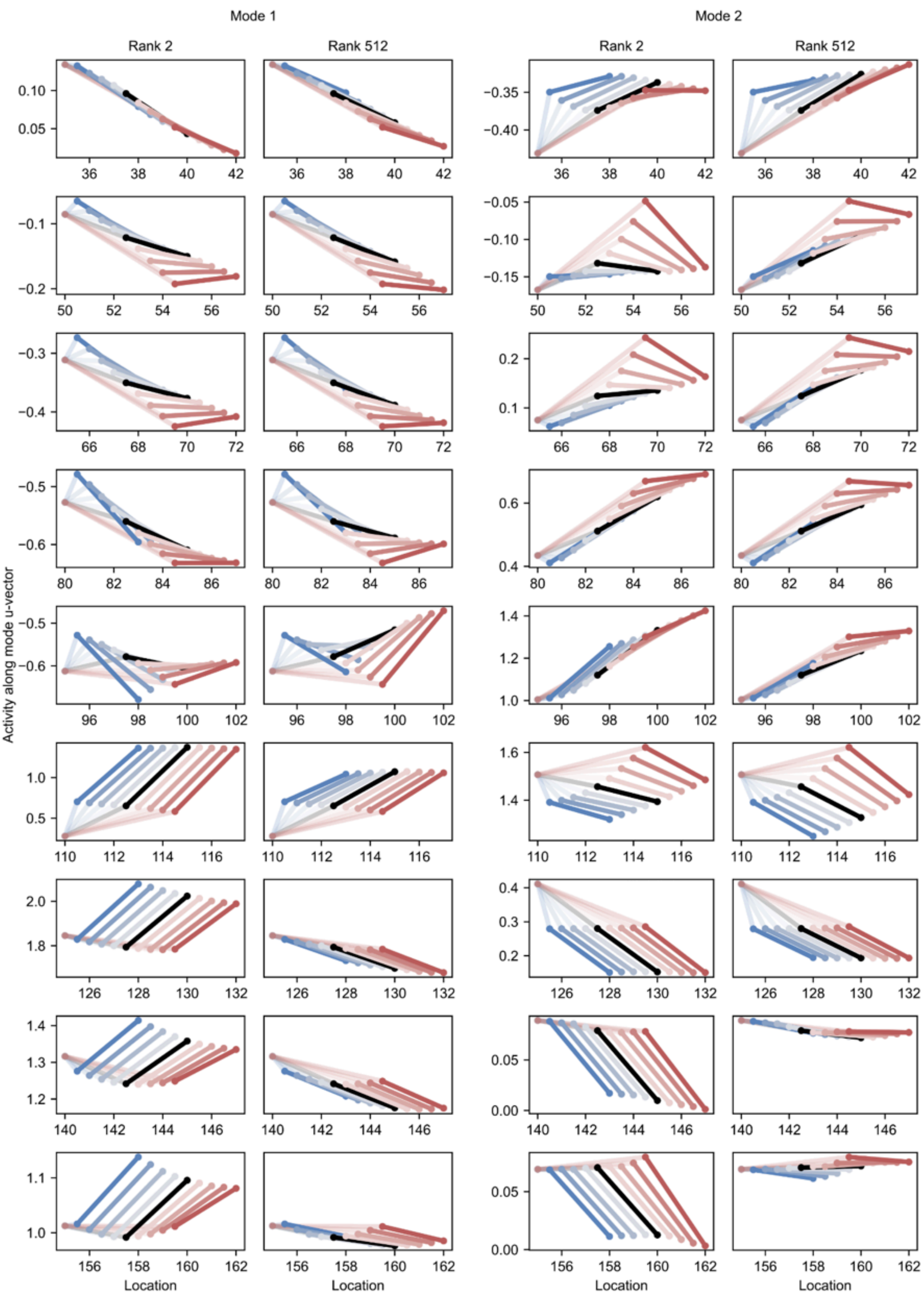

**Supplementary Fig. 28: Effects of low-rank truncation on single-timestep correction of location-estimation errors.** Single-step perturbation analysis from Fig. 4, repeated across multiple sampled track locations. Network states were perturbed with speed inputs from the tested range, then propagated for one RNN timestep using recurrent weights truncated to rank 2 or the full-rank control network (rank 512). Rows show different starting track locations. Columns show activity changes in mode 1 for rank 2 and full-rank networks, followed by activity changes in mode 2 for rank 2 and full-rank networks.

**Supplementary information related to Fig. 5**

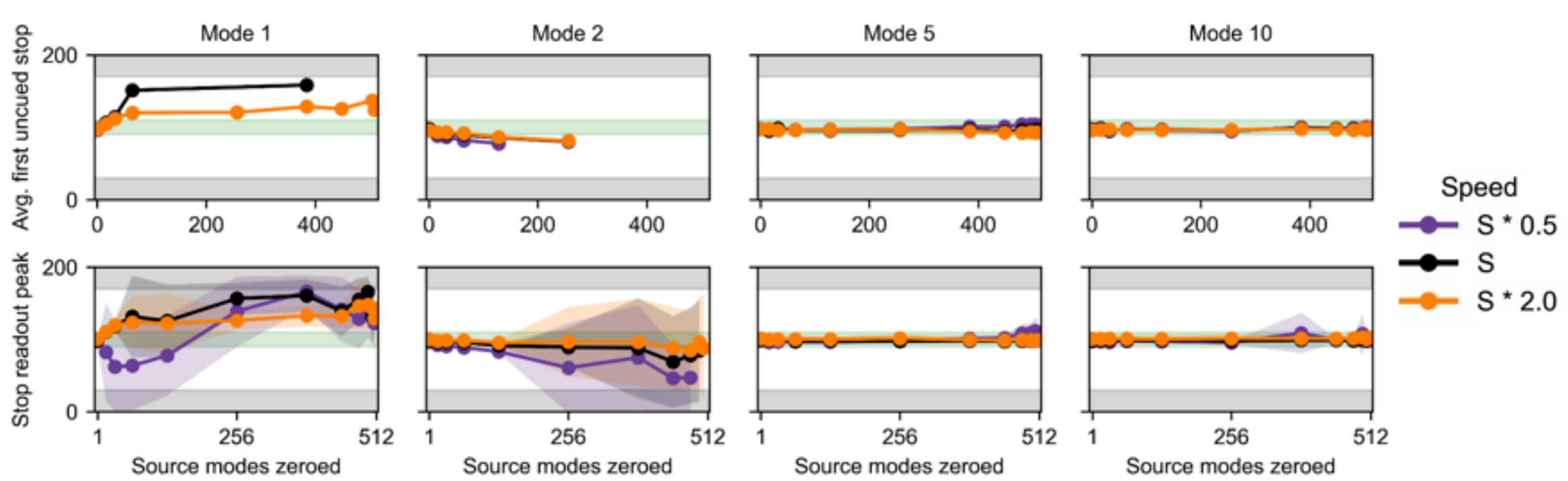

**Supplementary Fig. 29: Effects of removing nonself mode-to-mode inputs on stopping metrics.** Additional metrics for the mode-to-mode perturbations shown in Fig. 5. Columns show target modes 1, 2, 5, and 10. Top row shows average first-stop location on uncued trials, and bottom row shows the location of the stop-action readout peak, each as a function of the number of source modes zeroed. Colours indicate the speed distribution used for evaluation. Shaded regions for the stop readout peak indicate 1 s.d.

**Supplementary information related to Fig. 6**

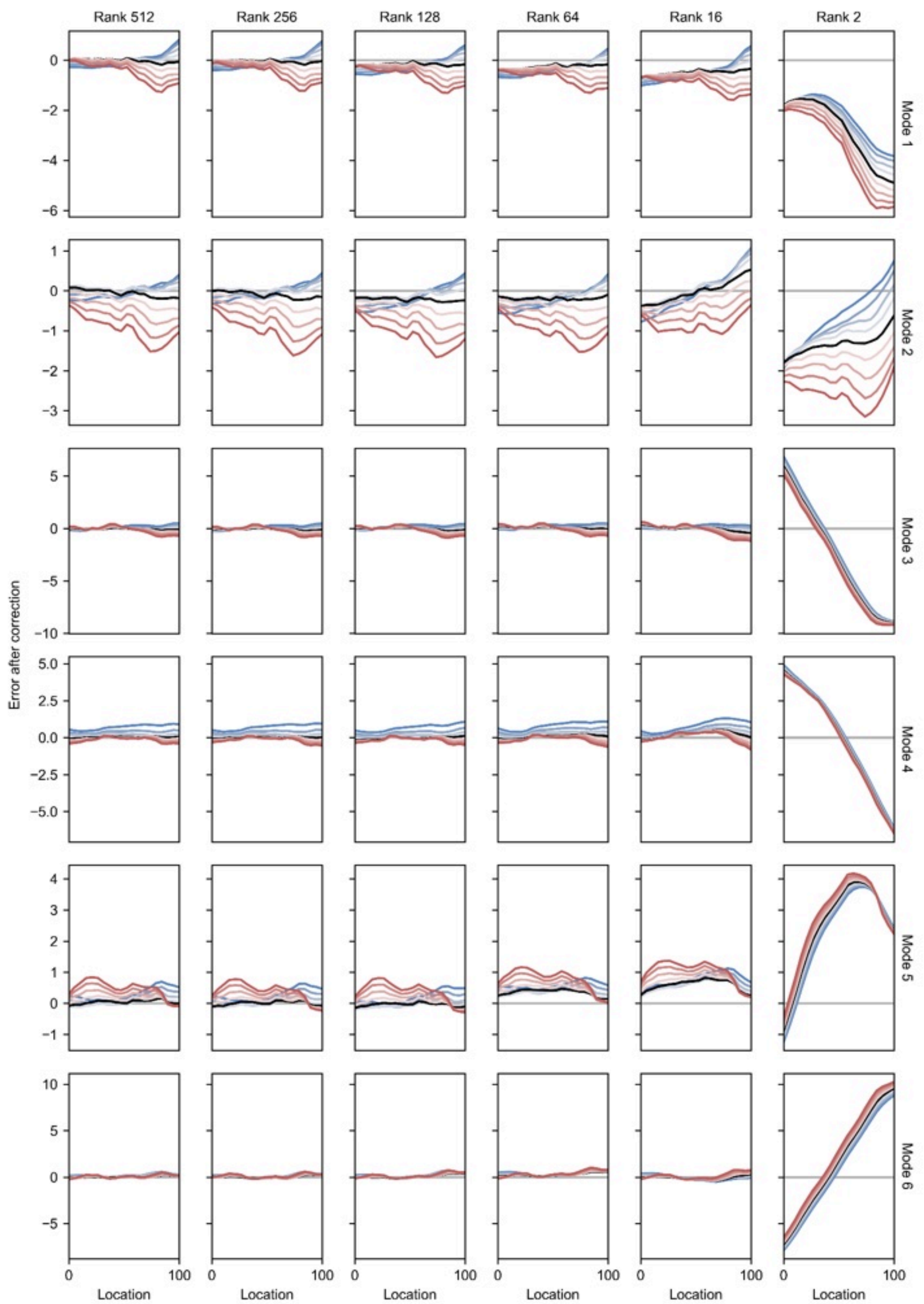

**Supplementary Fig. 30: Location estimation error compensation across low-rank approximations in the 2d task.** Location estimation error in the 2d task was computed as the difference between the expected activity indicated by the dashed line in Fig. 6e and the final network activity. Rows show errors across modes 1 and 2, and each corresponds to a rank-k approximation of the recurrent connectivity matrix. Colour indicates speed, with brighter red values corresponding to higher speed values and blue values corresponding to lower speeds.